%% file: main.tex
\documentclass[a4paper, twoside, 12pt]{report}

\usepackage{setspace}
\usepackage[english]{babel}
\usepackage[utf8x]{inputenc}
\usepackage[T1]{fontenc}

\usepackage[a4paper,top=3cm,bottom=2cm,left=3cm,right=3cm,marginparwidth=1.75cm]{geometry}

\usepackage{amssymb}
\usepackage{amsmath}
\usepackage{graphicx}
\usepackage[colorinlistoftodos]{todonotes}
\usepackage[colorlinks=true, allcolors=blue]{hyperref}
\usepackage{braket}
\usepackage[normalem]{ulem}

\allowdisplaybreaks

\title{Project Title}
\author{}

\begin{document}
\raggedbottom
\input{title/title.tex}

\input{Abstract/Abstract.tex}

\input{Acknowledgements/Acknowledgements.tex}

\tableofcontents
\listoffigures

\input{Introduction/Introduction.tex}

\input{Chapter1/Chapter1}

\input{Chapter2/Chapter2}

\input{Conclusions/Conclusions}

\input{appendix/appendix.tex}

\bibliographystyle{ieeetr}
\bibliography{bibs/sample}

\end{document}

%% file: title/title.tex
\begin{titlepage}

\newcommand{\HRule}{\rule{\linewidth}{0.5mm}} 


 

\center 


\textsc{\huge Master Thesis}\\[1.5cm] 
\textsc{\LARGE Universidad Arturo Prat}\\[0.5cm] 
\textsc{\Large Instituto de Ciencias Exactas y Naturales}\\[1.5cm] 

\makeatletter
\HRule \\[0.4cm]
{ \Huge \bfseries Aspects of Holographic Entanglement Entropy in Cubic Curvature Gravity}\\[0.4cm] 
\HRule \\[1.5cm]
 

 \Large
\emph{Student:}\\
Andrés Argandoña \\[1.2em]
\Large
\emph{Advisor:} \\
Prof. Ignacio Araya \\[1.2em] 
\Large
\emph{Co-Advisor:} \\
Prof. Georgios Anastasiou  \\[1.2em] 
\Large
\emph{Thesis committee:} \\
Prof. Rodrigo Olea, Prof. Cristóbal Corral, Prof. Diego Molina  \\[1.2em] 

\makeatother



{\Large \today}\\[2cm] 

\vfill 

\end{titlepage}

%% file: Abstract/Abstract.tex
\thispagestyle{plain}
\begin{center}
    \Large
    \textbf{Aspects of Holographic
Entanglement Entropy in Cubic
Curvature Gravity}
        
    \vspace{0.4cm}
    \large
    by Andrés Argandoña
        
    \vspace{0.9cm}
    \textbf{Abstract}
\end{center}

In this thesis we explore general aspects of the entanglement entropy (EE) for Conformal Field Theories (CFTs) dual to Cubic Curvature Gravity. We derived a covariant expression for the EE by using a scheme inherited from the bulk renormalization method through extrinsic counterterms. We evaluate this functional in  different entangling regions to calculate CFT data. In particular, we compute the $t_4$ coefficient of the 3-point function of the stress-tensor correlator by considering a deformed entangling region. We observe that there is a discrepancy between the outcomes attained through the employment of the EE functional for minimal and non-minimal splittings. We find  that only the $t_4$ obtained from the non-minimal functional agrees with previous results in the literature that were computed by splitting-independent CFT methods for specific theories such as the massless graviton case.

%% file: Acknowledgements/Acknowledgements.tex
\begin{titlepage}

\begin{center}
    \textsc{\LARGE Acknowledgment}\\
\end{center}

I would like to thank all those who in  the duality personal/academic life  (with lots of overlaps) have made this thesis possible. I would like to start by thanking my advisor Ignacio Araya and co-advisor Giorgos Anastasiou, for their support throughout the completion of this thesis. They were always willing to answer all of my questions. I also thank them for all the advice they have given me on academic, personal and other matters.

Another important participant in the development of this thesis (and in the process of my academic formation) is Rodrigo Olea. I thank him for welcoming me into his research group when I arrived to Chile and for always encouraging me to pursue an academic career even in times of crisis. 

I am deeply grateful to the physics department at ICEN, especially Cristóbal, Diego, José, Omar, Nelson Merino, and Patricio, for their invaluable assistance and support. To my classmates Carlos, Luis, Mayrim and Nelson Sánchez, thank you for the companionship and the experiences shared during my stay in Iquique.

These acknowledgements would not be complete without mentioning all the people in Peru whose great support I have always felt. I would like to thank Cristian, Fernando, Luis and Martin from the theoretical physics group at UNMSM. The discussions on physics and politics have always been fruitful. A big thank to my closest friends David, Jessica, Julio, Lale, Malu, Majo, Sandra and Sebastian for being there since before I decided to pursue a career in theoretical physics and for all your support along the way.

Special thanks to Teofilo Vargas, for introducing me to the world of theoretical physics and giving me the first glimpses into general relativity and field theory.

Last but not least, an immeasurable thanks to my family. To my parents, Gladys and Rolando, for always giving me the freedom to decide, to make mistakes and to learn. It is comforting to know that you will always be there for me no matter what. Also to my brother Omar and my sister Ara for teaching me the importance of dedication and concentration to achieve a goal.

Thanks to all of you, a part of this work is yours.

\end{titlepage}

%% file: Introduction/Introduction.tex
\chapter{Introduction}

Just as the study of  black body radiation led us to the formulation of quantum mechanics, the study of the statistical properties of black holes (BH) could help us to elucidate aspects of a theory of quantum gravity. These are objects where the effect of gravity is very intense and present interesting characteristics such as temperature and  emission of radiation \cite{Hawking:1974rv,Hawking:1975vcx}. They also have an associated entropy that is proportional to the area of its event horizon \cite{Hawking:1974rv, Bekenstein:1972tm,Bekenstein:1973ur,Bekenstein:1974ax}. The fact that the BH entropy  serves as an upper bound to the entropy that can be contained in a given region of spacetime \cite{Bekenstein:1980jp} \footnote{The formulation of this statement, called the Bekenstein bound, only works under certain assumptions such as weak gravity, spherical symmetry, etc. These conditions are violated by real physical systems and therefore a better formulation of a strong bound on the information content of spacetime regions was sought. Its present formulation is described by the covariant entropy bound \cite{Bousso:1999xy}.} inspired the formulation of the holographic principle \cite{tHooft:1993dmi,Susskind:1994vu}. It tells us that the fundamental degrees of freedom of a physical system in a given volume are encoded by the area that delimits that region.  In Ref. \cite{Susskind:1994vu} it was explained how this idea could be implemented in the context of string theory and finally in Refs. \cite{Maldacena:1997re, Witten:1998qj,Gubser:1998bc,Aharony:1999ti} the first concrete realization of the holographic principle was obtained in the famous anti-de Sitter/conformal field theory (AdS/CFT) correspondence\footnote{Also known as gauge/gravity duality. }. It states that in the limit of large $N$, a $d$-dimensionsal $SU(N)$ gauge theory with conformal symmetry is equivalently described by a superstring theory in a $(d+1)$-dimensional background of anti-de Sitter type (for more details go to section (\ref{HEE})). 

Since its formulation, the correspondence  has been an important step towards the understanding of the relationship between string theory, quantum field theory and general relativity. Moreover, it encounters applications in the context of  QCD \cite{Kruczenski:2003uq, Sakai:2005yt,Erlich:2005qh}, nuclear physics \cite{Mateos:2007ay}, non-equilibrium physics \cite{Keranen:2014lna}, quantum information \cite{Chen:2021lnq} and condensed matter physics \cite{Hartnoll:2007ai,Hartnoll:2007ih}. One of the most important claims made by the conjecture is that strongly coupled gauge theories can be studied using classical gravity. This property, has been exploited in many situation as a computational tool to translate difficult problems into simpler ones.  For a \textit{real world} example we can see that it turns out that one prediction of AdS/CFT is indeed close to the experimental results of the real quark–gluon plasma \cite{Kovtun:2004de}.

 On the other side of the story we have quantum entanglement, a property of many-body quantum systems (and quantum field theories) that will be described in great detail in sections (\ref{EE in QM}) and (\ref{EE in QFT}). The entanglement entropy, which is a specific measure of it, has been utilized as an order parameter to describe phase transitions and critical points in various phases of matter\cite{Kitaev:2005dm,Klebanov:2007ws}. In the 1980s and 1990s, several authors notice that the leading divergent term of the EE of a region of interest is proportional to its surface area \cite{Bombelli:1986rw, Srednicki:1993im}. As a result of this, it was proposed that the entropy of black holes might arise, at least in part, from the entanglement of quantum fields across the horizon \cite{tHooft:1984kcu, Susskind:1993ws, Callan:1994py}. Furthermore, a remarkable success on this topic was achieved in \cite{Strominger:1996sh} where, in the context of string theory, a microscopic derivation of the BH entropy for BPS black holes was provided.

Despite the many similarities, a concrete gravitational interpretation of entanglement entropy had not yet been formulated in the early 2000s. It was not until the second half of that decade, that S. Ryu and T. Takayanagui presented a proposal, based on holographic arguments, for this missing link \cite{Ryu:2006bv,Ryu:2006ef}. The conjecture, simply and elegantly, states that the EE for a CFT that admits a holographic description can be computed by finding a minimal area of a codimension-two hypersurface in the dual gravitational theory. From its formulation to the present day, much has been written on the subject. Ideas on how bulk geometry can arise from entanglement have been developed in Refs.\cite{VanRaamsdonk:2009ar, VanRaamsdonk:2010pw}, and beyond that it has been used as an easy way to obtain the EE for situations that would otherwise be intractable.

In recent years, higher curvature gravity theories (HCG) have been studied in the context of the AdS/CFT correspondence. Usually, they  appear as stringy or quantum correction in effective (super)gravity theories. A property of holographic CFTs that have Einstein gravity as their dual is that their central charges are equal $a=c$ \cite{Henningson:1998gx}. If we want to describe more general field theories such as strongly coupled CFTs with $a\neq c$, higher curvature terms have to be incorporated in the gravitational side of the description. This correction to the Einstein-Hilbert (EH) case have had non trivial consequences in the description of holographic fluids \cite{Brigante:2007nu} and in conformal collider physics \cite{Hofman:2008ar}. In addition, due to the limitations of working in theories that are dual to EH, the computation of EE has also been explored in more spicy CFTs dual to higher order curvature theories. In this thesis we will pay special attention to the case of cubic curvature gravity theories (CCG). Recently, a description for the EE of the CFT dual to these cubic theories was found  \cite{Caceres:2020jrf, Bueno:2020uxs}. In this project, we will see their description and consequences, as well as provide new results on their characterization. 

In the next chapter we present an overview of all the topics mentioned so far in this introduction. We start by giving a definition of entanglement in quantum mechanics (section (\ref{EE in QM})) and then we measure it using the bedrock concept of statistical mechanics: entropy. In section (\ref{EE in QFT}),  we  extend these concepts to the domain of quantum field theory by providing the appropriate technology to obtain the EE in such scenarios. After that, we will give a brief introduction to the AdS/CFT correspondence that will be used to demonstrate the Ryu-Takayanagi (RT) formula (see section \ref{HEE}). The EE carries universal information related to fundamental properties of the CFT. In order for this information to be extracted a renormalization procedure should be implemented. For this purpose, we will use a scheme inherited from the renormalized bulk action based on the addition of extrinsic curvature counterterms \cite{Olea:2005gb,Olea:2006vd} (Kounterterms (\ref{Kounterterms section})).

In chapter \ref{chapter 3}, we will start by determining  the renormalized CCG action using the Kounterterm method (section (\ref{Fixing the coupling in CCG})). In section (\ref{Renormalized HEE in CCG}) we obtain an expression for the renormalized EE in CCG.  On top of that, we give some remarks about the splitting problem \cite{Miao:2015iba,Miao:2014nxa} that arises in the determination of the EE. This is consequence of the several ways to do the regularization process (that give us different final results) of the conic singularities present in the computation of the EE in CCG\footnote{We will talk much more about this later.}. Then, we will test our proposal to  recover the universal part of the EE for spherical (section (\ref{First test: Hypersphere})) and cylindrical (section (\ref{Second test: Cylinder})) entangling regions. Finally, we consider harmonic deformations of spherical entangling surfaces to obtain the $C_{T}$ charge of the two-point function of the CFT stress-tensor, in terms of the lowest order term in the deformation of the entanglement density on the deformed region (section (\ref{Deformed sphere I})). We also explore the computation of higher correlators of the stress tensor from higher order terms in the deformation parameter (section (\ref{Deformed sphere II})).

In the last chapter we present the conclusions and final remarks. We will also discuss some open questions that could lead to future research work.

%% file: Chapter1/Chapter1.tex
\chapter{Holography, Entanglement Entropy and Renormalization}

\section{Entanglement Entropy in Quantum Mechanics}\label{EE in QM}

A notable aspect of quantum mechanics is the existence of entanglement. In classical mechanics, a system can be studied by dividing it into non-interacting individual parts and then reassembling them to describe the whole system. However, this intuition breaks down in quantum mechanics due to the existence  of the so-called entangled states which cannot be factored as a product of states in its local constituents. In that sense, having knowledge of the individual parts of a system does not necessarily allow to reconstruct the whole system \cite{Schrodinger:1935zz}.

To elucidate this property of quantum systems, let us consider a simple version of the Einstein-Podolsky-Rosen Paradox \cite{Einstein:1935rr} (EPR) proposed by Bohm \cite{bohm1951quantum} (Also known as EPRB). Suppose that a neutral pi meson decay into an electron and a positron\footnote{Bohm's original setup involved the dissociation of a diatomic molecule, where the total spin angular momentum is initially zero and remains so throughout the process. An example of such a process is the dissociation of an excited hydrogen molecule into a pair of hydrogen atoms without altering the initially zero total angular momentum.}:

\begin{equation}
    \pi^0\longrightarrow e^-+e^+. \label{decay pi meson}
\end{equation}

The neutral pion has spin zero so, by  conservation of angular momentum, the electron and positron should be in the singlet spin configuration:

\begin{equation}
    \ket{\Psi}=\frac{1}{\sqrt{2}}\ket{\uparrow}_{e^-}\otimes\ket{ \downarrow}_{e^+}-\frac{1}{\sqrt{2}}\ket{\downarrow}_{e^-}\otimes\ket{\uparrow}_{e^+}=\frac{1}{\sqrt{2}}\ket{\uparrow \downarrow}-\frac{1}{\sqrt{2}}\ket{\downarrow\uparrow}.\label{two spin}
\end{equation} 

If one measures the electron spin and finds that it has spin up, then the total state collapses into the state  $\ket{\uparrow \downarrow}$ , so one can automatically determine that the positron has spin down, and vice versa. This analysis should also work even if the measurement is taken when the electron and positron are separated by a very long distance. This violates the so-called locality condition which states that information cannot be transmitted faster than the speed of light. In fact, the EPR (EPRB) paradox was presented as an argument that quantum mechanics, was \textit{incomplete} as it was stated an therefore some sort of \textit{hidden variables} should be consider to complete the theory. We know now from the Bell's theorem \cite{Bell:1964kc} and its experimental test \cite{Aspect:1981nv,Handsteiner:2016ulx} that  any local hidden variable theory is not viable. It appears that nonlocality is a feature of \textit{entangled systems}.

Different measures of quantum entanglement have been proposed in order to apply them in extended quantum systems with many degrees of freedom (for an overview see \cite{Plenio:2007zz}). One of these measures in which we will be particularly interested is \textit{entanglement entropy}. To understand how entropy, which is a fundamental concept in statistical mechanics and thermodynamics, is deeply intertwined with entanglement, it is necessary to review some concepts.

For systems with many degrees of freedom it will not be always possible to characterize the whole system with only one quantum state $\ket{\Psi}$. Instead, we will have that a fraction of the members with relative population $p_1$ are
characterized by $\ket{\Psi_1}$, some other fraction with relative population $p_2$, by $\ket{\Psi_2}$ , and so on. In this case the system is said to be in a \textit{mixed state}. The relative populations, also known as probabilistic weights, should satisfy:

\begin{equation}
    \sum_i^{N}p_i=1. \label{sum of probability}
\end{equation}

If $p_k=1$ and $p_{i\neq k}=0$ then the system is in a \textit{pure state} characterize by  $\ket{\Psi_k}$. Now, suppose we want to measure a mixed state of a certain observable $\hat{O}$. If we carry out a lot of measurements we should be able to give an average :

\begin{equation}
    \langle \hat{O}\rangle=\sum_{i}p_i\langle \hat{O_i}\rangle=\sum_{i}p_i\bra{\Psi_i}\hat{O}\ket{\Psi_i}=\sum_{\phi}\sum_{i}p_i|\langle \phi| \Psi_i\rangle|^2\phi.\label{average}
\end{equation}

Notice that $\langle \hat{O_i}\rangle$ is the  usual quantum mechanical expectation value of $\hat{O}$ taken with respect to $\ket{\Psi_i}$. Also, we  introduced the eigenstate of the observable, that is $\hat{O}\ket{\phi}=\phi\ket{\phi}$. After doing so, we encounter two probabilistic concepts in the computation of the average of an observable; the first one, related to the classical probability $p_i$ that tells us the probability for finding in the ensemble a  state characterized by $\ket{\Psi_i}$; the other one, $|\langle \phi|\Psi_i\rangle|^2$ for the quantum-mechanical probability for state $\ket{\Psi_i}$ to be found in an eigenstate $\ket{\phi}$.

We can rewrite  Eq.\eqref{average} using a more general basis $\{\ket{\psi}\}$,

\begin{align}
    \langle \hat{O}\rangle&=\sum_{\psi}\sum_{\psi'}\sum_{i}p_i\langle\Psi_i|\psi'\rangle \bra{\psi'}\hat{O}\ket{\psi}\langle\psi|\Psi_i\rangle\nonumber\\
    &=\sum_{\psi}\sum_{\psi'}\langle\psi|\bigg(\sum_{i}p_i\ket{\Psi_i}\bra{\Psi_i}\bigg)|\psi'\rangle \bra{\psi'}\hat{O}\ket{\psi}\label{average and density matrix}\\
    &=\sum_{\psi}\langle\psi|\hat{\rho} \hat{O}\ket{\psi}=tr(\hat{\rho} \hat{O}) ,\nonumber
\end{align}

where we  have defined  the \textit{density matrix}:

\begin{equation}
    \hat{\rho}:=\sum_{i}p_i\ket{\Psi_i}\bra{\Psi_i}.\label{density matrix}
\end{equation}

We should notice that the trace is independent of the representation, Eq.\eqref{average and density matrix} can be evaluated using any convenient basis. 
Also, for a pure state Eq.\eqref{density matrix} takes the form $\hat{\rho}=\ket{\Psi}\bra{\Psi}$.

There are two important properties of the density matrix that we should take into account. First, it is an Hermitian operator, as is evident from its definition. Secondly, it satisfies the normalization condition, that is $tr(\rho)=1$. As an extra property one can check that the density matrix is a positive semi-definite operator, that is $\displaystyle{\hat{x}\hat{\rho}\hat{x}^T\geq 0}$ for all $\hat{x}\in \mathbb{R}^n$.

With these ingredients we can define the von Neumann entropy for a given density matrix $\hat{\rho}$:
\begin{equation}
    S(\hat{\rho})=-tr(\hat{\rho} \ln{\hat{\rho}}) .\label{von Neummann}
\end{equation}
If we choose a basis where the density matrix is diagonal, the von Neumman entropy could be expressed as $S(\hat{\rho})=-\sum_{i}p_iln(p_i)$. It can be easily observed that $S(\hat{\rho})\geq 0$ since $p_i$ is bounded by $0$ and $1$. The case where $S(\hat{\rho})= 0$ occurs if and only if $\rho$ is the density matrix of a pure state.

Another characteristic of the von Neumann entropy  is that it takes its maximum value when the probability distribution is constant. To show this, we should extremize Eq.\eqref{von Neummann} subject to the constraint $\sum_ip_i=1$.
\begin{align}
    \frac{\partial }{\partial p_j}\left( -\sum_{i}p_i\ln{(p_i)}-\lambda\left(\sum_ip_i-1\right)\right)=-\ln{(p_j)}-1-\lambda=0 .\label{maximal Entropy}
\end{align}

Therefore $p_j=e^{-(1+\lambda)}=cnst$ and from the constraint we conclude that $p_j=\frac{1}{N}$. After obtaining these results we can conclude that the von Neumann entropy is used as a measure of mixedness of a  quantum system.

Now that we have established the difference between a mixed state and a pure state, let's see how it relates to entanglement. To do so, consider a lattice model in a pure ground state $\ket{\Psi}$ in Hilbert space $\mathcal{H}_{tot}$ with a density matrix given by $\hat{\rho}_{tot}=\ket{\Psi}\bra{\Psi}$. Next, we divide the total system into two parts, one of which we call $A$ and its complement $\bar{A}=B$. This procedure allows us to express the total Hilbert space as a direct product of the two subspaces:
\begin{equation}
    \mathcal{H}_{tot}=\mathcal{H}_A\otimes\mathcal{H}_B. \label{Division of Hilbert space}
\end{equation}
Using the marginalization process, it is possible to define a reduced density matrix for the subsystem A (for example). This means that we will take the partial trace over the system $B$ to the total density matrix.
\begin{equation}
    \hat{\rho}_A:=tr_B(\hat{\rho}_{tot}). \label{reduced density matrix}
\end{equation}
Note that $\hat{\rho}_A$ does not depend on the choice of the basis for $\mathcal{H}_B$, it only depends on the regions we have chosen when dividing the total system. The observer, who has access only to subsystem $A$, would perceive the overall system as characterized by the reduced density matrix $\rho_A$. In other words, any  correlation function of the form $\hat{O}=\hat{O}_A\otimes\mathbb{I}_B$ could be computed  using  only the information of the reduced density matrix. 

The entanglement entropy of the subsystem $A$ is defined as the von Neumann entropy of the reduced density matrix $\hat{\rho}_A$:
\begin{equation}
    S(A):=-tr_A(\hat{\rho}_A\ln{(\hat{\rho}_A)}).
\end{equation}
Allow us to make a few comments on this definition before continuing. Entanglement is a shared property of the two subsystems $A$ and $B$, so it is natural that any measure of entanglement defined in one of the two regions coincides with the same measure in the other region. That is, as long as the complete system is in a pure state, the equation $S(A)=S(B)$ holds true\footnote{A formal proof of this assertion will be provided below.}. This implies that the entanglement entropy, unlike thermal entropy, is not extensive for pure states. 

To see how the EE of a quantum system is obtained, let's take as an example the process of Eq.\eqref{decay pi meson} where we get a two spin system. Let us define the electron and positron as subsystem $A$ and $B$ respectively and let its corresponding Hilbert spaces to have the basis  $\{\ket{\uparrow}_{A,B},\ket{\downarrow}_{A,B}\}$. The total Hilbert space is described by the tensor product of Eq.\eqref{Division of Hilbert space}, therefore its basis is expressed as:  $\{\ket{\uparrow}_A\otimes\ket{\uparrow}_B,\ket{\downarrow}_A\otimes\ket{\downarrow}_B,\ket{\uparrow}_A\otimes\ket{\downarrow}_B,\ket{\downarrow}_A\otimes\ket{\uparrow}_B\}=\{\ket{\uparrow\uparrow},\ket{\downarrow\downarrow},\ket{\uparrow\downarrow},\ket{\downarrow\uparrow}\}$. The total density matrix would be:
\begin{equation}
    \hat{\rho}=\ket{\Psi}\bra{\Psi}=\frac{1}{2}\left(\ket{\uparrow\downarrow}\bra{\uparrow\downarrow}-\ket{\uparrow\downarrow}\bra{\downarrow\uparrow}-\ket{\downarrow\uparrow}\bra{\uparrow\downarrow}+\ket{\downarrow\uparrow}\bra{\downarrow\uparrow}\right).
\end{equation}
We trace out the degrees of freedom of the positron to obtain the reduced density matrix of the electron:
\begin{equation}
   \hat{\rho}_A=\sum_{i=\uparrow,\downarrow}{}_{B}\langle i|\Psi\rangle\langle \Psi| i\rangle_B = \frac{1}{2}\left(\ket{\uparrow}_{A\: A} \bra{\uparrow}+\ket{\downarrow}_{A\: A}\bra{\downarrow}\right)=\sum_{i=\uparrow,\downarrow}p_i\ket{i}_{A\:A}\bra{i} .\label{reduced density matrix two spin}
\end{equation}
Let us take a closer look at this result. First, note that we start from a pure state, i.e., we know with certainty which quantum state describes the system. It should be noted that tracing out subsystem $B$ results in a reduced density matrix representing a mixed state, which indicates that we no longer know for sure which quantum state describes subsystem $B$. This certainly does not happen in classical statistics where if we start having all the knowledge of a system after a marginalization process we still have all the information of the subsystem we obtain. In fact this occurs, as we  said  above, due to  quantum entanglement a property that is not present in classical systems.

Now we calculate the EE:
\begin{equation}
  S(A)=-\sum_{i=1,2}p_i\ln{(p_i)}=\ln{2}.\label{EE two spin}
\end{equation}
This is  a case of maximum entropy because $p_1=p_2=\frac{1}{2}$. The impossibility of factorizing the initial pure state in a tensor product between a ket in  Hilbert space $\mathcal{H}_A$ and a ket in Hilbert space $\mathcal{H}_B$ produce a mixed state and therefore the EE is different than zero. This class of states that can not be factorized are called \textit{entangled states}. On the other hand, the ones that can be factorized are called \textit{separable states}. Understanding what does the EE measure we should properly formulate some definitions.

Setting the bases $\{\ket{i}_A, i=1,...,d_A\}$ and $\{\ket{\mu}_B, \mu=1,...,d_B\}$ for Hilbert space $\mathcal{H}_A$ and $\mathcal{H}_B$, respectively, then the most general form that a pure state can assume is :
\begin{equation}
    \ket{\Psi}=\sum_{i,\mu}C_{i\mu}\ket{i}_A\otimes\ket{\mu}_B. \label{pure state}
\end{equation}
If the coefficient matrix $C_{i\mu}$ can be expressed as $C_{i\mu}=C_{i}^AC^{B}_{\mu}$ then,
\begin{equation}
  \ket{\Psi}=\left(\sum_iC_i^A\ket{i}_A\right)\otimes\left(\sum_\mu C_\mu^B\ket{\mu}_B\right)=\ket{\Psi}_A\otimes\ket{\Psi}_B  \longrightarrow \hat{\rho}_A=\ket{\Psi}_{A\: A}\bra{\Psi}, \label{separable state}
\end{equation}
and therefore it is a separable state. On the other hand if $C_{i\mu}\neq C_{i}^AC^{B}_{\mu}$ the state described by Eq.\eqref{pure state} is an entangled state as it and can not be factorized as in Eq.\eqref{separable state}. However in can be expressed using the \textit{Schmidt decomposition} as,
\begin{equation}
    \ket{\Psi}=\sum_{k}^{min(d_A,d_B)}\sqrt{p_k}\ket{\psi_k}_A\otimes \ket{\psi_k}_B,
\end{equation}
where $\{\ket{\psi_k}_{A,B}\}$ is a new orthonormal base. The reduced density matrix for subsystem $A$ is $\hat{\rho}_A=\sum_{k}^{min(d_A,d_B)}p_k\ket{\psi_k}_{A A}\bra{\psi_k}$ and the corresponding EE is defined as:
\begin{equation}
   S(A)=-\sum_{k}^{min(d_A,d_B)}p_k\ln{(p_k)}.\label{Shannon Entropy}
\end{equation}
We  encountered this expression before, in the computation of the EE of the two spin system, where we use it because Eq.\eqref{reduced density matrix two spin} was already in the Schmidt decomposition form. Also, if we obtain the reduced density matrix for the region $B$ we will get $\hat{\rho}_B=\sum_{k}^{min(d_A,d_B)}p_k\ket{\psi_k}_{B B}\bra{\psi_k}$ and therefore  the expression for the EE would be exactly Eq.\eqref{Shannon Entropy} proving that  $S(A)=S(B)$.

From the same analysis  of Eq.\eqref{maximal Entropy} we can conclude that the EE is maximum when $p_i=\frac{1}{min(d_A,d_B)}$, so it only depends on the dimension of the Hilbert subspaces. Then, the maximum EE is
\begin{equation}
    S_{max}(A)=\ln{(min(d_A,d_B))} ,\label{Maximum EE}
\end{equation}
such that the constraint $S(A)<\ln{(min(d_A,d_B))}$ is always valid. Finding maximally entangled states is not as rare as it may seems. In fact, in Ref.\cite{Page:1993wv} it was shown that for a \textit{random pure state} in the limit of \textit{large} Hilbert spaces, the bound is almost saturated and therefore is essentially maximally entangled. We are going to come back to this later.

Using these definitions, we can conclude that the EE quantifies the extent to which a particular state deviates from a separable state. The maximum value of the EE is attained when the state is a superposition of all possible quantum states with equal weights.

Before we end this brief summary of  EE in quantum mechanics, we must introduce another measure of entanglement that will be useful in the thesis. Rényi entropy \cite{renyi1961measures} is defined as,
\begin{equation}
    S_n(A)=\frac{1}{1-n}\ln{(tr_A(\rho_A^n))}.\label{Rényi Entropy}
\end{equation}
with an associated  order parameter $n\in \mathbb{Z}^+$. We should note that it is possible to recover the EE from Eq.\eqref{Rényi Entropy} by making an analytical continuation of the parameter $n$ and then taking the limit $n\rightarrow 1$:
\begin{equation}
    S(A)=\lim_{n\rightarrow 1}S_n(A). \label{von Neumann from Rényi}
\end{equation}
In this sense, it is said that Rényi entropy is a generalization of von Neumann entropy.

\section{Entanglement Entropy in Quantum Field Theory} \label{EE in QFT}

The process described by Eq.\eqref{decay pi meson} is no longer useful if we want to study systems with many degrees of freedom.  To start exploring continuum systems like quantum field theories (QFT), it would be beneficial to first examine a more expansive lattice system (shown in Fig.(\ref{fig:lattice})), such as the spin chain. This is because many fascinating field theories can be derived by taking a continuum limit of such a lattice system. To begin, let us assume that the lattice spacing is $\delta$ and that each site is labeled with $\alpha$, which has a corresponding local Hilbert space $\mathcal{H_\alpha}$. In this way, a pure quantum state of the entire system can be represented as an element of the tensor product Hilbert space: 
\begin{figure}[t!]
  \centering
  \includegraphics[width=0.4\textwidth]{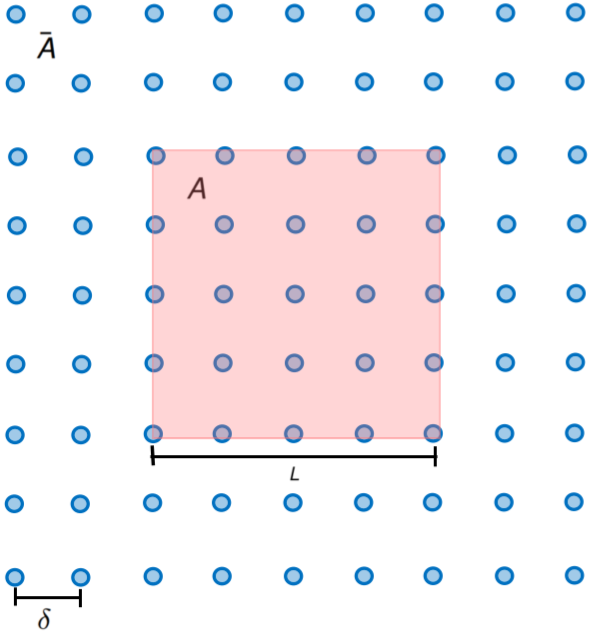}
  \caption{In a lattice system with lattice spacing $\delta$, a region $A$ of size $L$ is considered.}   
  \label{fig:lattice} 
  \end{figure}

\begin{equation}
    \mathcal{H}=\bigotimes_\alpha \mathcal{H}_\alpha.
\end{equation}
As before, we can divide the Hilbert space in a region $A$ and its complement $\bar{A}$:
\begin{equation}
  \mathcal{H}=\mathcal{H}_A\otimes \mathcal{H}_{\bar{A}}\:\:\:\text{where} \:\:\: \mathcal{H}_A=\bigotimes_{\alpha\in A} \mathcal{H}_\alpha\:\:;\:\:\mathcal{H}_{\bar{A}}=\bigotimes_{\alpha\in \bar{A}} \mathcal{H}_\alpha.
\end{equation}
Now lets try to give a heuristic estimation for the EE of region $A$. We assume that the characteristic size $L$ of the region $A$ is much larger than the lattice spacing $\delta$ (with an even larger total Hilbert space). As we have said before, $S(A)$ could be as large as Eq.\eqref{Maximum EE}. If we randomly select a state from the complete Hilbert space, it is possible that the bound could be reached. This means that if we assume $dim(A)<dim(B)$, then $S(A)\approx ln(dim(A))$. Suppose $N_A$ is the count of lattice sites in region $A$, and let's assume that all local Hilbert spaces $\mathcal{H_\alpha}$ have the same dimensions. In that case, $dim(A)$ will be equal to $(dim(\mathcal{H}_\alpha))^{N_A}$. This implies that:
\begin{equation}
    S(A) \propto N_A.
\end{equation}
We can see that if the number  of spatial dimensions of the system is $d-1$ then $N_A\sim \frac{L^{d-1}}{\delta^{d-1}}$ \footnote{To have a picture in mind consider a spin chain system in dimension $d=1+1$, in this case the number of sites would be the length of region A divided by the length separating each particle in the chain.}. This estimation tell us that for a random state the EE should be proportional to the volume, however it is a naive expectation as we are not usually interested in such types of states. In contrast, our focus is on the physical states of physical Hamiltonians, including ground states and minimally excited states, which preserve the concept of locality to some extent. We can understand these Hamiltonians as containing only nearest-neighbor interactions (e.g., the Ising model). If we compute the EE for the ground state of these physical systems, the locality of the Hamiltonian ensures that most of the entanglement will be limited to short distances. This implies that the entropy will be directly proportional to the number of bonds between adjacent sites severed by the boundary between regions $A$ and $\bar{A}$, and will therefore increase only in proportion to the area of this boundary, that is:
\begin{equation}
    S(A) \propto \frac{L^{d-2}}{\delta^{d-2}}. \label{area law discrete}
\end{equation}
In fact, this estimation turns out to be true for a large class of discrete physical systems \cite{Cramer:2005mx}. It can be seen that it is much smaller compared to the first approximation we give for a  random pure state. Therefore, we expect physical systems to have, in some sense, a small amount of entanglement. The Eq.\eqref{area law discrete} allows us to anticipate some aspects that could possibly arise when going to the continuous limit that describes quantum field theories. First, it can be predicted that in the limit of $\delta\rightarrow 0$, the EE will exhibit UV divergence (at least in dimensions greater than two, although it is also common for them to be divergent in two dimensions). Second, the leading order divergence must be proportional to the boundary of the region $A$, which is often referred to as the \textit{entangling surface}. It is possible to demonstrate that, in d-dimensional free field theories, the dominant divergent terms in the UV limit follow the \textit{Area Law} \cite{Srednicki:1993im,Bombelli:1986rw}:
\begin{equation}
    S(A)=\gamma \frac{Area(\partial\mathcal{A})}{\delta^{d-2}}+ \text{subleading terms,}
\end{equation}
here $\gamma$ is a constant that relies on the specifics of the field theory. By presenting general arguments, it can be shown that the arrangement of the subleading  terms relies on the intrinsic and extrinsic geometry of  $\partial A$. For states belonging to the Hilbert space of a QFT, the entropy $S(A)$ can be expressed as follows \cite{Liu:2012eea, Myers:2012ed}:
 \small\begin{equation}
  S(A) =
    \begin{cases}
     a_{d-2}\left(\frac{L}{\delta}\right)^{d-2} + a_{d-4}\left(\frac{L}{\delta}\right)^{d-4}+...+a_1\frac{L}{\delta}+(-1)^{\frac{d-1}{2}}S_A+O(\delta)& \mbox{odd $d$};\\
      a_{d-2}\left(\frac{L}{\delta}\right)^{d-2} + a_{d-4}\left(\frac{L}{\delta}\right)^{d-4}+...+a_1\frac{L}{\delta}+(-1)^{\frac{d-1}{2}}S_A \ln{(\frac{L}{\delta})}+O(\delta) & \mbox{even $d$}.\\
    \end{cases}     \end{equation}
The reason for the different behavior between odd and even QFTs can be attributed to the pattern of UV divergences within the theory. With the exception of $S_A$, which is universal, the majority of the coefficients $a_i$ in the expansion shown above are nonphysical and dependent on the regularization scheme employed. For CFTs this term captures useful information about the conformal anomalies in the theory and plays an important role in the entanglement-based results on renormalization group flow \cite{Casini:2004bw,Myers:2010tj,Casini:2012ei}.

Now that we have explored some aspects of EE in QFT it is time to see how we might actually compute it. We start by assuming that the background space is a globally hyperbolic $d$-dimensional Lorentzian spacetime $\mathcal{B}_d$, for example, we can take it to be a Minkowski spacetime $\mathbb{R}^{1,d-1}$. Global hyperbolicity allows us to choose consistently a Cauchy slice  $\Sigma_{d-1}$, that defines a moment of simultaneity in the QFT. In the case of Minkowski space it is a flat spacelike hyperplane that corresponds to a constant time slice. 

In the Heisenberg picture, if we have a complete set of commuting observables defined on $\Sigma_{d-1}$ then, by using the equations of motions (EOM), any other observable in $\mathcal{B}_d$ can be expressed in terms of those on $\Sigma_{d-1}$. This is a consequence of the commutativity property of spatially separated observables of relativistic QFTs. Therefore, in this setup the Cauchy slice plays the role that the spatial region played before. On this hypersurface is where we find the state of the system. It is possible to view the pure state as defined by the wavefunction $\Psi\left[\Phi(\vec{x})\right]$, where $\Phi(\vec{x})$ denotes the collective fields that describe the system and $\vec{x}$ represents a set of  coordinates that specify the spatial location on $\Sigma$. Assuming the Cauchy slice is held constant at $t=0$, the pure ground state can be expressed in the path integral representation as \cite{Pokorski:1987ed, Casini:2009sr}:
\begin{equation}
    \Psi\left[\Phi_0(\vec{x})\right]=\langle\Phi_0(\Vec{x}) |\Psi\rangle=\int_{t=-\infty}^{t=0, \Phi(t=0,\Vec{x})=\Phi_0(\Vec{x})} \left[ \mathcal{D}\Phi(t,\vec{x})\right]e^{-I_E[\Phi]}, \label{pure state path integral}
\end{equation}
where $I_E$ is the euclidean action of the theory and $\Phi_0(\Vec{x})$ is an eigenvector of the field operator $\hat{\Phi}(t,\Vec{x})$ at $t=0$ such that  $\hat{\Phi}(0,\Vec{x})\ket{\Phi_0(\Vec{x})}=\Phi_0(\Vec{x})\ket{\Phi_0(\Vec{x})}$. Its corresponding conjugate is given by the expression:
\begin{equation}
    \Psi^*\left[\Phi_0(\vec{x})\right]=\langle\Psi |\Phi'_0(\Vec{x})\rangle=\int^{t=\infty}_{t=0, \Phi(t=0,\Vec{x})=\Phi'_0(\Vec{x})} \left[ \mathcal{D}\Phi(t,\vec{x})\right]e^{-I_E[\Phi]}. \label{conjugate pure state path integral}
\end{equation}
We can find a pictorial representation of the wave function in Fig.(\ref{fig:groundstate}). The total density matrix will be defined similar as before only that now we include a partition function $Z$ to ensure that $tr(\hat{\rho})=1$:
\begin{equation}
    \hat{\rho}_{tot}=\frac{\ket{\Psi}\bra{\Psi}}{Z}\longrightarrow Z=\int \left[ \mathcal{D}\Phi(t=0,\vec{x})\right]\langle\Psi |\Phi_0\rangle\langle\Phi_0|\Psi\rangle .
\end{equation}
To obtain the reduced density matrix, first we should divide $\Sigma_{d-1}$ in two regions $A$ and its complement $\bar{A}=B$. To account for UV divergences, a regulator $\delta$ is introduced. Decomposing the total Hilbert space $\mathcal{H}$ of a QFT into $\mathcal{H}_A$ and $\mathcal{H}_B$, via a tensor product, is not always an easy task. For example, if the theory has gauge symmetries, it can be a challenge to set up this decomposition in a way that preserves gauge invariance. Nonetheless, for the purposes of this discussion, we assume that we can always define a reduced density matrix even in cases where the total Hilbert space cannot be expressed as $\mathcal{H}=\mathcal{H}_A\otimes\mathcal{H}_{B}$. To obtain the reduced density matrix for region $A$, we integrate the operator $\hat{\rho}$ over all states $\Phi(t=0,x\in B)=\Phi_B(x)$ as follows:
\begin{figure}[t!]
  \centering
  \includegraphics[width=0.8\textwidth]{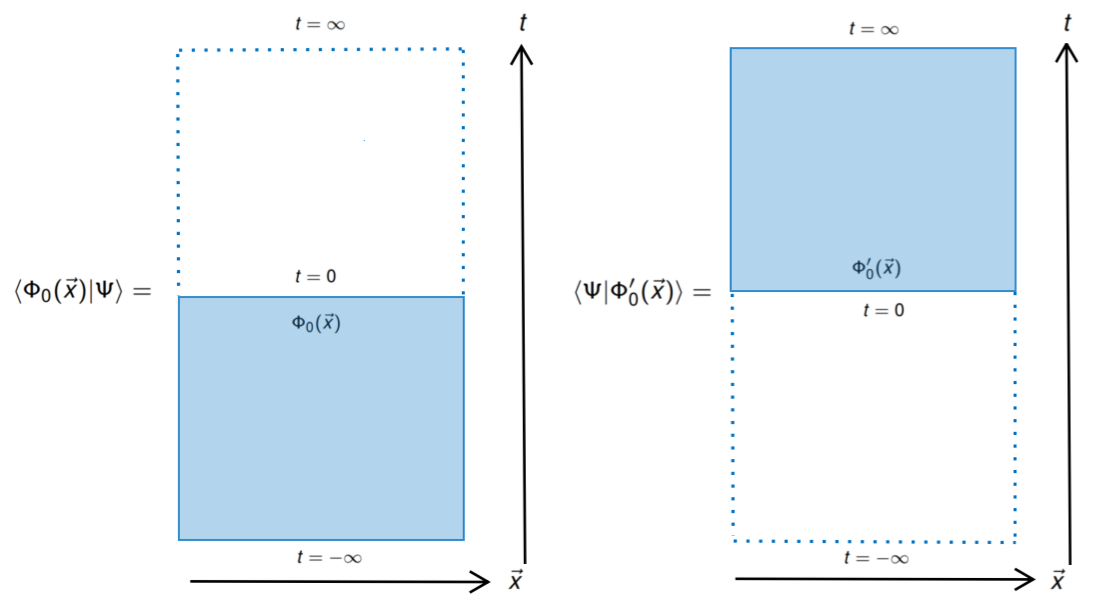}
  \caption{Path Integral representation of a wave function.}   
  \label{fig:groundstate} 
  \end{figure}

\begin{equation}
   \hat{ \rho}_A=\frac{1}{Z} \int \left[ \mathcal{D}\Phi_B(\vec{x})\right]\langle\Phi_B|\Psi\rangle\langle\Psi |\Phi_B\rangle .\label{reduced density matrix QFT}
\end{equation}
This is like gluing the edges of the two sheets of Fig.(\ref{fig:paste groundstate}) along B. The matrix elements of Eq.\eqref{reduced density matrix QFT}  may have two indices that specify their boundary conditions on $t=0^+$ and $t=0^-$ for the fields in region $A$:

\begin{figure}[t!]
  \centering
  \includegraphics[width=0.8\textwidth]{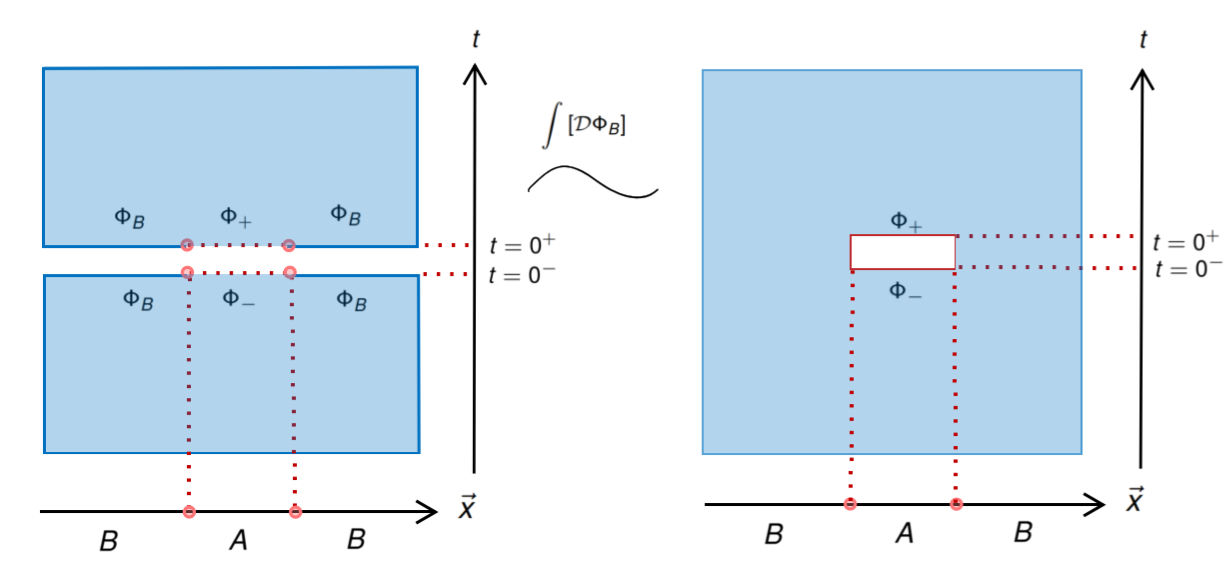}
  \caption{Representation of taking the partial trace over $B$ to obtain the reduced density matrix.}   
  \label{fig:paste groundstate} 
  \end{figure}

\begin{equation}
    \Phi\left(t=0^{+},x\in A\right)=\Phi_{+}\qquad \text{and} \qquad\Phi\left(t=0^{-},x\in A\right)=\Phi_{-},
\end{equation}
then
\begin{equation}
    \left[\rho_A\right]_{-+}=\bra{\Phi_-}\hat{\rho}_A\ket{\Phi+}=\frac{1}{Z}\int \left[\mathcal{D}\Phi_B(\vec{x})\right] \left(\bra{\Phi_-}\bra{\Phi_B}\right)\ket{\Psi}\bra{\Psi}\left(\ket{\Phi_B}\ket{\Phi_+}\right). \label{matrix component reduced density matrix in path integral formulation}
\end{equation}
The term $\langle \Phi_-,\Phi_B|\Psi\rangle$ is $\Psi[\Phi_0(\vec{x})]$ of Eq.\eqref{pure state path integral}, where the path integral for region $A$ is evaluated from $-\infty$ to $0^-$ . Similarly, the term $\langle \Psi | \phi_B,\phi+ \rangle$ is its conjugate and is evaluated from $0^+$ to $\infty$ for the region $A$ (see Eq.\eqref{conjugate pure state path integral}). These two expressions, along
with the integration of the fields in region $B$, is an integration over the
fields defined in the whole spacetime manifold with the exception of the points that
belongs to the region $A$ in $t = 0$. Therefore  the Eq.\eqref{matrix component reduced density matrix in path integral formulation} could be expressed as:
\begin{equation}
    \left[\rho_A\right]_{-+}=\frac{1}{Z}\int\left[\mathcal{D}\Phi(t,\vec{x})\right] e^{-I_E[\Phi]} \prod_{\vec{x}\in A}\delta(  \Phi(t=0^-,x)-\Phi_-)\delta(  \Phi(t=0^+,x)-\Phi_+).
\end{equation}
Before we continue with the computation of the EE lets mention some aspects about the causal structure where the sates and therefore the reduced density matrix is defined. 
We denote by $D\left[S\right]$ to the \textit{causal domain} of a subset $S\subseteq B_{d}$. Because $\Sigma_{d-1}$ is a Cauchy slice its causal domain is the entire manifold, that is $D\left[\Sigma\right]=B_d$. Now, the causal domain associated to  $A$ and its complement $\bar{A}$ are $D\left[A\right]$ and $D\left[\bar{A}\right]$ respectively. We have previously argued that any observable in $B_d$ can be defined in terms of the observables in the Cauchy slice and by using the equation of motion of the theory. Similarly, any observable defined in  $D\left[A\right]$ could be expressed in terms of the ones defined in region $A$. Hence, by using the reduced density matrix $\rho_A$, one can compute the expectation values of all the operators defined in  $D\left[A\right]$. This indicates that not only $A$, but also its entire causal domain $D[A]$, is associated with $\mathcal{H}_A$ and $\rho_A$. For instance, suppose we select another Cauchy Slice $\Sigma'$ and divide it into a region $A'$ and its complement. If $D[A]=D[A']$, then $\mathcal{H}_{A'}=\mathcal{H}_A$, $\hat{\rho}_{A'}=\hat{\rho}_{A}$, and consequently, $S(A)=S(A')$.

Having already made all the necessary observations about the reduced density matrix, it only remains to find out how to obtain the EE from it. The natural next step would be to take the logarithm of $\rho_A$, however this is a tricky step because taking the logarithm of a continuous operator entails several technical complications. Because of that, we will use a different strategy in which the EE is obtained by doing a proper analytical continuation of the Rényi entropy and then taking the limit of the order parameter  $n$ \footnote{From now on the replica parameter} to be 1 . This method is called the \textit{replica trick} and we will review it in below.

\begin{figure}[t!]
  \centering
  \includegraphics[width=0.6\textwidth]{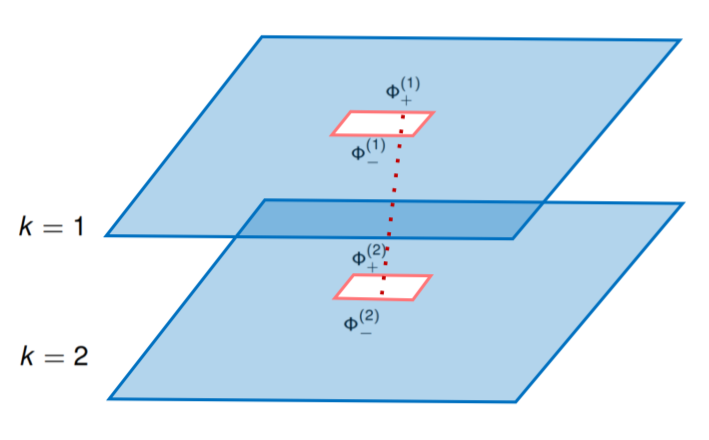}
  \caption{Replica trick for $n=2$}   
  \label{fig:replicatwo} 
  \end{figure}

First we need to obtain $\hat{\rho}_A^n$. To see how this goes, lets consider the case when $n=2$.
\begin{align}
    \left[\rho^2_A\right]_{-+}=
    \frac{1}{Z^2}\int\left[ d\Phi^{(1)}_+\right]\bra{\Phi_-}\hat{\rho}_A\ket{\Phi^{(1)}_+}\bra{\Phi^{(1)}_+}\hat{\rho}_A\ket{\Phi_+} .\label{reduced density matrix square}
\end{align}
What we have done is to introduce a basis $\ket{\Phi^{(1)}_+}$ for region $A$ in the upper limit $0^+$ and then use the completeness relation. Another way to express this, is to take two copies of the sheet of Fig.(\ref{fig:paste groundstate}) label by $k=1,2$ as it is shown in Fig.(\ref{fig:replicatwo}) where we have done the identification $\Phi^{(1)}_+=\Phi^{(2)}_-$ and we have call $\Phi_-=\Phi^{(1)}_-$ and $\Phi_+=\Phi^{(2)}_+$. Therefore Eq.\eqref{reduced density matrix square} can be rephrased as:
\begin{align}
      \left[\rho^2_A\right]_{-+}&= \frac{1}{Z^2}\int\left[ d\Phi^{(1)}_+\right]\delta(\Phi^{(1)}_+-\Phi^{(2)}_-)\bra{\Phi^{(1)}_-}\hat{\rho}_A\ket{\Phi^{(1)}_+}\bra{\Phi^{(2)}_-}\hat{\rho}_A\ket{\Phi^{(2)}_+}\\
      &=\frac{1}{Z^2}\int\left[ d\Phi^{(1)}_+\right]\delta(\Phi^{(1)}_+-\Phi^{(2)}_-)\times \left(\int\prod_{k=1}^2\left[\mathcal{D}\Phi^{(k)}\right] e^{-I_E[\Phi^{(k)}]}\delta(\Phi^{(k)}_{\mp})\right),
\end{align}
where
\begin{equation}
    \delta(\Phi^{(k)}_{\mp})= \prod_{\vec{x}\in A}\delta(  \Phi^{(k)}(t=0^-,x)-\Phi^{(k)}_-)\delta(  \Phi^{(k)}(t=0^+,x)-\Phi^{(k)}_+).
\end{equation}
Note that if we compute the trace of equation \eqref{reduced density matrix square}, we need to equate $\Phi^{(1)}_-$ with $\Phi^{(2)}_+$ and integrate over $\Phi^{(2)}_+$. This procedure effectively involves connecting the two remaining edges of the surface to each other, resulting in a two-sheeted surface with a branch cut along A that connects the two sheets.  Following the same steps we can arrive to a formula for $\hat{\rho}^n$:
\begin{equation}
     \left[\rho^n_A\right]_{-+}=\frac{1}{Z^n}\int\prod_{k}^{n-1}\left[ d\Phi^{(k)}_+\right]\delta(\Phi^{(k)}_+-\Phi^{(k+1)}_-) \left(\int\prod_{k=1}^n\left[\mathcal{D}\Phi^{(k)}\right] e^{-I_E[\Phi^{(k)}]}\delta(\Phi^{(k)}_{\mp})\right) .\label{reduced density matrix to the n}
\end{equation}
Similarly, taking the trace means to  set $\Phi^{(1)}_-=\Phi^{(n)}_+$ and running a integral over $\Phi^{(n)}_+$. In this case we are gluing $n$ copies of the same spacetime $B_d$ along $A$ using the identification $\Phi^{(k)}_+=\Phi^{(k+1)}_-$. In principle, this procedure constructs an n-fold cover $B_d^{(n)}$ of the original manifold. Taking the trace of the integral in Eq.\eqref{reduced density matrix to the n} give us  the partition function $Z[B_d^{(n)}]$ on this  n-fold cover. Therefore:
\begin{equation}
    tr_A(\hat{\rho}^n_A)=\frac{Z[B_d^{(n)}]}{(Z[B_d])^n}=\frac{Z_n}{Z^n}.
\end{equation}
We have rewritten $Z[B_d^{(n)}]$ as $Z_n$ where for $n=1$ we recover  the partition function on the original manifold. The path integral on $B_d^{(n)}$ could be computed using different methods such as a mode expansion, the  heat-kernel methods or any other that can allow you to solve the integral. It is important to realize that pasting over all the copies, induces a conical singularity in $B_d^{(n)}$ located in the codimension-two  entangling region $\partial A$ with a deficit angle $2\pi(1-n)$.

Now we can give and expression for the Rényi entropy:
\begin{equation}
    S_n(A)=\frac{1}{1-n}\ln{\left(\frac{Z_n}{Z^n}\right)}. \label{Renyi Entropy}
\end{equation}

If we want to compute EE from Rényi entropy, we need to perform an analytical continuation on the replica parameter. Although this may not always be possible, we can still use the replica trick to calculate the EE if we can demonstrate that it is well-behaved through physical reasoning.
Therefore:
\begin{equation}
    S(A)=\lim_{n\rightarrow 1} S_n(A)=-\lim_{n\rightarrow 1}\partial_n(\ln{(Z_n)}-n\ln{(Z)}) . \label{Replica trick}
\end{equation}
The procedure described above was used in Refs.\cite{Calabrese:2004eu, Calabrese:2005zw} to obtain the EE for relativistic field theories in $1+1$ dimensions. First they consider a quantum spin chain at the quantum critical point where the system undergoes  a \textit{continuum} phase transition. At the critical point, where the correlation length behaves as $\xi^{-1}=0$, the field theory is massless, and is a CFT. Exploiting the conformal symmetry, they arrive to  expressions for the EE for joint and disjoint intervals defining the subsystem $A$. The computation of EE in a thermal mixed state at a finite temperature is also examined by them.

The replica trick was also used to compute the EE for free massive scalar field theories in any dimensions (for a review see Ref.\cite{Casini:2009sr}). These models are the simplest that can be investigated, however, even in these cases arriving at an exact expression for the EE is not always possible. Few examples are available because difficult questions about the spectrum of certain integral operators, or equivalent questions expressed as functional integrals, need to be resolved. In addition, ensuring the uniqueness of the analytical continuation of the Rényi entropy is also not trivial and requires some additional information \cite{Castro-Alvaredo:2008usl}.

Everything described above suggests that, in general, it is a difficult task to obtain the EE except in exceptionally simple cases that usually involve non-interacting QFTs or the exploitation of some kind of symmetries in $1+1$ dimensions. In the next section we will see how the EE could be calculated by mapping the problem to a simpler one in a dual gravitational theory.

\section{Holographic Entanglement Entropy} \label{HEE}

In this section we will give a brief introduction to the AdS/CFT \cite{Maldacena:1997re,Witten:1998qj,Gubser:1998bc,Aharony:1999ti} duality, which is the fundamental framework in which the computation of the EE takes place. Afterward, we will follow the procedure of Ref.\cite{Lewkowycz:2013nqa} to compute the EE holographically and prove the RT formula (see Eq.\eqref{RT formula}). 

As previously mentioned, in  a quantum field theory that has a holographic dual description, it is possible to convert the task of determining the EE for a specific area in the QFT into a more straightforward problem on the corresponding gravitational side. As we have said before, for quantum field theories that accept dual holographic theories, it is possible to translate the problem of finding the EE for a given region in the QFT to a simpler problem on the corresponding gravitational side. Furthermore, if the dual theory is Einstein gravity, then the EE is given by the Ryu-Takayanagi formula \cite{Ryu:2006bv,Ryu:2006ef}, where the problem reduces to calculating the minimum area of a codimension-two surface anchored in the entangling surface located at the boundary where the field theory  live. In principle, this is more than a computational tool, as it relates the geometry of spacetime to the spatial EE of some region in the QFT and is therefore a key ingredient for understanding how the bulk geometric picture could emerge from entanglement \cite{VanRaamsdonk:2010pw}. However, for the purposes of this thesis, we will not address the latter aspects and  we will use the correspondence as an approach to obtain the EE.

Roughly speaking,  the AdS/CFT correspondence proposes that certain non-gravitational quantum field theories in $d$ dimensions are equivalent to a string theory with gravitational interaction in a $D=d+1$ dimensional (A)AdS background. To illustrate this idea, consider a simplified scenario where the QFT is characterized by two parameters, $\{c_{eff},\lambda\}$. The coupling parameter $\lambda$ measures the strength of the interaction among the elements, while the parameter $c_{eff}$ offers a measure of the effective number of degrees of freedom. In the \textit{planar limit} when the number of degrees of freedom tends to infinity, that is when $c_{eff}\rightarrow \infty$, we have that the string interaction in the gravity side becomes very weak and therefore can be treated as a classical dynamical system that involves strings. Furthermore, the strong coupling limit of the field theory, where the coupling parameter $\lambda$ approaches infinity, results in a truncation of the classical string theory, ultimately leading to the classical gravitational dynamics described by general relativity. This concept is commonly referred to as the strong-weak duality. If we combine the two limits  $c_{eff},\lambda \rightarrow \infty$ then we can gain information regarding the behavior of QFTs by performing calculations in the dual gravitational theory.

To give a concrete example lets take the setup of the original formulation of the AdS/CFT correspondence. This is a  $\mathcal{N}=4_{4d}$ $SU(N)$ Super Yang-Mills theory which arises from the low-energy behavior of open string theories on D-brane worldvolumes. Similar to its non-supersymmetric counterpart, this theory possesses exact conformal symmetry for any value of the coupling $g_{YM}$. In this scenario the theory is described by two parameters: the  t'Hooft coupling $\lambda=g_{YM}N$  and the number of degrees of freedom $c_{eff}\propto N^2-1$. In the planar limit, i.e., in the large limit of $N$, one obtains a classical string theory in $AdS_5\times\textbf{S}^5$. On the other hand, taking $\lambda\rightarrow\infty$CF reduces the dynamics to type IIB supergravity on an AdS background, which includes the dynamics of Einstein-Hilbert gravity as a subsector.  The pictorial way to view this duality is that of a theory of gravity defined in an AdS space with a CFT residing  on the boundary of it.

Beyond this particular example, there exists a plethora of CFTs in $d$ dimensions, with different degrees of supersymmetry, that accepts a classical gravity dual theory on a $AdS_{d+1}\times \Upsilon$, where $\Upsilon$ is a compact manifold. From now on we will only be interested in such CFT theories that in a planar and strongly coupled limit can be described using classical gravity in an asymptotically $AdS_{d+1}$ spacetime\footnote{From now on we will denote the bulk spacetime as $\mathcal{M}_{d+1}$.}. To proceed with our description, we need to establish a dictionary between the observables of the CFT on some timelike and globally hyperbolic background that we call $\mathcal{B}_d$ and the ones of  the dual gravitational theory in $\mathcal{M}_{d+1}$. The AdS/CFT correspondence establish an isomorphism between the sates of the Hilbert space in the CFT and the ones of the closed string Hilbert space. We are particularly interested in the so-called \textit{code subspace} of the CFT that have a geometric description in the dual theory side. One example is that the vacuum Hilbert space, which preserves the $SO(d,2)$ conformal symmetry, has a dual representation as the vacuum $AdS$ spacetime. On the other hand, thermal states of the $CFT_d$ can be mapped  to a  $AdS_{d+1}$ black hole spacetime.

Identifying the states of the QFT we want to examine allows us to obtain more general spacetimes that include matter. The AdS/CFT dictionary tell us that for every bulk field $\Phi$ there is a corresponding gauge invariant boundary operator $\hat{O}_{\Phi}$. Asymptotically AdS spaces have a boundary at spatial infinity, and one needs to impose appropriate boundary conditions there. In particular, bulk gauge fields correspond to boundary CFT sources (external currents). The partition function of the boundary CFT is then dual to a functional of the fields parametrizing the boundary values of the bulk fields. Let $Z_s(\Phi_0)$ be the supergravity partition function on  $\mathcal{M}_{d+1}$ computed with the boundary conditions $\Phi_0$ at the infinity boundary $B_d$. For example, in the planar limit, one computes $Z_s(\Phi_0)$  by simply extending $\Phi_0$ over $\mathcal{M}_{d+1}$ as a solution $\Phi$ of the classical supergravity equations, and then writing
\begin{equation}
    Z_s(\Phi_0)=\int_{\Phi=\Phi_0}{\mathcal{D}\Phi}e^{-I_S[\Phi]},
\end{equation}
where $I_s$ is the classical supergravity action. Following Ref.\cite{Witten:1998qj} the boundary values of the fields are identified with sources that couple to the dual operator:

\begin{equation}
    Z_s(\Phi_0)=\left\langle exp\left(\int_{\partial\mathcal{M}=B} \Phi_0\hat{O}_{\Phi}\right)\right\rangle_{CFT}. \label{GKP-W}
\end{equation}

Based on this, the AdS/CFT correspondence is state in the planar limit as $Z_{SUGRA}=Z_{CFT}$ for computational purposes. This last formula, also known as the GKP-W \footnote{S.S. Gubser, I.R. Klebanov, A.M. Polyakov and E. Witten. } relation, allow us to compute correlation functions on the CFT using the partition function of the supergravity theory with prescribed boundary conditions.  In the leading saddle point we can see from Eq.\eqref{GKP-W} that the onshell bulk partition function is equal to the  the generating functional of CFT correlation functions,
\begin{equation}
    I_{onshell}[\Phi]\approx -W_{CFT}[\Phi_0]\longrightarrow Z_{CFT}[\mathcal{B}]\approx e^{-I[\mathcal{M}]}, \label{GKP-W2}
\end{equation}
where we denote $\partial \mathcal{M}=\mathcal{B}$ just to be more general for future discussions. In the last analysis, the bulk field $\Phi$ labels the collection of matter fields as well as the metric of a given gravitational theory. Lets take as an example the case of the metric, then the one point correlation function of the stress energy momentum could be computed using just information of the classical gravitational on shell action as :
\begin{equation}
    G_{\mu\nu}\longrightarrow \langle T_{i j}\rangle=\frac{1}{\sqrt{g_{0}}}\frac{\delta I}{\delta g^{i j}_{0}}.
\end{equation}
So far we have given a  superficial review of the AdS/CFT correspondence, giving a practical look at the dictionary and also providing a computational tool such as the GKP-W relationship. Before using everything we have developed so far for the calculation of the EE, we will try to give an adequate description of the AdS spaces. An useful representations of  anti de Sitter spacetime is obtained by embedding a $D$ dimensional hyperboloid in a $D+1$ dimensional pseudo Euclidean flat spacetime that have double time coordinate and is described by:
\begin{equation}
    ds^2=-dX^{2}_{D}-dX^{2}_{0}+\sum_{i=1}^{D-1}dX^{2}_i . \label{D+1 flat space}
\end{equation}
The $AdS_{D}$ with radius $\ell$ is obtained by the embedding:
\begin{equation}
    X^{2}_{D}+X^{2}_{0}-\sum_{i=1}^{D-1}X^{2}_i=\ell^2  .\label{AdS embedding}
\end{equation}
Notice that this construction allow us to see that any element of the Lorentz group $SO(2,D-1)$ leaves invariant Eq.\eqref{D+1 flat space} and  Eq.\eqref{AdS embedding}. Also  the group $SO(2,D-1)$ have $D(D+1)/2$ Killing generators that are the maximal amount of generators that a $D$-dimensional space could have, therefore the  $AdS_D$ space is a maximally symmetric space with $SO(2,D-1)$ as is isometry group. Another important property of AdS spacetimes is that they are equipped with  a conformal boundary at hiperboloid infinity. To show this let's consider a rescaling of the coordinates of the form $X_\mu\rightarrow \Lambda X_\mu$ where $\Lambda\in \mathbb{R}$, then  we take $\Lambda\rightarrow\infty$, thus defining the boundary:
\begin{equation}
    X^{2}_{D}+X^{2}_{0}-\sum_{i=1}^{D-1}X^{2}_i=0. \label{conformal boundary}
\end{equation}
This is a conformal boundary because it admits not one but an equivalence class of metrics which are related via a conformal transformation. Now, if $X_0\neq0$ then after a rescaling and a division over $X_0$ we get the de Sitter space $-X_{D}^2+\sum_{i=1}^{D-1}X^{2}_i=1$ in $D-1$ dimensions that have topology $\mathbb{R}\times S^{D-2}$ \cite{Balasubramanian:1998sn}. On the other hand, if $X_0=0$ again after a rescale and a division over $X_{D}$ this time we obtain the unit sphere $\sum_{i=1}^{D-1}X^{2}_i=1$ in $D-2$ dimensions. We combine both results to arrive at a boundary with topology $S^1\times S^{D-2}$. Note that this boundary is invariant under $SO(1,D-1)$ since the isometry of AdS acts on it and, therefore, it is a maximally symmetric space. Moreover, the Eq.\eqref{conformal boundary} is invariant under $1$ dilatation and $D-1$ special conformal transformations as we can see by doing the rescale $\Lambda X_\mu$. This means that the symmetry group is the conformal group $\mathcal{C}(1, D-2)$ which possesses the Lorentz transformations as well as  angle-preserving transformations. The number of generators of the conformal group on the boundary are the same number of generators of the AdS isometry group in the bulk therefore the boundary symmetry group $\mathcal{C}(1, D-2)$ is said to be isomorphic to the bulk symmetry group $SO(2,D-1)$.

\begin{figure}[t!]
  \centering
  \includegraphics[width=0.4\textwidth]{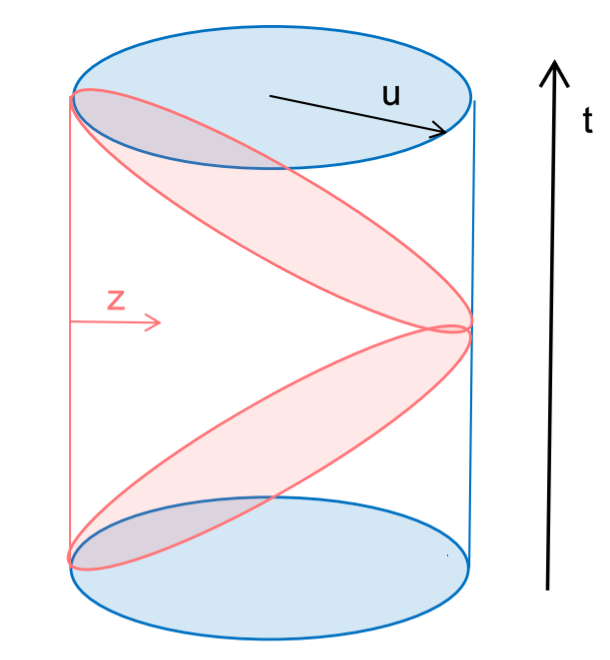}
  \caption{Poincaré patch in $AdS_3$}   
  \label{fig:Poincare patch} 
  \end{figure}

 To characterize the AdS space, it is necessary to define coordinates, and two types of them are particularly relevant: global coordinates and Poincaré coordinates. The global parametrization is achieved through the following transformations:
 
\begin{align}
    X_0&=\ell\cosh{\left(\frac{u}{\ell}\right)}\sin{t}\nonumber\\
    X_D&=\ell\cosh{\left(\frac{u}{\ell}\right)}\cos{t}\nonumber\\
    X_1&=\ell\sinh{\left(\frac{u}{\ell}\right)}\cos{\theta_1}\nonumber\\
     &\vdots\\
     X_{i}&=\ell\sinh{\left(\frac{u}{\ell}\right)}\sin{\theta_1}\hdots \sin{\theta_{i-1}}\cos{\theta_{i}}\nonumber\\
     &\vdots\nonumber\\
      X_{D-1}&=\ell\sinh{\left(\frac{u}{\ell}\right)}\sin{\theta_1}\sin{\theta_2}\hdots \sin{\theta_{D-3}}\sin{\theta_{D-2}},\nonumber
\end{align}
where $u\in [0,+\infty[$. They are global coordinates of the AdS space because all points of  Eq.\eqref{AdS embedding} are taken into account exactly once and therefore cover the whole space. The line element is given by:
\begin{equation}
    ds^2=-\ell^2\cosh^2{\left(\frac{u}{\ell}\right)}dt^2+du^2+\ell\sinh^2{\left(\frac{u}{\ell}\right)}d\Omega^2_{D-2}.
\end{equation}
The last term is the line element of a $D-2$ sphere. Notice that in this parametrization the boundary is located at $u\rightarrow\infty$, also notice that the timelike coordinate is and angular coordinate implying that the AdS space must have closed timelike curves. To avoid this we can unwrap the circle of the time coordinate and take a new coordinate $t\in  ]-\infty,+\infty[$ thus obtaining the universal cover of the AdS space.
Now, lets take another parametrization for Eq.\eqref{AdS embedding}:
\begin{align}
    y&=\ell ln\left(\frac{X_D+X_{D-1}}{\ell}\right)\nonumber\\
    t&=\frac{X_0}{X_D+X_{D-1}} \label{Poincare transformations}\\
    x_i&=\frac{X_i}{X_D+X_{D-1}}.\nonumber
\end{align}
We can see that  $y,t,x_i\in \mathbb{R}$, however this coordinates covers only  half of the AdS spacetime  since the argument in the logarithm of $y$ should satisfy $X_D+X_{D-1}>0$ in order to be well defined. To find the line element first we have to find the inverse transformations. After replacing Eq.\eqref{Poincare transformations} in Eq.\eqref{AdS embedding} we can manage to obtain:
\begin{align}
 X_0&=\ell t e^{\frac{y}{\ell}}\nonumber\\
    X_{D-1}&=\ell\sinh{\left(\frac{y}{\ell}\right)}-\frac{\ell}{2}e^{\frac{y}{\ell}}(\sum^{D-2}_{j=1}x_j^2+t^2)\\
    X_i&=\ell x_ie^{\frac{y}{\ell}}\nonumber\\
    X_{D}&=\ell\cosh{\left(\frac{y}{\ell}\right)}+\frac{\ell}{2}e^{\frac{y}{\ell}}(\sum^{D-2}_{j=1}x_j^2+t^2).\nonumber
\end{align}
Then the line element is given by :
\begin{equation}
    ds^2=\ell^2e^{\frac{2y}{\ell}}(-dt^2+\sum^{D-2}_{j=1}dx_j^2)+dy^2.
\end{equation}
The Poincaré patch is obtain by setting $r=\ell e^{\frac{2y}{\ell}}$:
\begin{equation}
    ds^2=r^2(-dt^2+\sum^{D-2}_{j=1}dx_j^2)+\frac{l^2dr^2}{r^2}.
\end{equation}
We can see that the conformally flat boundary in $r\rightarrow\infty$ is the Minkowski space $R^{D-2,1}$, we can also notice that there is a horizon at $r=0$ due to the fact that we cannot cover the whole AdS space. Usually it is useful to invert the $r$ coordinate as $z=\ell/r$ to yield another version of the Poincaré  coordinates,
\begin{equation}
     ds^2=\frac{\ell^2}{z^2}(-dt^2+dz^2+\sum^{D-2}_{j=1}dx_j^2).
\end{equation}
The $z$ coordinate is usually refer in the literature as the holographic coordinate. We can see that in this new type of coordinates the boundary is located at $z=0$ and that at each constant vale of $z$ we get a Minkowski spacetime. Also, now the Poincaré horizon will be located at the locus  $z\rightarrow \infty$. It is a degenerate Killing horizon. In Fig.(\ref{fig:Poincare patch}) you will find a representation of the domain covered by the Poincare patch with respect to the global AdS spacetime.

Now that we have provided an introduction to the AdS/CFT dictionary and have also described AdS spaces in greater depth, we may ask how EE is captured holographically. In what follows we will be interested in obtaining the EE for static time-independent states of holographic CFTs dual to Einstein gravity (A generalization for higher curvature theories will be provided subsequently)\footnote{For a discussion of arbitrary time dependent states you can follow the prescription of Hubeny, Rangamani, and Takayanagi (HRT) \cite{Hubeny:2007xt} and its derivation in \cite{Dong:2016hjy}.}. It is a good starting point for an explicit derivation, since the boundary manifold $\mathcal{B}_{d}$ where the CFT lives must have a  timelike Killing vector. This allow us to go from the Lorentzian signature to the Euclidean signature consistently in the path integral of the dual bulk theory. That is useful in two aspects: firstly, the path integral of quantum gravity in the Euclidean signature is reasonably well understood, and secondly, the boundary conditions required are straightforward to express and enforce.

We start by considering the Euclidean version of AdS space $\mathcal{M}_{d+1}$ whose boundary is $\partial\mathcal{M}_{d+1}=\mathcal{B}_{d}$. We are interested in finding the EE of a region $A$ of a Cauchy surface $\Sigma$ in $B_d$. To do so, we use the replica trick in the boundary field theory, as discussed in the previous section. By making use of the GKP-W relation we can rewrite  Eq.\eqref{Replica trick} in terms of the dual gravity solution $\mathcal{M}_{d+1}^{(n)}$ whose boundary is the n-fold cover $\partial\mathcal{M}^{(n)}_{d+1}=\mathcal{B}_d^{(n)}$ with conical singularity at the entangling region $\partial A$. This results in a well-defined boundary value problem for the Euclidean supergravity path integral. Assuming such a bulk space $\mathcal{M}_{d+1}^{(n)}$ exists, we can derive a holographic expression for the Rényi entropy as follows:
\begin{align}
    S_{n}(A)=\frac{1}{1-n}ln\left( \frac{Z_{CFT}[\mathcal{B}^{(n)}]}{(Z_{CFT}[\mathcal{B}])^n}\right)=\frac{1}{1-n}ln\left( \frac{Z_{SUGRA}[\mathcal{M}^{(n)}]}{(Z_{SUGRA}[\mathcal{M}])^n}\right).\label{Renyi in the bulk}
\end{align}

Note that the CFT path integral in the n-fold cover $\mathcal{B}^{(n)}_d$ is invariant under $\mathbb{Z}n$ symmetry due to the replica construction that has been employed. Now, we must assume that this symmetry extends into the bulk spacetime $\mathcal{M}^{(n)}{d+1}$, which is constructed by solving the field equations. This assumption restricts us to gravitational actions where the $\mathbb{Z}n$ symmetry can be consistently imposed. We should mention that the existence of an $\mathbb{Z}_n$ action  does not necessarily mean that the symmetry will act smoothly there. However the smoothness of $\mathcal{M}^{(n)}_{d+1}$ is required if we want that the entangling surface to extend into the bulk as a codimension-two surface  in the quotient space of  $\mathcal{M}^{(n)}_{d+1}$ \cite{Haehl:2014zoa}. Lets be more clear with this. Firstly, we construct the quotient space as $\hat{\mathcal{M}}^{(n)}_{d+1}=\mathcal{M}^{(n)}_{d+1}/\mathbb{Z}_n$. This quotient space has singularities of the orbifold type in the fixed locus points of $\mathbb{Z}_n$ that we will call $\gamma^{(n)}$. The request  that $\gamma^{(n)}$ is a codimension-two surface comes  from the fact that, at the boundary, the original n-fold cover  has the locus of fixed points of the $\mathbb{Z}_n$ symmetry  at the entangling region $\partial A$ which is a codimension-two surface there. Locally near the boundary one would see that the entangling region extends to the fixed points in the bulk and therefore we are tempted to see $\gamma^{(n)}$ as an analogous bulk dual to $\partial A$ that is  indeed anchored there. However, this local intuition is not valid at the global level. If we do not require that $\mathcal{M}^{(n)}_{d+1}$ is smooth then we can arrive at examples in which the fixed points on $\hat{\mathcal{M}}^{(n)}_{d+1}$ are not a codimension-two surface.

In order to simplify the analytical continuation of the Rényi entropy, we can adopt a gravitational approach by treating the surface $\gamma^{(n)}$ as a source of energy-momentum that causes deformation in the bulk spacetime $\mathcal{M}_{d+1}$, resulting in the formation of the quotient space $\hat{\mathcal{M}}^{(n)}_{d+1}$. This deformation can be attributed to a cosmic brane that carries a tension $T_n$, which vanishes when $n=1$. This allow us to state that the geometry  of $\hat{\mathcal{M}}^{(n)}_{d+1}$ could be computed by using the Einstein field equation with and energy momentum $T_{\mu \nu}^{brane}$ associated to this cosmic brane. In this interpretation, the analytical continuation of n is straightforward because there is no impediment for the brane tension to be defined with respect to a real parameter. With this in mind we can now take the limit  $n\rightarrow 1$ of Eq.\eqref{Renyi in the bulk}  to obtain the EE, moreover we consider the saddle point aproximation to arrive to the expression:
\begin{equation}
    S(A)=\lim_{n\rightarrow 1}S_n(A)\approx \lim_{n\rightarrow 1}\partial_n\left(I[\mathcal{M}^{(n)}]-nI[\mathcal{M}]\right).
\end{equation}
Using the replica symmetry, it is possible to express the on-shell bulk action for the orbifold as:
\begin{equation}
    I[\mathcal{M}^{(n)}]=n\hat{I}[\hat{\mathcal{M}}^{(n)}]. \label{replica per bulk action}
\end{equation}
Although $\mathcal{M}^{(n)}$ is a regular spacetime, $\hat{\mathcal{M}}^{(n)}$ contains a conical singularity with an opening angle of $2\pi(1-1/n)$ located at $\gamma^{(n)}$. Therefore, to ensure consistency, we must exclude the terms originating from the conical singularity in the right-hand side of Eq.\eqref{replica per bulk action}. Using the above relation the EE is written as:
\begin{equation}
    S(A)=\lim_{n\rightarrow 1} \partial_n\hat{I}[\hat{\mathcal{M}}^{(n)}]. \label{EE holographic}
\end{equation}
The derivative respect to $n$ could be rephrased as a variation of the bulk action. Then we would expect that after taking the limit  $n\rightarrow 1$ we would be doing a variation  of the action defined in $\hat{\mathcal{M}}^{(1)}=\mathcal{M}$  and therefore it would vanish because it is a regular solution of the EOMs. This naive expectation could not be further from the truth.  What happens is that removing the conic singularities induces a codimension-one boundary located around the locus of fixed points of the replica symmetry. We will call this boundary $\gamma^{(n)}(\epsilon)$ where the $\epsilon$ denotes that a tubular neighborhood around $\gamma^{(n)}$ has been taken to construct this boundary. This procedure will induce a Gibbons-Hawking type boundary term in $\gamma^{(n)}(\epsilon)$ (for EOM consistency) that will depend on the replica parameter $n$ and thus the variation of Eq.\eqref{EE holographic} will affect it. 

This approach, in principle, would be equivalent to consider from the beginning the contribution of the conic singularities since the calculation of the EE is located there as we have mentioned above.  To implement this strategy we must know how these brane contributions appear in the action. For this purpose we will consider the results of D. Fursaev, A. Patrushev, S. Solodukhin \cite{Fursaev:2013fta} where they studied the regularization of integral curvature invariants on manifolds with squashed cones singularities\footnote{It does not have U(1) symmetry.}:

\begin{align}
    \int_{\hat{\mathcal{M}}^{(n)}}d^{d+1}x\sqrt{G}R&= \int_{\hat{\mathcal{M}}^{(n)} \backslash \gamma^{(n)}}d^{d+1}x\sqrt{G}R\nonumber\\&+4\pi\left(1-\frac{1}{n}\right)\int_{\gamma^{(n)}}d^{d-1}y\sqrt{\sigma}+\hdots \label{Ricci scalar}\\
     \int_{\hat{\mathcal{M}}^{(n)}}d^{d+1}x\sqrt{G}R^2&= \int_{\hat{\mathcal{M}}^{(n)} \backslash \gamma^{(n)}}d^{d+1}x\sqrt{G}R^2\nonumber\\&+8\pi\left(1-\frac{1}{n}\right)\int_{\gamma^{(n)}}d^{d-1}y\sqrt{\sigma}R+\hdots\label{Ricci scalar squared}\\
     \int_{\hat{\mathcal{M}}^{(n)}}d^{d+1}x\sqrt{G}R_{\mu\nu}R^{\mu\nu}&= \int_{\hat{\mathcal{M}}^{(n)} \backslash \gamma^{(n)}}d^{d+1}x\sqrt{G}R_{\mu\nu}R^{\mu\nu}\nonumber\\
     &+4\pi\left(1-\frac{1}{n}\right)\int_{\gamma^{(n)}}d^{d-1}y\sqrt{\sigma}\left[ R^{a}_{a}-\frac{1}{2}\left(K^{a i}_i\right)^2\right]+\hdots\label{Ricci tensor }\\
     \int_{\hat{\mathcal{M}}^{(n)}}d^{d+1}x\sqrt{G}R_{\mu\nu\rho\sigma}R^{\mu\nu\rho\sigma}&=\int_{\hat{\mathcal{M}}^{(n)} \backslash \gamma^{(n)}}d^{d+1}x\sqrt{G}R_{\mu\nu\rho\sigma}R^{\mu\nu\rho\sigma}\nonumber\\
     &+8\pi\left(1-\frac{1}{n}\right)\int_{\gamma^{(n)}}d^{d-1}y\sqrt{\sigma}\left[ R^{a b}_{\:\:\: a b}-K^{a i}_{\:\: j} K^{j}_{a\:i}\right]+\hdots ,\label{Riemann Tensor}
\end{align}

where the dot points denotes higher order terms in $O\left(\left(1-\frac{1}{n}\right)\right)$. The worldvolume coordinates of  $\gamma^{(n)}$ are $y^i$ (with indices $i=1,...,d-1$) while the a,b indices are used to indicate the normal directions to this surface.

Having understood the regularization of curvature invariants in the presence of squashed cone singularities, the subsequent task is to examine the bulk-per-replica action $\hat{I}[\hat{\mathcal{M}}^{(n)}]$. This action differs from $I[\hat{\mathcal{M}}^{(n)}]$ due to the inclusion of an extra term arising from the singularity located at $\gamma^{(n)}$. For the case of Einstein gravity we can use the Eq.\eqref{Ricci scalar}  to obtain $\hat{I}[\hat{\mathcal{M}}]$ as :
\begin{equation}
    \hat{I}[\hat{\mathcal{M}}^{(n)}]=\frac{1}{16\pi G}\int_{\hat{\mathcal{M}}^{(n)}}d^{d+1}x\sqrt{G}R=I[\hat{\mathcal{M}}^{(n)}]+\frac{1}{4G}\left(1-\frac{1}{n}\right)\mathcal{A}[\gamma^{(n)}].\label{Action in replica orbifold}
\end{equation}

Note that from this equation we can read the exact form of the tension carried by the cosmic brane. That is:
\begin{equation}
    T_n=\frac{1}{4G}\left(1-\frac{1}{n}\right).
\end{equation}
If we replace $\hat{I}[\hat{\mathcal{M}}^{(n)}]$ by $I[\hat{\mathcal{M}}^{(n)}]$ in Eq.\eqref{EE holographic} and then take the derivative we see that :
\begin{align}
    \partial_n\hat{I}[\hat{\mathcal{M}}^{(n)}]&=\frac{\delta \Phi}{\delta n}\frac{\delta I[\hat{\mathcal{M}}^{(n)}]}{\delta \Phi}+\frac{1}{4n^2G}\mathcal{A}[\gamma^{(n)}]\\
    &\Longrightarrow S(A)=\frac{\mathcal{A}[\gamma]}{4G}, \label{ryu-takayanagi 1}
\end{align}
where we have denoted the collection of fields in the bulk by  $\Phi$. In addition, by imposing the equation of motion we note that $\frac{\delta I[\hat{\mathcal{M}}^{(n)}]}{\delta \Phi}=0$ and thus we obtain  Eq.\eqref{ryu-takayanagi 1}. On the other hand, we can argue that $\gamma^{(1)}=\gamma$ must have a minimal area. We can do this heuristically using the cosmic brane interpretation. When we take the limit $n \rightarrow 1$ the tension of the cosmic brane vanishes so there is no backreaction to the background bulk geometry $\hat{\mathcal{M}}^{(n)}$. This means that the position of the brane is fixed in  $\mathcal{M}$ and therefore must satisfy the equation of motion. That implies that the area variation must disappear so it is an extreme. Moreover, by choosing adapted coordinates at the conical singularity one can show that the trace of the extrinsic curvatures are zero (Appendix of \cite{Lewkowycz:2013nqa}) and therefore this extremum is a minimum. 

The Eq.\eqref{ryu-takayanagi 1} was first enunciated as a conjecture in  \cite{Ryu:2006bv}  inspired by  drawing an analogy with the entropy of black holes \cite{Bekenstein:1972tm,Hawking:1975vcx}. This conjecture, also known in the literature as the Ryu-Takayanagui (RT) conjecture, constitutes a major step towards the understanding of how regions of the bulk are related to regions in the CFT side of the correspondence. 

\begin{figure}[t!]
  \centering
  \includegraphics[width=0.3\textwidth]{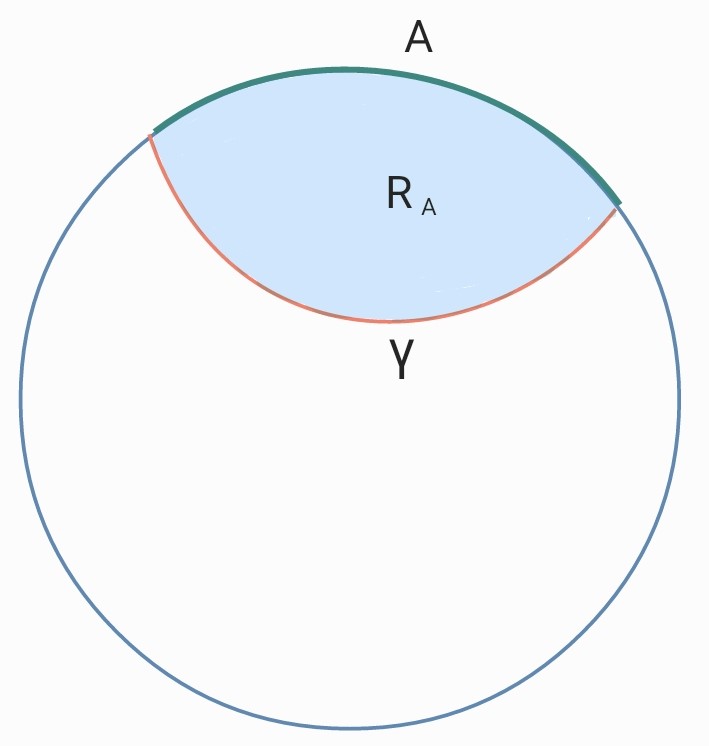}
  \caption{Homology condition: We can see that $\partial \mathcal{R}_A =A \cup \gamma$ }   
  \label{fig:homology} 
  \end{figure}

Before moving on to generalizations of the RT conjecture, lets make a comment in respect. In principle the conjecture is not complete unless we consider that the RT surface would satisfy the \textit{homology condition}.  In essence, this condition requires that $\gamma$ can be smoothly retracted to region $A$ in the boundary. From its construction we already know that $\gamma$ must be anchored to $\partial A$, however, the homology condition is much stronger as a statement. Let us consider a hypersurface of codimension-one  $\mathcal{R}_A\in \mathcal{M}$  whose boundary is defined as $\partial \mathcal{R}_A =A \cup \gamma $, if this hypersurface exists then we say that $\gamma$ is homologous to $A$. We can see a pictorial representation of the condition in Fig.(\ref{fig:homology}).

Now that we have finished our discussion of the Einstein-Hilbert  case, let us look at higher order curvature corrections to it. First we consider the case of quadratic gravity (QCG). The most generic Lagrangian containing curvature terms up to quadratic order  (without considering derivative of the Riemann) is :
\begin{equation}
    \mathcal{L}_{QCG}=R-2\Lambda+\lambda_1R^2+\lambda_2 R^{\mu \nu}R_{\mu \nu} +\lambda_{3}R^{\mu \nu \rho \sigma}R_{\mu \nu \rho \sigma}.
\end{equation}
We can follow the same recipe as before, but now we should consider in the replica-per-bulk action the contributions from this higher order terms. To do this we can use the regularization of higher order curvature invariants of Eq.\eqref{Ricci scalar squared}-\eqref{Riemann Tensor}. This will allow us to obtain the EE as \cite{Fursaev:2013fta}:
\begin{align}
    S(A)&=\frac{1}{4G}\left[\mathcal{A}[\gamma]+\int_{\gamma}d^{d-1}x\sqrt{\sigma} \left(2\lambda_1R+\lambda_2 \left( R^{a}_{a}-\frac{1}{2}\left(K^{a i}_i\right)^2\right) \right.\right.\\
   & \left.\left. +2\lambda_3\left( R^{a b}_{\:\:\: a b}-K^{a i}_{\:\: j} K^{j}_{a\:i}\right) \right)\right].\nonumber
\end{align}
 The procedure we have followed is useful so far, however, it will no longer be helpful if we wish to consider higher order contributions to the quadratic case. This is due to the fact that the direct evaluation of higher order curvature (cubic, quartic and so on) invariants on conically-singular manifolds is not known.  Therefore, in order to analyze the case of cubic curvature gravity (CCG) it will be necessary to start from the first prescription mentioned above,  i.e., we introduce a boundary that is close to the locus fixed points where the conditions are changing if we make a variation with respect to $n$. For the case of Einstein gravity, it is possible to choose  adapted  coordinates to the conical singularity to fix the location of $\gamma^{(n)}$ when $n$ approaches $1$ and thus obtain the aforementioned condition of minimum area $K^{a}=0$.
 
Following these steps, and with an increasing degree of difficulty, it is possible to arrive at an expression for the EE for a CFT that is dual to a higher curvature gravity (HCG) theories. The analysis was initially realized in Ref.\cite{Dong:2013qoa} and then in Ref.\cite{Camps:2013zua}, where for an arbitrary HCG lagrangian $\mathcal{L}_{HCG}=\mathcal{L}_{HCG}(R_{\mu\nu\rho\sigma})$  (without considering covariant derivatives of the curvature) they  obtain:
\begin{equation}
S_{HCG}(A)=\frac{1}{8G}\int_{\gamma} d^{d-1}y\sqrt{g}\left[\frac{\partial\mathcal{L}_{HCG}}{\partial R_{z\bar{z}z\bar{z}}}+\sum_{A}\left(\frac{\partial^2\mathcal{L}_{HCG}}{\partial R_{zizj}R_{\bar{z}k\bar{z}l}}\right)\frac{8K_{zij}K_{\bar{z}kl}}{q_A+1}\right]. \label{Dong}
\end{equation}
  Lets clarify some aspects of this equation: 
 \begin{itemize}
     \item $x^{\mu}$ are the coordinates of the  bulk manifold where the index $\mu$ runs from $0$ to $D=d+1$. The codimension-two hypersurface $\gamma$ where the integration of Eq.\eqref{Dong} is performed can be defined as an embedding $x^{\mu}=x^{\mu}(y^i)$ where $y^i$ are the coordinates of the worldvolume of the hypersurface ($i=1,...,d-1$).
     \item The embedding introduces two orthonormal vectors $n_a$ to $\gamma$.  The indices $a, b, c,$ and so on, will denote these two directions.
     \item As mentioned before, we choose a set of adapted coordinates  to $\gamma$  where $y^i$ are tangent to it and the complex coordinates $z$ and $\bar{z}$ are a combination of the normal coordinates such that:
     \begin{equation}
         G_{z\bar{z}}|_{\gamma}=\frac{1}{2}\:\:\: G_{ij}|_{\gamma}=g_{ij}.
     \end{equation}
     
    Typically, one chooses a  suitable set of local
coordinates in which  $\gamma$ is orthogonal to a plane $(\rho, \tau)$ where $\rho$ parameterize the minimal distance to the this hypersurface and $\tau$ is an angular parameter that has a range of $2\pi$. From here it is easy to see that the conical deficit at $\rho=0$ is $(1-\frac{1}{n})$. The complex coordinates are obtained just by a transformation $z=\rho e^{i\tau}$. In this construction the normal vectors have only components in these directions,
\begin{equation}
   n_1=\sqrt{\frac{z}{\bar{z}}}\partial_z+\sqrt{\frac{\bar{z}}{z}}\partial_{\bar{z}}\:\:\:\:;\:\:\:\: n_2=i\left(\sqrt{\frac{z}{\bar{z}}}\partial_z-\sqrt{\frac{\bar{z}}{z}}\partial_{\bar{z}}\right).
\end{equation}
\item  The extrinsic curvature of $\gamma$ along the normal direction is denoted by $K^a_{ij}$ and defined as $K^a_{ij}=e^\mu_i e^\nu_j \nabla_\mu (n^a)_\nu$, where $e^\mu_i = \partial_i x^\mu$ are tangent vectors. In  Eq.\eqref{Dong} the extrinsic curvatures appear as spacetime tensors of the form $K^{\mu}_{\nu \rho}=K^{a}_{i j} (n_a)^{\mu}e_{\nu}^{i}e_{\rho}^{i}$.

\item Note that the right-hand side of Eq.\eqref{Dong} is made up of a sum of two terms involving variations of the Lagrangian with respect to the Riemann tensor. The first term, which corresponds to first variations of $\mathcal{L}_{HCG}$, is the Wald's formula \cite{Wald:1993nt} that has been successfully used to generalized the Bekenstein-Hawking entropy formula for black hole solutions in more general theories of gravity. The second term, comprising second-order variations of $\mathcal{L}_{HCG}$ contracted with extrinsic curvature tensors, arises from a subtle contribution that is absent in Einstein gravity. As mentioned, to arrive to  Eq.\eqref{Dong} one must follow certain regularization process due to the conical singularity appearing in \eqref{EE holographic}. To address the singularity, one should first use adapted coordinates. Next, the curvature invariants of the Lagrangian must be expanded using a parameter $\epsilon=1-\frac{1}{n}$ that approaches zero as $n\rightarrow 1$. Initially, it may seem that only linear terms in the $\epsilon$ expansion should be considered when calculating the EE. However, in higher curvature theories, caution must be exercised as some seemingly second-order terms $O(\epsilon^2)$ can be enhanced to $O(\epsilon)$ after integration due to a possible logarithmic divergence at $\epsilon\rightarrow 0$. The anomalous term arises from this observation and corrects the Wald's formula.
 \end{itemize} 
Ideally, the regularization technique we choose should guarantee that the equations of motion of the theory are satisfied. Initially, Dong overlooked this aspect in Ref. \cite{Dong:2013qoa}. Nonetheless, it would be beneficial to outline his approach below to juxtapose it with the method that does take into account the EOMs.

The anomalous term specified in Eq.\eqref{Dong} should be a polynomial expression in curvature tensors, which can be broken down into three components: $R_{\alpha \beta i j}$, $R_{\alpha i \beta j}$, and $R_{i j k l}$ (where the $\alpha$ index represents the $z$ or $\bar{z}$ coordinates). These components can be further expanded using the extrinsic curvature $K^{a}_{ij}$, the lower-dimensional Riemann tensor $\mathcal{R}{ijkl}$, and auxiliary tensors such as $\Tilde{R}_{\alpha \beta i j}$, $\Tilde{R}_{\alpha i \beta j}$, and $Q_{\alpha \beta i j}$ (which are defined in Ref. \cite{Dong:2013qoa}, but are not essential for the present overview):

\begin{align}
      R_{\alpha \beta i j}&= \Tilde{R}_{\alpha \beta i j}+g^{kl}[K_{\alpha j k}K_{\beta i l}-K_{\alpha i k}K_{\beta j l}],\label{minexpan R}\\
      R_{\alpha i \beta j}&=\Tilde{R}_{\alpha i \beta j}+g^{kl}K_{\alpha j k }K_{\alpha i l } -Q_{\alpha \beta i j},\\
      R_{ijkl}&=\mathcal{R}_{ijkl}+G^{\alpha \beta}[K_{\alpha il}K_{\beta jk}-K_{\alpha ij}K_{\beta kl}].\label{Minimalexpan}
\end{align}

In this expansion, each term is represented by the label $A$, and the value of $q_A$ is determined by adding one for each instance of $Q_{zz ij}$ and $Q_{\bar{z} \bar{z} ij}$, and one-half for each instance of $K_{aij}$, $R_{abci}$, and $R_{aijk}$. After obtaining the $q_A$ for each term in the sum, we must multiply those terms by the weight $1/(q_A+1)$ as you can see in Eq.\eqref{Dong}. Once all terms have been expanded and weighted, equations \eqref{minexpan R}-\eqref{Minimalexpan} can be used to rewrite the entire expression in terms of the original curvature tensor.

The method we have explained is referred to as the \textit{minimal prescription}, and it was developed without enforcing the requirement that the regularized metric in the codimension-two hypersurface must satisfy the EOMs. It would be quite challenging to impose this condition, but there is a possible solution that involves regarding Einstein's equations as the conditions that must be met \cite{Miao:2015iba,Camps:2016gfs} (at least, at the perturbative level this is correct). We will call this procedure the \textit{non-minimal prescription} and it can be implemented in a simpler way by considering that, after the expansions of Eq.\eqref{minexpan R}-\eqref{Minimalexpan} had been done, the terms $R_{z\bar{z} z\bar{z}}$ and $Q_{z \bar{z} i j}$ can be written as :
\begin{align}
     R'_{z\bar{z}z\bar{z}}&=R_{z\bar{z}z\bar{z}}+\frac{1}{2}(K_{zij}K_{\bar{z}}^{ij}-K_z K_{\bar{z}}),\\
      Q'_{z\bar{z}ij}&=Q_{z\bar{z}ij}-K_{zi}^kK_{\bar{z}jk}-K_{zj}^kK_{\bar{z}ik}+\frac{1}{2}K_zK_{\bar{z}ij}+\frac{1}{2}K_{\bar{z}}K_{zij}.
\end{align}

To determine $q_A$ in this method, we add one half for each of the $K_{\alpha ij}$ and $R_{\alpha \beta \gamma i}$ terms and one for each of the $Q_{zz ij}$ and $Q_{\bar{z} \bar{z} ij}$ terms. We can then expand the equation, assign a weight of $(q_A+1)$ to each term, and proceed to express everything in terms of the curvature tensors, as previously described.

The existence of two different regularization methods resulting in two different outcomes for the EE is known as the \textit{splitting problem}. It can be shown that the two splittings coincide in the case of general quadratic curvature gravity \cite{Fursaev:2013fta} and also for Lovelock gravity \cite{deBoer:2011wk, Miao:2014nxa}. The first scenario where we can find a difference between them is when cubic curvature terms are involved. In Ref.\cite{Caceres:2020jrf} E. Cáceres, R. Castillo, A. Vilar starting with the cubic Lagrangian,
\begin{align}
        &\mathcal{L}_{CCG}=R-2\Lambda_0+\mu_1R_{\mu\:\:\nu}^{\:\:\rho\:\:\sigma}R_{\rho\:\:\sigma}^{\:\:\lambda\:\:\tau}R_{\lambda\:\:\tau}^{\:\:\mu\:\:\nu}+\mu_2R^{\mu\nu}_{\:\:\:\:\rho \sigma}R^{\rho \sigma}_{\:\:\:\:\lambda \tau}R^{\lambda\tau}_{\:\:\:\:\mu\nu}
        +\mu_3R^{\mu\nu\rho}_{\:\:\:\:\:\:\:\sigma}R_{\mu\nu\rho\tau}R^{\sigma\tau}\nonumber
        \\
        &+\mu_4R_{\mu\nu\rho\sigma}R^{\mu\nu\rho\sigma}R+\mu_5R^{\mu\rho}R^{\nu\sigma}R_{\mu\nu\rho\sigma}+\mu_6R_{\mu}^{\:\:\nu}R_{\nu}^{\:\:\rho}R_{\rho}^{\:\:\mu}+\mu_7R_{\mu\nu}R^{\mu\nu}R+\mu_8R^3 \label{CubicLagrangian}
\end{align}
determine the HEE functional for the two splittings as :\\
\textbf{Minimal Splitting}
\begin{equation}
  S_{CCG}^{min}=\frac{\mathcal{A}[\Sigma]}{4 G}-\frac{1}{8\pi G}\int_{\Sigma}d^{d-1}x\sqrt{\sigma}\left(S_{R^2}+S_{K^2R}+S_{K^4}^{min}\right), \label{HEE in CCG minimal}
\end{equation}
\textbf{Non-minimal Splitting}
\begin{equation}
  S_{CCG}^{non-min}=\frac{\mathcal{A}[\Sigma]}{4 G}-\frac{1}{8\pi G}\int_{\Sigma}d^{d-1}x\sqrt{\sigma}\left(S_{R^2}+S_{K^2R}+S_{K^4}^{non-min}\right),\label{HEE in CCG nonminimal}
\end{equation}
where
 \begin{align}
         S_{R^2}&=-6\mu_8R^2-2\mu_7\left(R_{\mu\nu}R^{\mu\nu}+R_{a}^{\:\:\:a}R\right)-3\mu_6R_{a\mu}R^{a\mu}\label{SR2}\\
         &-\mu_5\left(2R_{\mu\nu}R_{a}^{\:\;\mu a\nu}-R_{ab}R^{ab}+R_{a}^{\:\:a}R_{b}^{\:\:b}\right)\nonumber\\
         &-2\mu_4\left(R_{\mu\nu\rho\sigma}R^{\mu\nu\rho\sigma}+2RR_{ab}^{\:\:\:\:ab}\right)-\mu_3\left(R_{a\mu\nu\rho}R^{a\mu\nu\rho}-4R_{a\mu}R_{b}^{\:\:a b\mu}\right)\nonumber\\
         &-6\mu_2R_{a b \mu \nu}R^{a b \mu \nu}+3\mu_1\left(R_{a\mu b\nu}R^{a \nu b\mu}-R_{a\mu\:\:\nu}^{\:\:\:\:\:a}R_{b}^{\:\:\mu b\nu}\right),\nonumber\\
        S_{K^2R}&=\mu_7K_aK^aR+\frac{3}{2}\mu_6K_aK^aR_b^{\:\:b}+2\mu_5K_aK^a_{ij}R^{ij} \label{SK2R}\\
        &-\frac{1}{2}\mu_5K_aK^aR_{bc}^{\:\:\:\:bc}+4\mu_4K_{aij}K^{aij}R\nonumber\\&+2\mu_3K_{aik}K^{a\:\:k}_{\:\:j}R^{ij}+\mu_3K_{aij}K^{aij}R_{b}^{\:b}+2\mu_3K_aK_{ij}^aR_b^{\:\:i bj}\nonumber\\&+12\mu_2K_{aik}K^{a\:\:k}_{\:\:j}R_{b}^{\:\:ibj}+3\mu_1K_{aij}K^a_{\:\:kl}R^{ikjl}-\frac{3}{2}\mu_1K_{aij}K^{aij}R_{bc}^{\:\:\:\:bc}\nonumber\\&+6(2\mu_2+\mu_1)K_{aik}K_{bj}^{\:\:\:\:k}R^{abij},\nonumber
 \\
        S_{K^4}^{min}&=\frac{1}{2}\mu_7K_aK^aK_bK^b+\mu_5K_aK^a_{\:\:ij}\left(K_bK^{bij}-2K_{b\:\:k}^{\:\:i}K^{bjk}\right) \label{SK4 minimal}\\
        &-\frac{1}{4}(6\mu_7+3\mu_6\nonumber-8\mu_4)K_aK^aK_{bij}K^{bij}\nonumber\\
        &-\frac{1}{2}(12\mu_4+\mu_3-3\mu_1)K_{aij}K^{aij}K_{bkl}K^{bkl}\nonumber\\
        &-\frac{3}{2}(4\mu_2+3\mu_1)K_{ai}^{\:\:\:\:j}K_{bj}^{\:\:\:k}K^{a\:\:l}_{\:\:k}K^{b\:\:i}_{\:\:l}+(-2\mu_3+3\mu_1)K_{ai}^{\:\:\:j}K^{a\:\:k}_{\:\:j}K_{bk}^{\:\:\:l}K^{b\:\:i}_{\:\:l},\nonumber\\
        S_{K^4}^{non-min}&=+\frac{1}{4}(\mu_5-3\mu_1)K_aK^aK_{bij}K^{bij}-\frac{1}{4}\mu_5K_aK^aK_bK^b \label{SK4 nonminimal}\\
        &+(\mu_3-6\mu_2)K_aK^a_{\:\:ij}K_{b\:\:k}^{\:\:i}K^{bjk}-\mu_3K_aK^a_{\:\:ij}K_bK^{bij}\nonumber\\
        &+\frac{3}{4}\mu_1K_{aij}K^{aij}K_{bkl}K^{bkl}+\frac{3}{2}\mu_1K_{aij}K_{bkl}K^{bij}K^{akl}\nonumber\\
        &-\frac{3}{2}(4\mu_2+3\mu_1)K_{ai}^{\:\:\:\:j}K_{bj}^{\:\:\:k}K^{a\:\:l}_{\:\:k}K^{b\:\:i}_{\:\:l}+3(4\mu_2+\mu_1)K_{ai}^{\:\:\:j}K^{a\:\:k}_{\:\:j}K_{bk}^{\:\:\:l}K^{b\:\:i}_{\:\:l}.\nonumber
\end{align} 

It is worth noting that the discrepancy between the two splittings is evident in the quartic contractions of the extrinsic curvature, which comes from the anomalous term. Conversely, the quadratic curvature term is identical in both splittings. The reason for this is that it arises only from the Wald's part in the EE which is the same in both prescriptions. The last term that remains to be analyzed is the one containing quadratic terms in the extrinsic curvature. Although this term originates from the anomalous part, the result is the same for both prescriptions after the rewriting and counting procedures are applied.

To conclude this section, we should mention that the determination of the correct splitting remains an open question in the case of cubic and higher curvature gravity. Moreover, it is possible that the two splittings presented above are incorrect and a new prescription not yet known is needed.

\section{Kounterterms}\label{Kounterterms section}

One of the entries of the AdS/CFT dictionary is that ultraviolet (UV) effects on the CFT side correspond to infrared (IR) effects on the gravity theory \cite{Susskind:1998dq}. In particular, the correlation functions of the CFT usually suffer from UV divergences so this UV/IR duality tells us that we can renormalize them by dealing with the IR divergences of the on-shell gravitational action.

In Refs.\cite{Bianchi:2001kw,deHaro:2000vlm} it was developed the formalism of holographic renormalization. It is a systematic method in which all divergences of the on-shell bulk action can be cancelled out by adding covariant local boundary counterterms. The form of these counterterms can be worked out from the resolution of the equation of motion near the boundary \cite{Bianchi:2001kw}. In the pure gravity case, the counterterms are described by  local functionals of the boundary metric, the intrinsic curvature of the boundary and its covariant derivatives.The method has been used with success in many situations \cite{Skenderis:2002wp,Bianchi:2001kw,Bianchi:2001de}, however it gets complicated in higher dimensions and for higher curvature gravity theories. The fact that in these scenarios the number of possible counterterms increases drastically, lets us wonder if there exists a different sort of terms whose expressions can be worked out in any dimensions. 

  A different renormalization scheme, that depends on the extrinsic curvature, was introduce in \cite{Olea:2005gb} for the  renormalization of the gravitational action in AdS spaces of even dimensions. Later on, in Ref.\cite{Olea:2006vd}, a procedure for the odd-dimensional cases was found and the concept of Kounterterms was introduced. In the following lines we will review the proposal for the case of Einstein gravity.

Consider an asymptotically locally AdS bulk spacetime (ALAdS)  $\mathcal{M}$, the renormalized action, following the Kounterterms method, is given by :
\begin{equation}
    I_{EH}^{ren}=I_{EH}+I_{KT}
    =\frac{1}{16\pi G}\left(\int_{\mathcal{M}} d^{d+1}x\sqrt{-G}(R-2\Lambda_0)+c^{EH}_d\int_{\partial \mathcal{M}}d^{d}x\sqrt{-h} B_d\right). \label{Renormalized EH action}
\end{equation}
where:
\begin{equation}
  c_d^{EH} =
    \begin{cases}
      \frac{(-1)^{\frac{d+1}{2}}\ell_0^{d-1}}{(\frac{d+1}{2})(d-1)!} & \mbox{if $d$ is odd};\\
      \frac{(-1)^{\frac{d}{2}}\ell_0^{d-2}}{2^{d-3}d((\frac{d}{2}-1)!)^2} & \mbox{if $d$ is even},\\
    \end{cases}   \label{coupling Einstein Hilbert}    
\end{equation}
and the Kounterterms can be expressed as:
\begin{equation}
   B_d =
    \begin{cases}
      &-(d+1)\int_0^1ds\delta^{a_1...a_{d}}_{b_1...b_{d}}K^{b_1}_{a_1}\left(\frac{1}{2}\mathcal{R}_{a_2 a_3}^{b_2 b_3}-s^2K^{b_2}_{a_2}K^{b_3}_{a_3}\right)\times ...\\ &...\times\left(\frac{1}{2}\mathcal{R}_{a_{d-1} a_{d}}^{b_{d-1} b_{d}}-s^2K^{b_{d-1}}_{a_{d-1}}K^{b_{d}}_{a_{d}}\right)  \qquad\qquad\qquad\qquad\qquad\quad \mbox{odd $d$};\\
     & -d\int_0^1ds\int_0^sdt\delta^{a_1...a_{d-1}}_{b_1...b_{d-1}}K^{b_1}_{a_1}\left(\frac{1}{2}\mathcal{R}_{a_2 a_3}^{b_2 b_3}-s^2K^{b_2}_{a_2}K^{b_3}_{a_3}+\frac{t^2}{l_{0}}\delta^{b_2}_{a_2}\delta^{b_3}_{a_3}\right)\times ...\\ &...\times\left(\frac{1}{2}\mathcal{R}_{a_{d-2} a_{d-1}}^{b_{d-2} b_{d-1}}-s^2K^{b_{d-2}}_{a_{d-2}}K^{b_{d-1}}_{a_{d-1}}+\frac{t^2}{l_{0}}\delta^{b_{d-2}}_{a_{d-2}}\delta^{b_{d-1}}_{a_{d-1}}\right) \quad\qquad\mbox{even $d$ }.
    \end{cases}   \label{Kounterterms}  
\end{equation}
By using Eq.\eqref{Renormalized EH action}, it has been possible to replicate the correct asymptotic charges and the thermodynamic aspects of  ALAdS manifolds \cite{Olea:2005gb,Aros:1999kt,Mora:2004rx}. Note that the usual Gibbons-Hawking boundary term is not included in this formulation. The reason is that the Kounterterms do the work of defining a well-posed variational problem (with the standard  Dirichlet boundary condition for the CFT metric, $\delta g^{(0)}=0$) as well as rendering the action finite.

 An interesting aspect of the Kounterterm scheme is that it can be more easily related to what is known in the mathematical literature as \textit{renormalized volume}. In Ref.\cite{Anastasiou:2018mfk} it was shown that in $D=4$ and $D=6$ the  renormalized euclidean EH action can be used to reproduce known results of the renormalized volume of (A)AdS Einstein manifolds \cite{Anderson2000,Chang:2005ska}. Moreover, they proposed that the relation between the two, for even $D=d+1$ bulk dimension could be written as:
\begin{equation} 
 I_{EH}^{ren}=\left(-\frac{2d}{\ell^2_0}\right)\frac{1}{16\pi G}Vol^{ren}(\mathcal{M}).
\label{renormalizedVolumen}
\end{equation}
An intuitive way to see how this proportionality constant appears is to consider  the case of Einstein-AdS spacetimes where $R=-\frac{d(d+1)}{l^2_0}$ and $\Lambda_0=-\frac{d(d-1)}{2l^2_0}$. Substituting it into Eq.\eqref{Renormalized EH action} we obtain that the EH Lagrangian becomes $R-2\Lambda=-2d/\ell_0^2$ and therefore the renormalized volume reads:
\begin{equation}
    Vol^{ren}(\mathcal{M})=Vol(\mathcal{M})-\frac{\ell^2_0}{2d}c_d^{EH}\int_{\partial \mathcal{M}}d^{d}x\sqrt{-h} B_d.
\end{equation}
Based on the previous considerations we are aiming in renormalizing arbitrary HCG actions. It is supposed that for a general HCG theory of gravity in lower dimensions $D\leq 5$, renormalization is achieved by adding extrinsic counterterms to the action as before but with a different coupling \cite{Araya:2021atx}:
  \begin{equation}
    I_{HCG}^{ren}=I_{HCG}+\frac{c^{HCG}_d}{16\pi G}\int_{\partial \mathcal{M}}d^{d}x\sqrt{-h} B_d.\label{RenormalizedHCG}
\end{equation}
Now, we consider the on-shell higher curvature Lagrangian for the maximally symmetric case to be $\mathcal{L}_{HCG}=\mathcal{L}_{HCG}(\ell_0(\ell_{eff},\{\alpha_i\}),\ell_{eff},\{\alpha_i\})$ where $\ell_{eff}$ is the effective AdS radius and $\{\alpha_i\}$ are the couplings of the theory. Then, replacing this Lagrangian on Eq.\eqref{RenormalizedHCG} we will get:
  \begin{equation}
    I_{HCG}^{ren}=\frac{1}{16\pi G} \mathcal{L}_{HCG}\left(Vol(\mathcal{M})+\frac{c_d^{HCG}}{\mathcal{L}_{HCG}}\int_{\partial \mathcal{M}}d^dx\sqrt{-h} B_d\right).\label{HCGren}
\end{equation} 
The renormalization is achieved when $c_d^{HCG}=-\frac{\ell^2_{eff}}{2d}\mathcal{L}_{AdS}c_d^{EH}$ because in that case, the parenthesis of  Eq.\eqref{HCGren} would be just the renormalized volume. In principle, it is necessary to use the EOMs for each particular theory to obtain the relation between $\ell_0$ and $\ell_{eff}$. However, there is an easier way to accomplish this using only the Lagrangian information. We consider the EOM of an arbitrary HCG theory \cite{Padmanabhan:2010zzb}:
 \begin{equation}
    \mathcal{E}^{\mu}_{\nu}=P^{\rho \lambda}_{\alpha \nu}R^{\alpha \mu}_{\rho \lambda}-\frac{1}{2}\delta^{\mu}_{\nu}\mathcal{L}_{HCG}-2\nabla^{\alpha}\nabla_{\rho}P^{\rho \mu}_{\nu \alpha}=0,\label{EOM}
 \end{equation}
 where $P^{\rho \mu}_{\nu \alpha}=\frac{\partial \mathcal{L}}{\partial R_{\rho \mu}^{\nu \alpha}}$. Using the solution of pure AdS we found that :
 \begin{equation}
    \nabla_{\rho}P^{\rho \mu}_{\nu \alpha}=0\qquad\qquad
    R^{\alpha \mu}_{\rho \lambda}=\frac{2\Lambda_{eff}}{d(d-1)}\delta^{\alpha \mu}_{\rho \lambda}\qquad\qquad
    P^{\alpha \mu}_{\rho \lambda}=\frac{2e}{d(d-1)}\delta^{\alpha \mu}_{\rho \lambda}.
\end{equation} 
 where $\Lambda_{eff}=-\frac{d(d-1)}{2\ell_{eff}^2}$ is the effective cosmological constant. The factor $e$ depends on the specific theory and appears because one can notice that $P^{\rho \mu}_{\nu \alpha}=\frac{\partial \mathcal{L}}{\partial R_{\rho \mu}^{\nu \alpha}}$ have the same symmetries as the Riemann tensor and therefore in the maximally symmetric ansatz should be proportional to $ \delta_{\rho \mu}^{\nu \alpha}$ being $e$ that proportionality constant. Combining the previous formulas, one expresses the Lagrangian as $ \mathcal{L}_{HCG}=\frac{16e\Lambda_{eff}d}{d^2(d-1)^2}$. Following Ref.\cite{Bueno:2016ypa} we  obtain the relation:
\begin{align}
  \frac{d\mathcal{L}_{HCG}}{d\Lambda_{eff}}=\frac{\partial\mathcal{L}_{HCG}}{\partial R_{\rho \mu}^{\nu \alpha}}\frac{d R_{\rho \mu}^{\nu \alpha}}{d\Lambda_{eff}}&=P^{\rho \mu}_{\nu \alpha}\frac{2}{d(d-1)}\delta^{\nu\alpha}_{\rho \mu}\nonumber=\frac{1}{\Lambda_{eff}}\frac{8e\Lambda_{eff}d(d+1)}{d^2(d-1)^2} \\
  \Longrightarrow &  \Lambda_{eff}\frac{d\mathcal{L}_{HCG}}{d\Lambda_{eff}}=\frac{d+1}{2}\mathcal{L}_{HCG}.\label{Relation}
 \end{align} 
We should note that the right hand side of this last equation does not depend on $\ell_0$ since by taking the derivative with respect to $\Lambda_{eff}$ we have gotten rid of it. If we substitute Eq.\eqref{Relation} in the definition of the coupling for a CCG theory we  obtain a formula in which the coupling can be calculated directly from the HCG Lagrangian. Then the coupling for the general higher curvature theory is :
\begin{equation}
      c_d^{HCG}=-\frac{\ell^2_{eff}\Lambda_{eff}}{d(d+1)}\frac{d\mathcal{L}_{HCG}}{d\Lambda_{eff}}c_d^{EH}=\frac{\ell^3_{eff}}{2d(d+1)}\frac{d\mathcal{L}_{HCG}}{d\ell_{eff}}c_d^{EH}.\label{HigherCurvaturecoupling}
\end{equation}
This coupling ensures that the divergences in the action, for the maximally-symmetric solution, are eliminated. Once  fixed, it can be verified whether the same boundary term is effective for AlAdS solutions other than the pure AdS configuration. Although the procedure has been performed considering even bulk dimension, the same coupling is assumed to work for odd dimensions but considering the definition of $c_d^{EH}$ in those dimensions (see Eq.\eqref{coupling Einstein Hilbert}).

We will not elaborate much more on this because in the present work we will be more interested  in the study of the implementation of this renormalization method in codimension-two for the extraction of the universal part of the EE. So, for a more extensive and detailed explanation of this renormalization scheme (and its generalization to HCG) we suggest the readers the bibliography \cite{Araya:2021atx,Anastasiou:2019ldc,Anastasiou:2020zwc} .

\section{Renormalized Entanglement Entropy} \label{renormalized EE}

The EE for QFTs is a UV divergent quantity, characterized by the scaling of the leading divergence with the area  of the entangling region. In the case of discrete lattice systems, there is no divergence as the UV cutoff scale is naturally set by the lattice spacing. However, if we are interested in entanglement entropy in the context of field theory, the divergence poses a problem unless a renormalization procedure is considered.

First attempts on resolving the issue where made in  \cite{Liu:2012eea,Klebanov:2012yf,Liu:2013una}  by using entanglement differentiation with respect to geometric parameters characterizing the entangling region. A particular problem of this procedure is that it depends on the geometry of the entangling region, and therefore it is hard to implement in situations where the shape of the region itself is being varied. 

Later on it was shown, in \cite{Taylor:2016aoi}, that EE renormalization can be implemented from the standard holographic renormalization counterterms for asymptotically locally anti-de Sitter (AlAdS) spacetimes. This task is carried out by evaluating the typical counterterms for Einstein gravity on the
singular conic boundary  spacetime, which is conformal to the replica CFT manifold.

In this project, however, we will focus on an HEE  renormalization procedure inherited from the bulk renormalization by Kounterterms \cite{Anastasiou:2018mfk,Anastasiou:2019ldc, Anastasiou:2017xjr,Anastasiou:2018rla}. To implement it, we need to see how the renormalized action behaves in presence of squashed-cone singularities.

For even  bulk dimensional manifolds with boundary we know that the Euler-Theorem is stated as,

\begin{equation}
    \int_{\mathcal{M}}d^{d+1}x\mathcal{E}_{d+1}=\left(4\pi\right)^{(d+1)/2}\left(\frac{d+1}{2}\right)!\chi(\mathcal{M})+\int_{\partial \mathcal{M}}d^{d}xB_d .\label{Euler theorem}
\end{equation}

where $\mathcal{E}_{d+1}$ is the Euler density in $d+1$ dimensions and $B_d$  is the Chern form, which appears as the correction to the Euler characteristic $\chi$ in a manifold with boundary. In Ref.\cite{Fursaev:2013fta} it was proven that the Euler density self replicates for $D=4$ in the presence of squashed cone singularities:

\begin{equation}
    \int_{\hat{\mathcal{M}}^{(n)}}d^{3}x\mathcal{E}_{4}= \int_{\hat{\mathcal{M}}^{(n)} \backslash \gamma^{(n)}}d^{4}x\mathcal{E}_{4}+8\pi \left(1-\frac{1}{n}\right)\int_{\gamma^{(n)}}d^{2}x\mathcal{E}_{2} .\label{self replicate gauss bonnet}
\end{equation}

In fact, the Euler density in four dimension is just the Gauss-Bonnet term $GB=R_{\mu\nu\rho\sigma}R^{\mu\nu\rho\sigma}-4R_{\mu \nu}R^{\mu\nu}+R^2$ and therefore you can use Eq.\eqref{Ricci scalar squared}-\eqref{Riemann Tensor} to prove Eq.\eqref{self replicate gauss bonnet}. In the same reference, it was also shown that the Euler characteristic satisfies: 
\begin{equation}
    \chi_4\left(\hat{\mathcal{M}}^{(n)}\right)= \chi_4\left(\mathcal{M}^{(n)} \backslash \gamma^{(n)}\right)+ \left(1-\frac{1}{n}\right) \chi_2\left(\gamma^{(n)}\right).\label{self replicate euler characteristic}
\end{equation}
Using Euler's theorem, and considering Eq.\eqref{self replicate euler characteristic} and Eq.\eqref{self replicate gauss bonnet}, it can be shown that the Chern form must also admit this self-replication property:
\begin{equation}
    \int_{\partial \hat{\mathcal{M}}^{(n)}}d^{3}x B_3=\int_{\partial \hat{\mathcal{M}}^{(n)} \backslash \partial\gamma^{(n)}}d^{3}x B_3+8\pi\left(1-\frac{1}{n}\right)\int_{\partial \gamma^{(n)}}dx B_1.
\end{equation}
If we assume that this property  holds for arbitrary even dimension \footnote{In fact, this statement have been proven where the singularity have $U(1)$ symmetry \cite{Fursaev:1995ef}.} we get:
\begin{equation}
      \int_{\partial \hat{\mathcal{M}}^{(n)}}d^{d}x B_d=\int_{\partial \hat{\mathcal{M}}^{(n)} \backslash \partial\gamma^{(n)}}d^{d}x B_d+4\pi\left(\frac{d+1}{2}\right)\left(1-\frac{1}{n}\right)\int_{\partial \gamma^{(n)}}d^{d-2}x B_{d-2}.
\end{equation}
 It is worth-mentioning that the Euler density is only defined in even bulk dimensions and therefore Eq.\eqref{Euler theorem} cannot be extended to odd dimensions. Moreover, in odd  bulk dimension the Kounterterms are not related to the Chern form. However, even though $B_d$ corresponds to different geometrical objects in even and odd dimension (as you can see in their definition in Eq.\eqref{Kounterterms}), it can be shown that for both cases the following equation is satisfied \cite{Anastasiou:2019ldc}:
\begin{equation}
    \int_{\partial \hat{\mathcal{M}}^{(n)}}dx^{d}B_d=\frac{1}{n}\int_{\partial \mathcal{M}^{(n)} \backslash \gamma^{(n)}}d^{d}x B_d+4\pi\bigg\lfloor \frac{d+1}{2}\bigg\rfloor\left(1-\frac{1}{n}\right)\int_{\partial \gamma^{(n)}}d^{d-2}x B_{d-2}.
\end{equation}
This last equation tells us that if we use the renormalized action to obtain the EE in Eq.\eqref{EE holographic} then an extra term in codimension-three arises that extracts the universal part of the EE, and reads:
\begin{equation}
 S_{KT}=\frac{c_d^{HCG}}{4G}\bigg\lfloor \frac{d+1}{2}\bigg\rfloor\int_{\partial \gamma}d^{d-2}x B_{d-2}.
\end{equation}
 
With this last equation we have finished reviewing the material required for understanding this work. Our results concerning entanglement entropy in cubic curvature gravity theories will be presented in the following chapter. In section (\ref{Fixing the coupling in CCG}) we find the renormalized action of CCG gravity. Then in (\ref{Renormalized HEE in CCG}) we proposed an expression for the universal part of the HEE. In (\ref{First test: Hypersphere}) and (\ref{Second test: Cylinder}) we test our results for a spherical and cylindrical entangling region and from that we obtain the type-A and type-B anomaly coefficients for  this particular type of CFTs. We obtain our last results in (\ref{Deformed sphere I})  and (\ref{Deformed sphere II}) where we consider a deformed sphere to obtain the $C_t$ and the $t_4$ for a CFT dual to a general cubic curvature gravity.

%% file: Chapter2/Chapter2.tex
\chapter{Results on HEE in CCG} \label{chapter 3}

\section{Fixing the coupling in CCG} \label{Fixing the coupling in CCG}
 
In this section  we are going to follow the procedure described in section (\ref{Kounterterms section}) to fix the coupling for the Kounterterm that will render the generic cubic curvature gravitational action finite (without considering terms of covariant derivatives of the curvature tensor).  The prescription tells us that now we have to evaluate the Lagrangian of Eq.\eqref{CubicLagrangian} in the pure $AdS_{d+1}$ solution and then take the derivative with respect to the effective radius to find the coupling that will achieve the renormalization of the action. In this ansatz the cubic curvature invariants become:

\begin{minipage}{.47\textwidth}
 \begin{align}
&R_{\mu\:\:\nu}^{\:\:\rho\:\:\sigma}R_{\rho\:\:\sigma}^{\:\:\lambda\:\:\tau}R_{\lambda\:\:\tau}^{\:\:\mu\:\:\nu}=-\frac{d(d^2-1)}{\ell^6_{eff}}\label{maximal Riemmann}\\
 &R^{\mu\nu}_{\:\:\:\:\rho \sigma}R^{\rho \sigma}_{\:\:\:\:\lambda \tau}R^{\lambda\tau}_{\:\:\:\:\mu\nu}=-\frac{4d(d+1)}{\ell^6_{eff}}\\
 &R^{\mu\nu\rho}_{\:\:\:\:\:\:\:\sigma}R_{\mu\nu\rho\tau}R^{\sigma\tau}=-\frac{2d^2(d+1)}{l^6_{eff}}\\
 &R_{\mu\nu\rho\sigma}R^{\mu\nu\rho\sigma}R=-\frac{2d^2(d+1)^2}{\ell^6_{eff}}
\end{align}
\end{minipage}
\begin{minipage}{.47\textwidth}
\begin{align}
  &R^{\mu\rho}R^{\nu\sigma}R_{\mu\nu\rho\sigma}=-\frac{d^3(d+1)}{\ell^6_{eff}}\\
  & R R_{\mu}^{\:\:\nu}R_{\nu}^{\:\:\rho}R_{\rho}^{\:\:\mu}=-\frac{d^3(d+1)}{\ell^6_{eff}}\\
  &R_{\mu\nu}R^{\mu\nu}R=-\frac{d^3(d+1)^2}{l^6_{eff}}\\
  &R^3=-\frac{d^3(d+1)^3}{\ell^6_{eff}}.\label{maximal Ricci}
\end{align}
\end{minipage}\\

Notice that the cosmological constant does not depend explicitly on $\ell_{eff}$. However, after imposing the equation of motion it is possible to see that the cosmological constant is a polynomial on $1/\ell_{eff}$ with coupling dependent coefficients. After that one can replace the expression obtained in the Lagrangian to eliminate the $\ell_0$ dependence of it. This process is very messy and large so in section (\ref{Kounterterms section}) we provided a shortcut to computed it. We start by replacing Eq.\eqref{maximal Riemmann}-\eqref{maximal Ricci} in Eq.\eqref{CubicLagrangian}:
 \begin{align}
 \mathcal{L}=&-\frac{d(d+1)}{\ell^2_{eff}}-2\Lambda_0-\frac{d(d+1)}{\ell^6_{eff}}[\mu_1(d-1)+4\mu_2+2\mu_3d+2\mu_4d(d+1)\nonumber\\&+\mu_5d^2+\mu_6d^2+\mu_7d^2(d+1)+\mu_8d^2(d+1)^2].  \end{align}
Taking the derivative with respect to $\ell_{eff}$ the cosmological constant vanishes and therefore we do not have to use the EOMs explicitly to fix the coupling. Taking the derivative we get:
 \begin{align}
   \frac{d\mathcal{L}}{d\ell_{eff}}=&\frac{2d(d+1)}{\ell^3_{eff}}+\frac{6d(d+1)}{\ell^7_{eff}}[\mu_1(d-1)+4\mu_2+2\mu_3d+2\mu_4d(d+1)\nonumber\\&+\mu_5d^2+\mu_6d^2+\mu_7d^2(d+1)+\mu_8d^2(d+1)^2].
 \end{align}
After replacing Eq.\eqref{HigherCurvaturecoupling}, we obtain the expression for the coupling of this particular theory for even and odd  bulk dimension as:
\begin{equation}
  c_d^{CCG} =
    \begin{cases}
     a_{d} \frac{(-1)^{\frac{d+1}{2}}\ell_{eff}^{d-1}}{(\frac{d+1}{2})(d-1)!} & \mbox{if $d$ is odd};\\
     a_{d} \frac{(-1)^{\frac{d}{2}}\ell_{eff}^{d-2}}{2^{d-3}d((\frac{d}{2}-1)!)^2} & \mbox{if $d$ is even,}
    \end{cases}   \label{coupling HEE in CCG}  
\end{equation}
where
\begin{align}
 a_d=&1+\frac{3}{\ell^4_{eff}}\big(\mu_1(d-1)+4\mu_2+2\mu_3d+2\mu_4d(d+1)+\mu_5d^2\nonumber\\&+\mu_6d^2 +\mu_7d^2(d+1)
 +\mu_8d^2(d+1)^2\big).\label{ad}
\end{align}
A nontrivial test is that after fixing the Lagrangian couplings to correspond to Lovelock theory in $d=6$, the Eq.\eqref{ad} reproduce the result of Refs.\cite{Anastasiou:2021jcv, Kofinas:2007ns}. Now, we assume that the same coupling that achieve the renormalization of pure AdS also renormalize any other AlAdS action (in $D\leq 5$). Therefore, we proposed that the renormalized gravitational action is given by :
\begin{equation}
    I_{CCG}^{ren}=\frac{1}{16\pi G}\int_{\mathcal{M}}d^{d+1}x\sqrt{|G|}\mathcal{L}_{CCG}+\frac{c^{CCG}_d}{16\pi G}\int_{\partial \mathcal{M}}d^{d}x\sqrt{|h|} B_d.\label{RenormalizedCubicCurvatureAction}
\end{equation}
Now that we have our renormalized gravitational action, in the next section, we will exploit the self-replication property of the Kounterterms to obtain an expression for the $S_{KT}$, which will isolate the universal component of the EE.

\section{Renormalized HEE in CCG}\label{Renormalized HEE in CCG}

In section (\ref{renormalized EE}) we explore a procedure inherited from the bulk renormalization by Kounterterms to cancel the divergences appearing in the EE. This procedure have already been used in the context of higher curvature gravity theories. In Ref.\cite{Anastasiou:2021swo} a prescription was given for a generic quadratic curvature theory and in Ref.\cite{Anastasiou:2021jcv} the authors achieve the renormalization of the EE in Lovelock gravity. It should be noted that in these two instances, the splitting problem (outlined in section (\ref{HEE})), did not arise, implying that the minimal and non-minimal prescriptions yielded the same HEE functional. In contrast, for CCG, as stated in Ref. \cite{Caceres:2020jrf}, two distinct functionals would emerge. Based on what we have developed until now, we proposed that the universal part of the EE for the two splitting can be written as:

\textbf{Minimal Splitting}

\begin{equation}
   S_{CCG}^{Univ-min} =S_{CCG}^{min}+S_{KT}.\label{minimal Spliting}
\end{equation}

\textbf{Non-minimal Splitting}

\begin{equation}
  S_{CCG}^{Univ-non-min} =S_{CCG}^{non-min}+S_{KT},\label{non-minimal Spliting}
\end{equation}

where the expressions for $S_{CCG}^{min}$ and $S_{CCG}^{Univ-non-min}$ can be found in Eq.\eqref{HEE in CCG minimal} and Eq.\eqref{HEE in CCG nonminimal}. In this case the codimension-three Kouterterm is :
\begin{equation}
    S_{KT}=\frac{c_d^{CCG}}{4G}\bigg\lfloor \frac{d+1}{2}\bigg\rfloor\int_{\partial\Sigma}d^{d-2}x\sqrt{\tilde{\sigma}}B_{d-2}.\label{HEE in CCG Kounterterm}
\end{equation}

It is easy to see where this expression comes from. One can use the replica trick for the renormalized bulk action of Eq.\eqref{RenormalizedCubicCurvatureAction}. The divergent part can be evaluated using Dong's functional \cite{Dong:2013qoa}  (minimum) or Camp's functional  \cite{Camps:2013zua,Camps:2016gfs} (non-minimum) as was done in the paper mentioned above  (E. Caceres, et.al. \cite{Caceres:2020jrf}) and the contribution of the Kounterterms can be obtained as in section  (\ref{renormalized EE}) but with the CCG coupling of Eq.\eqref{coupling HEE in CCG}. As mentioned before, the differences between the two splittings come from the term $S_{K^4}$ that depends exclusively of contractions of four extrinsic curvature tensor as you can see in Eq.\eqref{SK4 minimal} and Eq.\eqref{SK4 nonminimal}. If we consider the Fefferman-Graham (FG) coordinates for the bulk metric  \cite{fefferman1985elie} where the conformal boundary is located at $z=0$ we can easily find that, near the boundary, the extrinsic curvature of the codimension-two surface scales as $O(z)$\footnote{This is the case for the extrinsic curvature expressed with one contravariant and one covariant index. The reason to be interested in this particular form of the extrinsic curvature is that any term in $S_{K^4}$ can be expressed as contraction of this object.} meaning that the  $S_{K^4}$ should scales as $O(z^4)$ for the two splittings. Considering that the squared root of the determinant of the induced metric of the cod-2 embedding scales as $O(z^{-(d-1)})$ we  notice that the term in the FG expansion that contributes to the logarithmic anomaly of the universal EE in even dimension is  $z^{d-2}$. We can see that for $d=2$ and $d=4$ the $S_{K^4}$ term will not contribute and therefore the universal part for the two splittings will be the same. In odd dimensional CFTs the universal part is finite, therefore the upper limit in the integration of $z$ of the $S_{K^4}$ will always contribute to it meaning that we can always find a difference between the the universal part of each splitting. As a side note, we should notice that we are not going to see a difference in the divergence structure unless we go to $d=7$ dimensions or higher due to the fact that $\sqrt{\sigma}S_{K^4}=O(z^{d-5})$.

Finally, it should be noted that this renormalization method allows us to determine the Weyl anomaly, present in even-dimensional CFTs, which is characterized by the coefficient that multiplies the logarithmic divergence of the EE. Considering an entanglement region $A$, whose width is described by $L$, we would obtain that the associated renormalized EE is $S^{\text{Univ}} \left[A\right]=c_0\ln(\frac{L}{\delta})$, where: 

\begin{equation}
    c_0=(-1)^{d/2+1}2A\chi[\partial \Sigma]-\sum_{i}C_i\left(\partial_n\int\limits_{\mathcal{M}_n}I_i\right) \,.\label{central charges}
\end{equation}

The first coefficient on the right-hand side of Eq.\eqref{central charges} is the type-$A$ central charge, which is multiplied by the Euler characteristic denoted by $\chi[\partial \Sigma]$. The second part is formed by the coefficients $C_i$, which represent the type-$B$ central charges, and the $I_i$, which are the conformal invariants in that dimension. The  $A$ charge will be the only relevant factor for universal EE when a ball-shaped entanglement region is chosen, while for a cylinder-shaped region, only the $B$ charge will survive.

The upcoming analysis will involve extracting the universal component of the EE for highly-symmetric shapes in a CFT that has a CCG bulk dual. This will be achieved through the use of the renormalization scheme outlined in Eqs.\eqref{minimal Spliting} and \eqref{non-minimal Spliting}. With this, we determine the $A$ and $F$ charges \cite{Jafferis:2011zi,Klebanov:2011gs}. These two quantities are expected to characterize the degrees of freedom of the theory and to decrease monotonically along the renormalization group flow \cite{Myers:2010xs,Myers:2010tj}. Next, in section \ref{Second test: Cylinder}, we compute the EE for a cylindrical region in a four-dimensional CFT and, from it, extract the type-$B$ charge.

\section{First test: Hypersphere} \label{First test: Hypersphere}

We want to obtain the EE for the vacuum state for an entangling region $A$ that has the shape of an hypersphere. The corresponding bulk of this state is  pure $AdS_{d+1}$, and its metric in Poincaré coordinates is:

\begin{equation}
    ds^2=\frac{\ell^2_{eff}}{z^2}\left(d\tau^2+dz^2+dr^2+r^2d\Omega_{d-2}^2\right), \label{Poincare}
\end{equation}
 
 where $\Omega_{d-2}$ represents the angular directions of an $\mathbb{S}^{d-2}$ sphere. Now, the region of interest in a Cauchy slice $\tau=cte$ of the CFT is going to be described by:
 \begin{equation}
     A:\left[\tau=cte, z=\delta \:\:and\:\: r\leq R\right], \label{Entangling A}
 \end{equation}
with entangling surface located in $r=R$ and also we have introduce a cutoff at $z=\delta$. The bulk extremal surface could be found by assuming that the $r$ coordinate extends to the bulk and due to the radial symmetry of the entangling surface the embedding should have the form $r=f(z)$. Replacing in the EOMs of the theory one obtains a differential equation for $f(z)$ whose solutions reads  $f(z)=\sqrt{R^2-z^2}$. Therefore the embedding\footnote{Here we change the notation a bit and call $\Sigma$ to the hypersurface of codimension two in the bulk where the EE calculation is located. The induced metric would be $\sigma$ with indices $\bar{\mu},\bar{\nu}$. } is a spherical hemisphere of same radius: 
  \begin{equation}
     \Sigma:\left[\tau=cte,  r^2+z^2=R^2\right] \label{Entangling region bulk}
 \end{equation}

 where $\partial\Sigma= \partial A$ meaning that is anchored the entangling regions.
 In Poincare coordinates, the induced metric reads:
 \begin{equation}
    ds_{\Sigma}^2=\sigma_{\bar{\mu}\bar{\nu}}dy^{\bar{\mu}}dy^{\bar{\nu}}=\frac{\ell^2_{eff}}{z^2}\left(\frac{R^2}{R^2-z^2}dz^2+(R^2-z^2)\Omega_{d-2}^2\right) ,\label{induced metric}
\end{equation}
 with normal vectors
 \begin{equation}
     n^{(1)}_{\mu}=\left(\frac{\ell_{eff}}{z},0,0,...,0\right) \quad;\quad n^{(2)}_{\mu}=\left(0,\frac{\ell_{eff}}{\sqrt{r^2+z^2}},\frac{\ell_{eff}r}{z\sqrt{r^2+z^2}},...,0\right). \label{normal vector sphere}
 \end{equation}
Due to the fact that the extrinsic curvature can be formulated in terms of a Lie derivative with respect to the normal vectors, we can see that all components of $ K_{\bar{\mu}\bar{\nu}}^{(1)}$ and $ K_{\bar{\mu}\bar{\nu}}^{(2)}$ must vanish.  This is because the normal vectors to this hypersurface are  Killing vectors of the space. From this observation we realize that in this particular case the minimum and non-minimum splitting will be equal because $S_{K^4}^{min}=S_{K^4}^{non-min}=0$. Moreover the term $S_{K^2R}$ would also be zero and therefore the EE would be:
\begin{equation}
    S_{CCG}^{min} = S_{CCG}^{non-min}= S_{CCG} =\frac{\mathcal{A}[\Sigma]}{4 G}-\frac{1}{8 G}\int_{\Sigma}d^{d-1}x\sqrt{\sigma}S_{R^2}.
\end{equation}
To find the term $S_{R^2}$ we have to compute all the squared curvature invariants appearing in Eq.\eqref{SR2} using the metric of Eq.\eqref{Poincare} and the normal vectors of \eqref{normal vector sphere}:

\begin{minipage}{.47\textwidth}
 \begin{align}
& R^2=\frac{d^2(d+1)^2}{\ell^4_{eff}}\label{R2}\\
    &R_{\mu\nu}R^{\mu\nu}=\frac{d^2(d+1)}{\ell^4_{eff}}\\
    &R_{a}^{\:\:\:a}R=\frac{2d^2(d+1)}{\ell^4_{eff}}\\
    &R_{a\mu}R^{a\mu}=\frac{2d^2}{\ell^4_{eff}}\\
   & R_{\mu\nu}R_{a}^{\:\;\mu a\nu}=\frac{2d^2}{\ell^4_{eff}}\\
  & R_{ab}R^{ab}=\frac{2d^2}{\ell^4_{eff}}\\
   & R_{a}^{\:\:a}R_{b}^{\:\:b}=\frac{4d^2}{\ell^4_{eff}}
\end{align}
\end{minipage}
\begin{minipage}{.47\textwidth}
\begin{align}
  &R_{\mu\nu\rho\sigma}R^{\mu\nu\rho\sigma}=\frac{2d(d+1)}{\ell^4_{eff}}\\
    &RR_{ab}^{\:\:\:\:ab}=\frac{2d(d+1)}{\ell^4_{eff}}\\
   & R_{a\mu\nu\rho}R^{a\mu\nu\rho}=\frac{4d}{\ell^4_{eff}}\\
    &R_{a\mu}R_{b}^{\:\:a b\mu}=\frac{2d}{\ell^4_{eff}}\\
    &R_{a b \mu \nu}R^{a b \mu \nu}=\frac{4}{\ell^4_{eff}}\\
   & R_{a\mu b\nu}R^{a \nu b\mu}=\frac{2d}{\ell^4_{eff}}\\
   & R_{a\mu\:\:\nu}^{\:\:\:\:\:a}R_{b}^{\:\:\mu b\nu}=\frac{2(2d-1)}{\ell^4_{eff}}.\label{R last}
\end{align}
\end{minipage}\\

Replacing Eqs.\eqref{R2}-\eqref{R last} in Eq.\eqref{SR2} gives:
 \begin{align}
      S_{R^2}=&-\frac{6}{\ell^4_{eff}}\left(\mu_1(d-1)+4\mu_2+2\mu_3d+2\mu_4d(d+1)+\mu_5d^2+\mu_6d^2\right.\nonumber\\& \left. +\mu_7d^2(d+1)+\mu_8d^2(d+1)^2\right) .
 \end{align}
Notice that this term does not depend on the coordinates and therefore its contribution to the EE   would be proportional to the area of the codimension-two hypersurface. Therefore the whole EE would be just proportional to the area and we can see that the proportionality constant is related to $a_d$ of Eq.\eqref{coupling HEE in CCG}. Thus:
 \begin{equation}
    S_{CCG}^{Univ} =\frac{a_d\mathcal{A}[\Sigma]}{4 G}  .
 \end{equation}
If we add the EE Kounterterm of Eq.\eqref{HEE in CCG Kounterterm} and express $c_d^{CCG}=a_dc_d^{EH}$ we can obtain an equation for the Universal part of the EE, as shown below: 
\begin{equation}
     S_{CCG}=\frac{a_d}{4 G}\left(\mathcal{A}[\Sigma]+c_d^{EH}\bigg\lfloor \frac{d+1}{2}\bigg\rfloor\int_{\partial\Sigma}d^{d-2}x\sqrt{\tilde{\sigma}}B_{d-2}\right)=\frac{a_d}{4G}\mathcal{A}^{Univ}[\Sigma].\label{renormalized EE for a sphere}
\end{equation}
It can be observed from that the universal HEE for this region is proportional to the universal part of the minimum surface area. The explicit form of $\mathcal{A}^{Univ}[\Sigma]$ has already been obtained in \cite{Anastasiou:2017xjr, Anastasiou:2018rla} (Also see Appendix of \cite{Anastasiou:2021swo}) :
\begin{equation}
\mathcal{A}^{Univ}[\Sigma]=
    \begin{cases}
      (-1)^{\frac{d-1}{2}}\frac{2^{d-1}\pi^{\frac{d-1}{2}}\ell_{eff}^{d-1}}{(d-1)!} & \mbox{if $d$ is odd};\\

      (-1)^{\frac{d}{2}-1}\frac{2\pi^{\frac{d-1}{2}}\ell_{eff}^{d-1}}{(\frac{d}{2}-1)!}\log{\left(\frac{2R}{\delta}\right)} & \mbox{if $d$ is even}.\\
    \end{cases}   \label{Area universal}    
\end{equation}
Having successfully isolated the universal part of the EE, it is time to ask ourselves what information we can obtain from it. From \cite{Myers:2010xs,Myers:2010tj} we  note that the candidates to the $C$-function  for even and odd CFTs can be read from Eq.\eqref{renormalized EE for a sphere}. This quantities are important because they have been  conjectured to decrease along the renormalization group (RG) flow  \cite{Zamolodchikov:1986gt,Jafferis:2011zi,Klebanov:2011gs,Cardy:1988cwa,Komargodski:2011vj}. In this particular scenario we have that :
\begin{equation}
S_{CCG}^{Univ}[\Sigma]=
    \begin{cases}
      (-1)^{\frac{d-1}{2}}F & \mbox{if $d$ is odd};\\
     
      (-1)^{\frac{d}{2}-1}4 A \log{\left(\frac{2R}{\delta}\right)}& \mbox{if $d$ is even,}\\
    \end{cases}     
\end{equation}
where
\begin{equation}
    F=a_d\frac{2^{d}\pi^{\frac{d-1}{2}}\ell_{eff}^{d-1}}{8G(d-1)!}  \quad;\quad A=a_d\frac{\pi^{\frac{d-1}{2}}\ell_{eff}^{d-1}}{8G(\frac{d}{2}-1)!} .\label{anomaly sphere}
\end{equation}

The Eq.\eqref{anomaly sphere} are the corresponding $F$-quantity  and type-$A$ charge of this theory. It is worth noting that these results are identical to those of Einstein-AdS gravity, except for an overall factor $a_d$ which depend on the CCG couplings.

\section{Second test: Cylinder}\label{Second test: Cylinder}
Analogous to the methodology used in the previous section, let us begin by considering the Euclidean space $AdS_5$ expressed in polar coordinates as:
\begin{equation}
    ds^2=\frac{\ell^2_{eff}}{z^2}\left(d\tau^2+dz^2+dx_3^2+dr^2+r^2d\theta^2\right) \label{Poincare cylinder}
\end{equation}
We define the entanglement region at a cutoff  $z = \delta$ and at a fixed time $\tau=0$. We take $r \leq l$ and $x_3\in [0,H]$ to describe a cylinder of length $H$, radius $l$ and axis along coordinate $x_3$. One can parametrize  the bulk extremal surface as $r=f(z)$ and use the EOM of CCG to obtain a differential equation for it. We show that, near the boundary,  the ansatz given in  Ref.\cite{Anastasiou:2021swo} also solves perturbatively the differential equation coming from the CCG EOM and therefore  the embedding can be cast as:
  \begin{equation}
     \Sigma:\left[\tau=cte,  r=l\left(1-\frac{z^2}{4l^2}+O(z^4)\right)\right]. 
 \end{equation}
Using this embedding the induced metric is:
 \begin{equation}
    ds_{\Sigma}^2=\frac{\ell^2_{eff}}{z^2}\left(\left(1+\frac{z^2}{4l^2}+O(z^4)\right)dz^2+dx_3^2+\left(l-\frac{z^2}{4l}+O(z^4)\right)^2d\theta^2\right) ,\label{induced metric cylinder}
\end{equation}
 where the normal vectors to the hypersurface are given by:
 \begin{equation}
     n^{(1)}_{\mu}=\left(\frac{\ell_{eff}}{z},0,0,0,0\right)\quad;\quad n^{(2)}_{\mu}=\left(0,\frac{\ell_{eff}}{\sqrt{4l^2+z^2}},0,\frac{2\ell_{eff}l}{z\sqrt{4l^2+z^2}},0\right).\label{normal vectors cylinder}
 \end{equation}
 With  the induced metric and the normal coordinates we have all the ingredients to obtain all the terms appearing in the computation of the EE in CCG. We will present the results of the quantities involved (see Appendix \ref{Explicit calculation: EE for the cylinder} for detailed calculation) in the following:
 \begin{align}
     \frac{\mathcal{A}[\Sigma]}{4G} &=\frac{\pi H}{4l G}\ell^3_{eff}\left(\frac{l^2}{\delta^2}-\frac{1}{4}\ln\left(\frac{l}{\delta}\right)\right)\label{area cylinder}\\
      S_{R^2}=&-\frac{6}{\ell^4_{eff}}\left(3\mu_1+4\mu_2+8\mu_3+40\mu_4+16\mu_5+16\mu_6 +80\mu_7+400\mu_8\right) \\
      S_{K^2R}=& -\frac{z^2}{l^2\ell^4_{eff}}(40\mu_4+8\mu_3+12\mu_2-3\mu_1)+O(z^4)\\
      S_{K^4}^{min}=&S_{K^4}^{non-min}=O(z^4). \label{SK4 for the cylinder}
 \end{align}
 As you can see again, the term $S_{R^2}$ is a constant and can come out of the integration. Moreover, if we put it together with the contribution of the usual Einstein part we will obtain and overall factor $a_4$ multiplying the area. Note also that the terms $S_{\text{K}^4}$ which, in principle, should produce two different functionals, coincide in the normalizable order. If we replace Eqs.\eqref{area cylinder}-\eqref{SK4 for the cylinder} into the HEE functional for CCG we obtain:
\begin{equation}
  S_{CCG}=\frac{\pi H}{4l G}\ell^3_{eff}\left(\frac{a_4l^2}{\delta^2}-\frac{1}{4}b_4\ln\left(\frac{l}{\delta}\right)\right)+O(\delta),\label{HEE cylinder}\\
\end{equation}
where:
\begin{align}
 a_4&=1+\frac{3}{\ell^4_{eff}}\left(3\mu_1+4\mu_2+8\mu_3+40\mu_4+16\mu_5+16\mu_6 +80\mu_7+400\mu_8\right)\label{a4}\\ 
    b_4&=a_4-\frac{4}{\ell^2_{eff}}\left(40\mu_4+8\mu_3+12\mu_2-3\mu_1\right).\label{b4}
\end{align}
Note that in Eq.\eqref{HEE cylinder} we obtain the anomalous logarithmic term, typical of the even dimensional CFTs but in addition appears another divergent term proportional to $a_4$. To subtract this last term we have to add the corresponding counterterm $S_{KT}$ for this theory. From Eq.\eqref{HEE in CCG Kounterterm} we can see that:
 \begin{align}
  S_{KT}&=\frac{c_4}{4G}\bigg\lfloor\frac{4+1}{2}\bigg\rfloor\int_{\partial\Sigma}dx^2\sqrt{\tilde{\sigma}}B_2=-\frac{2c_4\pi H }{G}\frac{\ell_{eff}l}{\delta^{2}}+O(\delta).\nonumber
 \end{align}
 We can express $c_4=a_dc_4^{EH}$ and from Eq.\eqref{coupling Einstein Hilbert} we know tat $c_4^{EH}=\frac{\ell^2_{eff}}{8}$. Therefore:
 \begin{equation}
  S_{KT}=  -\frac{a_4\pi H }{4G}\frac{\ell^3_{eff}l}{\delta^{2}}+O(\delta).
 \end{equation}
 This term is exactly the same (with negative sign) as the first part of the right-hand side of Eq.\eqref{HEE cylinder}. Therefore, it does the job of recovering the universal part:
 \begin{equation}
     S_{CCG}^{Univ}=S_{CCG}+S_{KT}=-b_4\frac{\pi H\ell^3_{eff}}{16l G}\ln\left(\frac{l}{\delta}\right).
 \end{equation}
The type-B anomaly coefficient of a holographic CFT dual to CCG can be inferred from this result. In $d=4$ the universal term  of the EE is associated to c (the type-B anomaly) by the following equation \cite{Hung:2011xb,Bhattacharyya:2014yga}:
 \begin{equation}
      S_{CCG}^{Univ}=-\frac{cH}{2l}log\left(\frac{l}{\delta}\right).
 \end{equation}
Then, we can see in our result that the type-B anomaly, for this particular theory, is $c=b_4\frac{\pi\ell^3_{eff}}{8G}$.

  \section{Deformed sphere I: $C_T$} \label{Deformed sphere I}

The coefficients of the contact-term expansion of stress-tensor correlators are crucial properties of CFTs. These coefficients are inherent to the definition of the theory and have diverse interpretations. For example, the coefficient $C_{\text{T}}$, which determines the unitarity of the theory, is associated with the two-point function and is subject to a positivity constraint \cite{Osborn:1993cr}. Moreover, the coefficients $C_{\text{T}}$, $t_\text{2}$, and $t_\text{4}$, which are related to the three-point function, control the energy flux that an observer located in a specific direction at null infinity receives \cite{Osborn:1993cr,Erdmenger:1996yc,Hofman:2008ar}. In this and the next section we will give a recipe to obtain the mentioned quantities from the universal part of the  EE. We will first review the method for the Einstein Hilbert case and then we will move on to the CCG case.

Lets start, by mentioning some results concerning the EE of a spherical entangling region. In Ref.\cite{Casini:2011kv} it was shown that by making a sequence of conformal transformations it was possible to relate the vacuum state of the original geometry to a thermal state placed on an $\mathbb{S}_d$ background\footnote{This sequence of transformations was first performed by Casini, Huerta and Myers, and is therefore called the CHM map.}. Using that map, the EE of that region was related to the free energy $F=-ln(Z[\mathbb{S}_d])$ of a CFT residing in $\mathbb{S}_{d}$. Then,  the AdS/CFT dictionary was used to translate this problem to the one of finding the horizon entropy of a certain topological black hole.

On the other hand, in  Ref.\cite{Allais:2014ata}  it was conjectured that the sphere minimizes the universal contribution to the entanglement entropy, among all possible entanglement regions with sphere topology, for the ground state of a CFT. To notice this, the authors make an analysis  in $d=3$ dimensions of a deformed spherical entangling region with deformation parameter $\epsilon$ and then they generalized their results to any dimension. Furthermore, in Ref.\cite{Mezei:2014zla}, Mezei shows that the $C_T$ charge appears at order $O(\epsilon^2)$ in the expansion of the renormalized EE around the deformation parameter. Therefore, up to second order in $\epsilon$, the EE  reads as:

\begin{equation}
    S^{Univ}_{EH}(\mathbb{S}_{\epsilon})= S^{Univ,(0)}_{EH}(\mathbb{S})+\epsilon^2 S^{Univ,(2)}_{EH}(\mathbb{S})+O(\epsilon^4).
\end{equation}

where $S^{Univ,(0)}_{EH}(\mathbb{S})$  is the renormalized EE of the unperturbed sphere and the term $S^{Univ,(2)}_{EH}(\mathbb{S})\propto C_T$. The heuristic argument that can be used to understand this proportionality, is that the $S^{Univ,(2)}$ term can be though as obtained by making a variation of the free energy (considering the relationship with the EE given by the CHM map). This variation, as you can see, give us the two correlation function of the stress energy momentum tensor and therefore it should be proportional to $C_T$.

Now that we have made this remarks its time to obtain the $C_T$ for a $CFT_3$ dual to CCG, however it would be useful to review first the computation for the EH case. For that we will follow the Ref.\cite{Anastasiou:2020smm}.

The Poincaré-AdS4 spacetime can be expressed in polar coordinates as follows:

\begin{equation}
    ds^2=\frac{\ell^2}{z^2}\left(d\tau^2+dz^2+dr^2+r^2d\phi^2\right). \label{Poincare CT}
\end{equation}

We want to compute the ground state EE in the dual $CFT_3$ for a region that is a deformed sphere with a parameter deformation $\epsilon$.  If the corresponding theory dual to the CFT is Einstein-AdS gravity, we can obtain the entanglement entropy using the RT formula:
\begin{equation}
    S_{EH}=\frac{\mathcal{A}[\Sigma_{RT}]}{4G}.\label{RT formula}
\end{equation}

The results of Refs.\cite{Allais:2014ata,Anastasiou:2020smm} have shown that the embedding function for the RT surface is expressed as follows:
\begin{equation}
    \Sigma_{RT}: \:r=\sqrt{1-z^2}\left(1+\epsilon \sum_{l}\left(\frac{1-z}{1+z}\right)^{l/2}\frac{1+lz}{1-z^2}(a_l\cos{(l\phi)}+b_l\sin{(l\phi}))+O(\epsilon^2)\right).\label{embedding deformed sphere}
\end{equation}

We can use this embedding to obtain the induced metric and with it compute the minimal surface to obtain the EE. Following these steps and after renormalization, we obtain that the universal part of the EE is:
\begin{equation}
    S_{EH}^{Univ}=-\frac{\pi \ell^2}{2G}\left(1+\epsilon^2\sum_{l}\frac{l(l^2-1)}{4}(a_l^2+b_l^2)+O(\epsilon^4)\right).\label{EE deformed sphere for EH}
\end{equation}
The quadratic term in the perturbation of the HEE,  indicates the susceptibility of entanglement to changes in the shape of the entangling region. This term holds universal information, thanks to the coefficient $C_T$ that appears in the two-point correlation function of the stress tensor. Indeed, the subleading term of Eq.\eqref{EE deformed sphere for EH}, can equivalently be written as:
\begin{equation}
     S_{EH}^{Univ,(2)}=\frac{\pi^4 C_T}{24}\sum_{l}l(l^2-1)(a_l^2+b_l^2)\rightarrow  C_T=\frac{3\ell^2}{\pi^3G} .\label{SEE eps}
\end{equation}
Now we want to compute the EE for a CFT with a general Cubic Curvature Gravity as a dual theory. We are going to fix the  cod-2 brane at the position of the RT surface. This is analog as assuming the Einstein-Hilbert EOMs for determining the adapted coordinates for the foliation with respect to the cod-2 brane. Then, we introduce the corresponding embedding into the correct entropy functional for the theory. At least, for small values in the CCG theory couplings, we know that this prescription should work at the perturbative level.

Using the embedding given by Eq.\eqref{embedding deformed sphere}, we can compute the relevant terms appearing in  the HEE in CCG functional, these are:
\begin{align}
    S_{R^2}&=-\frac{6 \left(2 \mu_\mathrm{1}+4 \mu_\mathrm{2}+6 \mu_\mathrm{3}+24 \mu_\mathrm{4}+9 \mu_\mathrm{5}+9 \mu_\mathrm{6}+36 \mu_\mathrm{7}+144 \mu_\mathrm{8}\right)}{\ell_{eff}^{4}}\label{SR2 deformed CT}\\
    S_{K^2R}&= \frac{12z^4 (\mu_1 -4 \mu_2 -2 \mu_3 -8 \mu_4)}{\ell_{eff}^4(1-z^2)^2}\sum_{l}\left(\frac{1-z}{z +1}\right)^{l}(a_l^{2}+b_l^{2}) l^{2} (l^2 -1)^{2} \epsilon^{2}\\
    &+\mathrm{O}\! \left(\epsilon^{3}\right)\nonumber\\
    S_{K^4}^{min}&=S_{K^4}^{non-min}=\mathrm{O}\! \left(\epsilon^{4}\right).\label{SK4 deformed CT}
\end{align}
We can see from here that at least at order $\epsilon^2$ the two splittings coincide, that is $S_{CCG}^{min}=S_{CCG}^{non-min}=S_{CCG}$.  By replacing in the functional Eq.\eqref{HEE in CCG minimal} or Eq.\eqref{HEE in CCG nonminimal} and after integration (See Appendix \ref{Explicit calculation: EE for the Deformed sphere}) we obtain, up to order $\epsilon^2$ :
\begin{align}
   S_{CCG}=&-\frac{a_3\pi \ell_{eff}^2}{2G}\left(1-\frac{1}{\delta}\right)-\frac{\pi \ell_{eff}^2}{8G}\sum_l\bigg[(a_l^2+b_l^2)l(l^2-1)\bigg(1\label{HEE in CCG deformed sphere for CT}\\&+\frac{3}{\ell_{eff}^4}(4\mu_1-4\mu_2+2\mu_3+8\mu_4
   +9\mu_5+9\mu_6+36\mu_7+144\mu_8)\bigg)\nonumber\\
   &-\frac{a_3}{\delta}(a^2+b^2)l^2\bigg]\epsilon^2\nonumber+O(\epsilon^3).\nonumber 
\end{align}
The Kounterterm is constructed by setting  $z=\delta$ and then use it to compute the chern-form in one dimension $B_1=-\frac{2}{\ell_{eff}}+O(\epsilon^3,\delta)$. Then, we replace it in the formula for the Kounterterm \eqref{HEE in CCG Kounterterm} (with the identification $c_4=\frac{a_3\ell_{eff}^2}{4}$):
 \begin{equation}
  S_{KT}=  -\frac{\pi \ell_{eff}^2 a_3}{2G}\frac{1}{\delta}\left(1+\sum_l\frac{l^2}{4}(a_l^2+b_l^2)\epsilon^2\right)+O(\epsilon^3,\delta) . 
 \end{equation}

 The previous expression, with an overall minus sign, is equal to the divergent part of Eq.\eqref{HEE in CCG deformed sphere for CT}. Thus, when we sum $S_{CCG}$ and $S_{KT}$, the divergence cancels out successfully, and we recover the universal part:
 \begin{align}
     S_{CCG}^{Univ}&=-\frac{\pi \ell_{eff}^2}{2G}\bigg[a_3+\sum_l\frac{1}{4}(a_l^2+b_l^2)l^2(l^2-1)\bigg(1+\frac{3}{\ell_{eff}^4}(4\mu_1-4\mu_2+2\mu_3\nonumber\\&+8\mu_4+9\mu_5+9\mu_6+36\mu_7+144\mu_8)\bigg)\epsilon^2\bigg] +O(\epsilon^4).\label{EE in CCG universal for deformed sphere}
 \end{align}
 By examining the $\epsilon^2$ term of the expression given above, we can determine the $C_T^{CCG}$ for the CCG theory. We assume that the polynomial in $l$ is the same for EH and CCG, as it should be independent of the specific theory and only depend on the geometry of the chosen entanglement region. We compare the order $\epsilon^2$ term of Eq.\eqref{EE in CCG universal for deformed sphere} with Eq.\eqref{SEE eps} and observe that: 
\begin{equation}
     S_{CCG}^{Univ,(2)}=\frac{\pi^4 C_T^{CCG}}{24}\sum_{l}l(l^2-1)(a_l^2+b_l^2), \label{univ deformed sphere CT}
\end{equation}
where:
\begin{equation}
    C_T^{CCG}=\bigg(1+\frac{3}{\ell_{eff}^4}(4\mu_1-4\mu_2+2\mu_3+8\mu_4+9\mu_5+9\mu_6+36\mu_7+144\mu_8)\bigg)C_T.\label{Ct CCG}
\end{equation}
Therefore, we find that the $C_T^{CCG}$ is equivalent to the $C_T$ of the Einstein-AdS, but multiplied by a coefficient that depends on the couplings of the CCG theory. The Eq.\eqref{Ct CCG} is consistent with what was found in Ref.\cite{Bueno:2020uxs}, albeit using a different computational method.

\section{Deformed sphere II: $t_4$} \label{Deformed sphere II}

From the previous section, we can observe that Eq.\eqref{univ deformed sphere CT} does not have contributions from the splitting-dependent terms. Nevertheless, as per Eq.\eqref{SK4 deformed CT}, these terms should appear at the next subleading order, i.e. in $S_{\text{CCG}}^{Univ,\text{(4)}}$. To calculate this, we need to consider higher orders in the $\epsilon$ expansion of the embedding function in Eq.\eqref{embedding deformed sphere}. Using the heuristic interpretation described at the beginning of the previous section, we can relate the coefficients of this higher powers in the $\epsilon$ expansion with higher point functions of the stress tensor and therefore use it to compute the coefficients ($C_T,t_2, t_4$, etc) that determine them.
 
Again we use the embedding obtained for the case of Einstein gravity. For computational purposes, we will suppress the solution of type $sin(l\phi)$ in Eq.\eqref{embedding deformed sphere}. Then, up to second order in the expansion around the deformation parameter, we obtain:
 \begin{align}
    \Sigma_{RT}: \:\:r&=\sqrt{1-z^2}\bigg[1+\epsilon \sum_{l}\left(\frac{1-z}{1+z}\right)^{l/2}\left(\frac{1+lz}{1-z^2}\right)\cos{(l\phi)}\nonumber\\
    &+\epsilon^2\sum_{l}\left(\frac{1-z}{1+z}\right)^{l}\left(\frac{1}{4\left(1-z^{2}\right)^{2}}\right)\bigg(\left(1+2lz+\left(3l^{2}-2\right)z^{2}+2l\left(l^{2}-1\right)z^{3}\right)&\nonumber\\
    &+\left(2l\left(l^{2}-4\right)z^{3}+\left(3l^{2}-5\right)z^{2}+8lz+4\right)\cos\left(2l\phi\right)\bigg)+O(\epsilon^4)\bigg].\label{embedding deformed sphere eps2}
\end{align}
As far as we are aware, the explicit expression for the $R_{22}$ term in Ref.\cite{Allais:2014ata}, which we are providing here, constitutes a new and original result. To ensure consistency, we will now compute the universal HEE for the case of Einstein gravity. We will use the embedding \eqref{embedding deformed sphere eps2} in the RT formula \eqref{RT formula}, resulting in:
 \begin{equation}
    S_{EH}^{Univ}=-\frac{\pi \ell^{2}}{2G}\left(1+\sum_{l}\frac{1}{4} l\left(l^{2}-1\right) \epsilon^{2}-\sum_{l}\frac{ l\left(23 l^{6}-246l^{4}+63 l^{2}-2\right)}{4\left(64l^{2}-16\right)} \epsilon^{4}+O(\epsilon^6)\right).\label{EE deformed sphere for EH epsilon4}
\end{equation}
 We know that for CFTs in d=3 dual to  Einstein Gravity, $t_4=0$ and $t_2=0$ \cite{Mezei:2014zla} so the term appearing in $\epsilon^4$ should be proportional just to  $C_T^{EH}$. We can expressed Eq.\eqref{EE deformed sphere for EH epsilon4} as:
 \begin{equation}
    S_{EH}^{Univ}=-\frac{\pi \ell^{2}}{2G}-\frac{\pi^{4}C_T^{EH}}{24}\sum_{l}l\left(l^{2}-1\right) \epsilon^{2}+\frac{\pi^{4}C_T^{EH}}{24}\sum_{l}\left(\frac{ l\left(23 l^{6}-246l^{4}+63 l^{2}-2\right)}{16\left(4l^{2}-1\right)}\right) \epsilon^{4} .\label{EE deformed sphere for EH epsilon4 in terms of CT}
\end{equation}
For reasons that will be described below, let's assign a name to the polynomial on $l$ that multiplies $C_T$ at the order $\epsilon^4$:
\begin{equation}
    P_1(l)=\sum_{l}\left(\frac{\pi^4 l\left(23 l^{6}-246l^{4}+63 l^{2}-2\right)}{384\left(4l^{2}-1\right)}\right)\label{P1}
\end{equation}
Now, to compute the EE in CCG, we employ the same RT surface embedding. As stated in Ref.\cite{Bueno:2020uxs}, the use of this surface for evaluating the HEE functional for the deformed sphere is valid up to the leading order in the higher curvature couplings.  We proceed to find all the quantities involved up to order $\epsilon^4$:
 \begin{align}
     S_{R^2}&=-\frac{6 \left(2 \mu_\mathrm{1}+4 \mu_\mathrm{2}+6 \mu_\mathrm{3}+24 \mu_\mathrm{4}+9 \mu_\mathrm{5}+9 \mu_\mathrm{6}+36 \mu_\mathrm{7}+144 \mu_\mathrm{8}\right)}{\ell_{eff}^{4}}\\
     S_{K^2R}&=\frac{12z^{4}\left(\mu_\mathrm{1}-4 \mu_\mathrm{2}-2 \mu_\mathrm{3}-8 \mu_\mathrm{4}\right)}{\ell_{eff}^4\left(1-z^{2}\right)}\sum_{l}\left(\frac{1-z}{1+z}\right)^{l} l^{2} \left(l^{2}-1\right)^{2} \epsilon^{2}\nonumber\\
     &+\frac{12z^{4}\left(\mu_\mathrm{1}-4 \mu_\mathrm{2}-2 \mu_\mathrm{3}-8 \mu_\mathrm{4}\right)}{\ell_{eff}^{4}\left(1-z^{2}\right)^{3}}\sum_{l}\left(\frac{1-z}{1+z}\right)^{\frac{3 l}{2}}l^{2}\left(l^{2}-1\right) \left(2 l^{4} z^{2}-2 l^{3} z-14 l^{2} z^{2}\right.\nonumber\\
     &\left.+8 l^{2}+2 l z+3 z^{2}+1\right) \cos\left(l \phi\right)\epsilon^3\nonumber\\
     &-\frac{12\mathit{lz}^{3}\left(\mu_\mathrm{1}-4 \mu_\mathrm{2}-2 \mu_\mathrm{3}-8 \mu_\mathrm{4}\right)}{\ell_{eff}^{4}\left(1-z^{2}\right)^{4} }\sum_{l}\left(\frac{1-z}{1+z}\right)^{2 l}\bigg(\left(l^{2}-1\right)\big(l \left(6 l^{4}-23 l^{2}-1\right) z^{5}\nonumber\\
     &+3 \left(11 l^{4}-34 l^{2}+5\right) z^{4}+2 l \left(15 l^{4}-29 l^{2}-4\right) z^{3}+2l^{2}\left(14 l^{2}+4\right) z^{2}\\
     &+l\left(53l^{2}-3\right) z-l^{4}+22 l^{2}-3\big) \cos\left(l \phi\right)^{4}+\frac{1}{2}\left(l^{2}-1\right)\big(2 l \left(14 l^{6}-88 l^{4}+111 l^{2}-16\right) z^{5}\nonumber\\
     &-\left(8 l^{6}+56 l^{4}-139 l^{2}+9\right) z^{4}+2 l \left(19 l^{4}-34 l^{2}+18\right) z^{3}-2 \left(5 l^{6}-5 l^{4}+24 l^{2}-3\right) z^{2}\nonumber\\
     &-2 l \left(2 l^{4}+17 l^{2}+5\right) z+2 l^{4}-29 l^{2}+3\big)\cos\! \left(l \phi\right)^{2}+\frac{l\left(1-z^{2}\right)}{4}\big(\big(8 l^{8}-112 l^{6}\nonumber\\
     &+332 l^{4}-168 l^{2}+21\big) z^{2}+4 l \left(4 l^{6}-17 l^{4}+22 l^{2}-9\right) z\nonumber\\
     &+8 l^{6}-142 l^{4}+68 l^{2}-15\big)\bigg)+O(\epsilon^5)\nonumber\\
     S_{K^4}^{min}&=\frac{3 z^{8}\left(-4 \mu_\mathrm{2}+\mu_\mathrm{1}-2 \mu_\mathrm{3}-8 \mu_\mathrm{4}\right)}{\ell_{eff}^{4}\left(1-z^{2}\right)^{4}} \sum_{l}\left(\frac{1-z}{1+z}\right)^{2 l} l^{4} \left(l^2-1\right)^{4} \epsilon^{4}+O(\epsilon^5)\label{SK4min t4}\\
      S_{K^4}^{non-min}&=\frac{6 z^{8}\left(\mu_\mathrm{1}+2 \mu_\mathrm{2}\right)}{\ell_{eff}^{4}\left(1-z^{2}\right)^{4}} \sum_{l}\left(\frac{1-z}{1+z}\right)^{2 l} l^{4} \left(l^2-1\right)^{4} \epsilon^{4}+O(\epsilon^5).\label{SK4nonmin t4}
 \end{align}
 We can see that the last two equation for $S_{K^4}^{min}$ and the $S_{K^4}^{non-min}$  are not the same at order $\epsilon^4$ and therefore we will have a difference between the splittings. Notice that this is the first scenario where we can find a difference between the minimal and non-minimal prescription. Now, we replace this terms in the functional of the EE. We already know that the leading and subleading order will be the same of Eq.\eqref{EE in CCG universal for deformed sphere}. On the other hand, by seeking the polynomial $P_1(l)$ given by Eq.\eqref{P1}, we can notice that there is a term proportional to $C_T^{CCG}$ at order  $\epsilon^4$ as well as other term that is supposed to be proportional to $t_4$ (the $t_2$ always vanishes in this dimension): 
\begin{align}
   S_{\text{CCG}}^{\text{Univ,min}}&=-\frac{\pi \ell_{\text{eff}}^{2}}{2G}a_3-\epsilon^2C_{\text{T}}^{\text{CCG}}\sum_l\frac{\pi^{4}}{24}l\left(l^{2}-1\right) \nonumber\\
   &+\epsilon^4\bigg(C_{\text{T}}^{\text{CCG}}  P_{1}\left(l\right)-\sum_l\frac{135 \pi l^{3} \left(l^{2}-1\right)^{3} \left(\mu_\mathrm{1}-4 \mu_\mathrm{2}-2 \mu_\mathrm{3}-8 \mu_\mathrm{4}\right)}{256 G \ell_{\text{eff}}^{2} \left(4 l^{2}-1\right) \left(4 l^{2}-9\right)}\bigg) +O(\epsilon^6) \,, \label{SCCGmin t4}\\
    S_{\text{CCG}}^{\text{Univ,non-min}}&=-\frac{\pi \ell_{\text{eff}}^{2}}{2G}a_3-\epsilon^2C_{\text{T}}^{\text{CCG}}\sum_l\frac{\pi^{4}}{24}l\left(l^{2}-1\right)  \nonumber\\
   &+\epsilon^4\bigg(C_{\text{T}}^{\text{CCG}} P_{1}\left(l\right)-\sum_l\frac{135 \pi l^{3} \left(l^{2}-1\right)^{3} \left(\mu_\mathrm{1}+2 \mu_\mathrm{2}\right)}{128 G \ell_{\text{eff}}^{2} \left(4 l^{2}-1\right) \left(4 l^{2}-9\right)}\bigg) +O(\epsilon^6) \,.
   \label{SCCGnonmin t4}
\end{align}

It is not surprising to find that the polynomial $P_1(l)$ appears in the last two equations, as it relies solely on geometric factors and is independent of the specific theory at hand. However, it is important to note that this only holds true for CFTs duals to EH gravity with perturbative corrections, as we are employing the RT surface for the computation of the entanglement entropy. For more generic CFTs, the aforementioned assertion may not hold and $P_1(l)$ could no longer be theory independent.

The discrepancy between Eq.\eqref{SCCGmin t4} and Eq.\eqref{SCCGnonmin t4}, at order $\epsilon^4$, can be traced back to the coupling dependent parts of Eq.\eqref{SK4min t4} and Eq.\eqref{SK4nonmin t4}, respectively. Surprisingly, the coupling combination $(\mu_\mathrm{1}+2 \mu_\mathrm{2})$ that appears in the non-minimal prescription agrees with the coupling dependent part of the $t_4$ charge obtained for a CFT that is dual to a CCG theory (see Ref.\cite{Sen:2014nfa,Chu:2016tps}). Thus, the order $O(\epsilon^4)$ in Eq.\eqref{SCCGnonmin t4} can be expressed as a linear combination of the $C_{T}$ and $t_{4}$ charges of CCG \footnote{In Ref.\cite{Bueno:2015ofa}, it was shown that in generic CFTs, the term $O(\epsilon^4)$ of the HEE cannot be completely determined by $C_{T}$ and $t_{4}$ when corner contributions are taken into account.}.

To establish the normalization and obtain the $t_4$, we consider the massless limit of CCG, which was analyzed in Ref.\cite{Li:2019auk}. When demanding that the massive modes are decoupled from the particle spectrum of the theory, we arrive at the equations (in $d=3$):
\begin{align}
    &12\mu_{7}+9\mu_{6}+5\mu_{5}+48\mu_{4}+16\mu_{3}+24\mu_{2}-3\mu_{1}=0 \,,\label{1st constranint}\\
    &432\mu_{8}+120\mu_{7}+36\mu_6+32\mu_{5}+16\mu_{4}+28\mu_{3}+6\mu_{1}+24\mu_{2}=0 \,.\label{second constraint}
\end{align}
By solving this set of equations, we can represent the coupling constants $\mu_1$ and $\mu_2$ in relation to the remaining couplings:
\begin{align}
    \mu_1&=-48\mu_8-12\mu_7-3\mu_6-3\mu_5-\frac{16}{3}\mu_{4}-\frac{4}{3}\mu_3 \,,\label{mu1}\\
    \mu_2&=-6\mu_8-2\mu_7-\frac{3}{4}\mu_6-\frac{7}{12}\mu_5-\frac{8}{3}\mu_4-\frac{5}{6}\mu_3 \,.\label{mu2}
\end{align}
Replacing these relations in $(\mu_\mathrm{1}+2 \mu_\mathrm{2})$  we  recover the coupling dependence of the $t_4$  which was previously determined in Ref.\cite{Li:2019auk}. This is\footnote{This result was obtained without the need to deal with two different splittings.}: 
\begin{equation}
    t_4^{\text{CCG}-massless}=
   - \frac{120(360\mu_8+96\mu_7+27\mu_6+25\mu_5+64\mu_4+18\mu_3)}{\ell_{\text{eff}}^4}+\mathcal{O}(\mu^2) \,.
    \label{masslesst4}
\end{equation}
In contrast to the non-minimal counterpart, substituting Eq. \eqref{mu1} and Eq. \eqref{mu2} into the expression $\left(\mu_\mathrm{1}-4 \mu_\mathrm{2}-2 \mu_\mathrm{3}-8 \mu_\mathrm{4}\right)$, that appears in the minimal entropy functional given in Eq. \eqref{SCCGmin t4}, will not yield Eq. \eqref{masslesst4} . This is further evidence of its inconsistency.

For the massless case, it is easy to see that the polynomial of the multipole momenta accompanying the $t_4$ is:
\begin{equation}
     P_{2}(l)=\sum_l \frac{\pi^4}{2048}\frac{l^{3} \left(l^{2}-1\right)^{3}}{ \left(4 l^{2}-1\right) \left(4 l^{2}-9\right)} \,.\label{P2}
\end{equation}
In the generic CCG theory, $P_{2}(l)$ should also appear multiplying the $t_4$. Therefore, by keeping track of it in the Eqs.\eqref{SCCGmin t4} and \eqref{SCCGnonmin t4}, we obtain:
\begin{align}
   S_{\text{CCG}}^{\text{Univ, min}}=&-\frac{\pi \ell_{\text{eff}}^{2}}{2G}a_3-\epsilon^2C_{\text{T}}^{\text{CCG}}\sum_l\frac{\pi^{4}}{24}l\left(l^{2}-1\right)\nonumber \\
   &+\epsilon^{4}C_{\text{T}}^{\text{CCG}}\left[P_{1}\left(l\right)-t_4^{CCG-min}P_{2}\left(l\right)\right] +\mathcal{O} \left(\epsilon^6 \right)\,,\\
    S_{\text{CCG}}^{ \text{Univ, non-min}}=&-\frac{\pi \ell_{\text{eff}}^{2}}{2G}a_3-\epsilon^2C_T^{CCG}\sum_l\frac{\pi^{4}}{24}l\left(l^{2}-1\right) \nonumber
    \\&+\epsilon^{4}C_{\text{T}}^{\text{CCG}}\left[P_{1}\left(l\right)-t_4^{CCG-non-min}P_{2}\left(l\right)\right] +\mathcal{O}\left(\epsilon^6 \right)\,,
    \label{SCCGfinal}
\end{align}
\normalsize
where we can identify the $t_4$ charges for each of the splittings. These are:
\begin{align}
    t_4^{\text{CCG-min}}&=\frac{360\left(\mu_\mathrm{1}-4 \mu_\mathrm{2}-2 \mu_\mathrm{3}-8 \mu_\mathrm{4}\right)}{\ell_{\text{eff}}^{4}\left(1+\frac{3}{\ell_{\text{eff}}^4}(4\mu_1-4\mu_2+2\mu_3+8\mu_4+9\mu_5+9\mu_6+36\mu_7+144\mu_8)\right)}\,,\label{t4 min}\\
   t_4^{\text{CCG-non-min}}&=
    \frac{720(\mu_1+2\mu_2)}{\ell_{\text{eff}}^4\left(1+\frac{3}{\ell_{\text{eff}}^4}(4\mu_1-4\mu_2+2\mu_3+8\mu_4+9\mu_5+9\mu_6+36\mu_7+144\mu_8)\right)}\,.\label{t4 nonmin}
\end{align}
Therefore, the non-minimal prescription is the one that matches the $t_4$ value derived by Sen and Sinha in Ref.\cite{Sen:2014nfa} for graviton perturbations on a shockwave background. Moreover, we can contrast the non-minimal splitting of $t_4$ in Eq.\eqref{t4 nonmin} with its equivalent expression for Einsteinian Cubic Gravity (ECG), as presented in Ref.\cite{Bueno:2018xqc}. To define this theory the CCG couplings should be taken as:
\begin{equation}
  \mu_1=-\frac{3}{2}\mu  \:\: ; \:\: \mu_2=-\frac{\mu}{8} \:\:  ;\:\:\mu_3= 0 \:\: ; \:\: \mu_4= 0 \:\: ; \:\:\mu_5= \frac{3}{2}\mu  \:\: ;\:\:  \mu_6= -\mu \:\: ;\:\:\mu_7= 0  \:\:;\:\:\mu_8=0\,.
\label{ECG_couplings}
\end{equation}
Thus, we obtain that the value of $t_4$ for this theory,
\begin{equation}
     t_4^{\text{ECG-non min}}=-\frac{1260\mu}{\ell_{\text{eff}}^{4}\bigg(1-\frac{3}{\ell_{\text{eff}}^4}\mu\bigg)}\,,
\end{equation}
is in full agreement with the given reference.

%% file: Conclusions/Conclusions.tex
\chapter{Conclusions}
In this chapter we will give a summary of the main results of this thesis, as well as give indications for future research. The main conclusions are as follows:
\begin{enumerate}
    \item We have used the maximally symmetric ansatz of global AdS spacetime to fix the coupling needed to achieve the renormalization of the bulk action of an arbitrary HCG theory in the context of the Kounterterm renormalization scheme. We determined the coupling needed for the renormalization of the CCG theory using also this method. Using the self-replication property of the Kounterterms in codimension-2 submanifolds,  we proposed a renormalized Holographic EE formula for a CFT dual to CCG theory.
    
    \item  To test our proposal we  choose  two different entangling shapes: a hypersphere region in arbitrary CFT dimension and a cylinder in $d=4$ dimensions. For the case of the sphere, due to the symmetries that it possesses, it was possible to obtain a non-perturbative solution of the embedding of the codimension-two surface. We replace the solution of the embedding in our proposal to extract the universal part of the entanglement entropy. We were able to see that in this scenario the divergences were successfully cancelled. From this expression we were able to read off the F-quantity and the type-A anomaly of this theory. We note that the discrepancy with the values for CFTs dual to Einstein gravity is only of an overall coupling dependent factor.  
    
    For the cylindrical entangling region, we work perturbatively to obtain an embedding near the boundary. Using this and the Kouterterm method, we were able to read of the coefficient of the type-B anomaly from the renormalized EE.

    \item We observed that for an arbitrary entangling region, the extrinsic curvature (with one covariant and one contravariant index) scales in terms of the holographic coordinate as $O(z)$ so the $S_{K^4}$ term producing the difference between the splittings should scale as $O(z^4)$. With this in mind, we were able to show that for $d=2$ and $d=4$ it is impossible to distinguish between the minimal and non-minimal prescriptions from the universal part of the EE. This occurs because the part of the EE corresponding to $S_{K^4}$ scales as $O(z^{d-5})$, so it will not contribute to the universal EE.

    \item  Finally, we consider a deformed sphere as an entangling region in a $CFT_3$ with the deformation parameter being $\epsilon$. Using the CHM map it is possible to relate the EE of a ball-shaped region of flat space to the free energy of a thermal CFT in a spherical background. The perturbation around the $\epsilon$ parameter can be considered as a variation of the free energy. Therefore, it is possible to relate the correlation functions of the energy momentum tensor to the terms of the series expansion of the EE for the deformed sphere.

    The coefficient of the two-point correlation function $C_T$ is related to the $S^{(2)}_{CCG}$ term in the expansion of the EE around $\epsilon$. The fourth-order term of the expansion is then related to $C_T$ and $t_4$ which are the coefficients appearing in the calculation of the fourth-point correlation function of the momentum energy-momentum tensor.
    
    To obtain the $C_T$ coefficient for a general CCG theory we use the result of Einstein gravity to fix the polynomial in $l$, and therefore we can then read $C_T$ off from the $S^{(2)}_{\epsilon}$ term. In Einstein gravity the $t_4$ is zero and therefore, in this case, we can not use it to fix the polynomial in $S^{(4)}_{CCG}$. 

    Thus, we use the massless limit of Ref.  \cite{Li:2019auk} to obtain the polynomial \eqref{P2} in the Fourier expansion. By tracking it in the entropy functionals of Eqs.\eqref{SCCGmin t4} and \eqref{SCCGnonmin t4} we were able to arrive at two different expressions for $t_4$ (one for the minimal and the other for the non-minimal). Our conclusion is that only the non-minimal prescription is consistent. As a consistency check we computed the $t_4$ for the Einsteinian cubic gravity case and saw that it matches with the results obtained in Ref.\cite{Bueno:2018xqc}. 

    \end{enumerate}

In the process of conducting this thesis we realized, based on a power counting argument, that the Kounterterm method also works to achieve the renormalization of the EE for any higher curvature gravity theory in $D=4$. This  has already been discussed in  Ref.\cite{Anastasiou:2022pzm} in which the author of this thesis was involved.

To conclude with this thesis, we will mention a possible avenue for future research.  As previously mentioned, the appearance of CFT coefficients of the stress tensor in the expansion of the HEE around the deformation parameter suggests the possibility of extending the CHM map. This means that the HEE calculated on the deformed spherical surface could be represented as the partition function on a similarly deformed thermal sphere. It would be intriguing to derive a comprehensive mapping between the distorted Euclidean-sphere manifold and the perturbed spherical entangling surface in terms of their respective deformation parameters.

%% file: appendix/appendix.tex
\appendix
\chapter{Explicit calculation: EE for the cylinder}\label{Explicit calculation: EE for the cylinder}

In this Appendix we will approach the calculation of the Entanglement Entropy for the vacuum state of a cylindrical region of a CFT which, on the gravitational side, is characterized by a Cubic Curvature theory. We start by considering the metric of Eq.\eqref{Poincare cylinder}:
\begin{equation}
    ds^2=\frac{\ell^2_{eff}}{z^2}\left(d\tau^2+dz^2+dx_3^2+dr^2+r^2d\theta^2\right) \label{Poincare for cylinder appendix}
\end{equation}
A cylindrical region in the CFT in a Cauchy slice is defined in the following way:
 \begin{equation}
     A:\left[\tau=cte, z=\delta \:\:and\:\: r\leq l\right] 
 \end{equation}
Therefore, the entangling surface is obtain at constant radius $l$:
  \begin{equation}
     \partial A:\left[\tau=cte, z=\delta \:\:and\:\: r=l\right] 
 \end{equation}
In the bulk, the extremal surface that is anchored to $\partial A$, in the near-boundary region, is described by the embedding:
 \begin{equation}
     \Sigma:\left[\tau=cte,  r=l\left(1-\frac{z^2}{4l^2}+O(z^4)\right)\right]
 \end{equation}
 With this we compute the induced metric $\sigma$ for the codimension-two hypersurface:
 \begin{equation}
    ds_{\Sigma}^2=\frac{\ell^2_{eff}}{z^2}\left(\left(1+\frac{z^2}{4l^2}+O(z^4)\right)dz^2+dx_3^2+\left(l-\frac{z^2}{4l}+O(z^4)\right)^2d\theta^2\right) 
\end{equation}
 therefore the square root of the determinant of the induced metric is :
 \begin{equation}
     \sqrt{\sigma}=\frac{\ell^3_{eff}l}{z^3}-\frac{1}{8}\frac{\ell^3_{eff}}{lz}+O(z) \label{detsigma}
 \end{equation}
and using the Eq.\eqref{normal vectors cylinder} of the normal vectors  we can compute the extrinsic curvatures as:
 \begin{align}
     K_{\:\:\:\bar{\mu}}^{(1)\:\bar{\nu}}&=0\label{intrinsic curvature 1}\\
      K_{\:\:\:\bar{z}}^{(2)\:\bar{z}}&=-\frac{z^3}{8\ell^3_{eff}l}+O(z^5)\\
       K_{\:\:\:\bar{x_3}}^{(2)\:\bar{x_3}}&=-\frac{z}{2\ell_{eff}l}+\frac{z^3}{16\ell^3_{eff}l}+O(z^5)\\
        K_{\:\:\:\bar{\theta}}^{(2)\:\bar{\theta}}&=\frac{z}{2\ell_{eff}l}+\frac{3z^3}{16\ell^3_{eff}l}+O(z^5)
      \label{intrinsic curvature 2}
 \end{align}
Now using \eqref{detsigma} we can obtain the Area of $\Sigma$:
\begin{align}
   \frac{\mathcal{A}[\Sigma]}{4G}&=\frac{1}{4G}\int_{\Sigma}d^3x\left(\frac{\ell^3_{eff}l}{z^3}-\frac{1}{8}\frac{\ell^3_{eff}}{lz}+O(z)\right)\nonumber\\
   &=\frac{2\pi H}{4G}\int_{\delta}^{z_{max}}dz\left(\frac{\ell^3_{eff}l}{z^3}-\frac{1}{8}\frac{\ell^3_{eff}}{lz}+O(z)\right)\nonumber\\
   &=\frac{\pi Hl}{2G}\int_{\delta/l}^{z_{max}/l}d\rho\left(\frac{\ell^3_{eff}}{l^2\rho^3}-\frac{1}{8}\frac{\ell^3_{eff}}{l^2\rho}+O(z)\right)\nonumber \\
   &=\frac{\pi H}{2G}\frac{\ell^3_{eff}}{l}\left(-\frac{1}{2\rho^2}-\frac{1}{8}\ln(\rho)\right)_{\delta/l}^{z_{max}/l} \nonumber\\
   &=\frac{\pi H}{4l G}\ell^3_{eff}\left(\frac{l^2}{\delta^2}-\frac{1}{4}\ln\left(\frac{l}{\delta}\right)\right)
\label{Area cylinder}
\end{align}
where the upper limit has been neglected because it only contributes to the finite part, which in $d=4$ is not universal. Now, for the cubic part of Eq.\eqref{minimal Spliting} and Eq.\eqref{non-minimal Spliting} and using the results for the extrinsic curvature we obtain:
 \begin{align}
      S_{R^2}=&-\frac{6}{\ell^4_{eff}}\left(3\mu_1+4\mu_2+8\mu_3+40\mu_4+16\mu_5+16\mu_6 +80\mu_7+400\mu_8\right) \\
      S_{K^2R}=& -\frac{z^2}{l^2\ell^4_{eff}}(40\mu_4+8\mu_3+12\mu_2-3\mu_1)+O(z^4)\\
      S_{K^4}^{min}=&S_{K^4}^{non-min}=O(z^4)
 \end{align}

 Replacing all this results in the entropy for CCG \eqref{minimal Spliting} or \eqref{non-minimal Spliting} we get:
 \begin{align}
    S_{CCG} =&\frac{\mathcal{A}[\Sigma]}{4 G}-\frac{1}{8 G}\int_{\Sigma}d{3}x\sqrt{\sigma}\bigg[\frac{6}{\ell^4_{eff}}\left(3\mu_1+4\mu_2+8\mu_3+40\mu_4+16\mu_5\right.\nonumber\\
    &\left.+16\mu_6 +80\mu_7+400\mu_8\right)-\frac{z^2}{l^2 \ell^4_{eff}}(40\mu_4+8\mu_3+12\mu_2-3\mu_1)+O(z^4)\bigg]\nonumber \\
    &=\frac{a_4\mathcal{A}[\Sigma]}{4 G}+\frac{1}{8G}\int_{\Sigma}dx^{3}\left(\frac{\ell^3_{eff}l}{z^3}+\frac{1}{8}\frac{\ell^3_{eff}}{lz}\right)\nonumber\\
    &\times\left(\frac{z^2}{l^2 \ell^4_{eff}}(40\mu_4+8\mu_3+12\mu_2-3\mu_1)\right)+O(z)\nonumber\\
    &=\frac{a_4\mathcal{A}[\Sigma]}{4 G}+\frac{1}{8G}\int_{\Sigma}dx^{3}\left(\frac{\ell^3_{eff}l}{z^3}+\frac{1}{8}\frac{\ell^3_{eff}}{lz}+O(z)\right)\nonumber\\
    &\times\left(\frac{z^2}{l^2 \ell^4_{eff}}(40\mu_4+8\mu_3+12\mu_2-3\mu_1)+O(z^4)\right)\nonumber\\
    &=\frac{a_4\mathcal{A}[\Sigma]}{4 G}+\frac{\pi H\left(40\mu_4+8\mu_3+12\mu_2-3\mu_1\right)}{4Gl\: \ell_{eff}}\int_{\delta}^{z_{max}}dz \left(\frac{1}{z}+O(z)\right)\nonumber\\
    &=\frac{a_4\mathcal{A}[\Sigma]}{4G}+\frac{\pi H(40\mu_4+8\mu_3+12\mu_2-3\mu_1)}{4Gl\: \ell_{eff}}\ln{\left(\frac{l}{\delta}\right)}+O(\delta)
 \end{align}
Now we replace the area formula  \eqref{Area cylinder} in this last equation:
\begin{align}
  S_{CCG}&=\frac{a_4\pi H}{4l G}\ell^3_{eff}\left(\frac{l^2}{\delta^2}-\frac{1}{4}\ln\left(\frac{l}{\delta}\right)\right)\nonumber\\
  &+\frac{\pi H\left(40\mu_4+8\mu_3+12\mu_2-3\mu_1\right)}{4Gl\: \ell_{eff}}\ln{\left(\frac{l}{\delta}\right)}+O(\delta)\nonumber\\
  &=\frac{\pi H}{4l G}\ell^3_{eff}\bigg(\frac{a_4l^2}{\delta^2}-\frac{1}{4}\bigg(a_4-\frac{4}{\ell^2_{eff}}\left(40\mu_4+8\mu_3+12\mu_2-3\mu_1\bigg)\bigg)\ln\left(\frac{l}{\delta}\right)\right)\nonumber\\
  &=\frac{\pi H}{4l G}\ell^3_{eff}\left(\frac{a_4l^2}{\delta^2}-\frac{1}{4}b_4\ln\left(\frac{l}{\delta}\right)\right)
\end{align}
where $b_4$ and $a_4$ is defined in Eq.\eqref{a4} and Eq.\eqref{b4}. As we have mention before, this last expression has a divergent term that is not part of the logarithmic divergences and therefore we need to use the Kounterterms to get rid of it. For that we start with  \eqref{Poincare for cylinder appendix} and take  $z=\delta$ because the Kounterterms are  defined at the boundary:
  \begin{equation}
    ds_{\partial\Sigma}^2=\tilde{\sigma}_{\bar{a}\bar{b}}dY^{\bar{a}}dY^{\bar{b}}=\frac{\ell^2_{eff}}{\delta^2}\left(dx_3^2+\left(l-\frac{\delta^2}{4l}\right)^2d\theta^2\right) \label{induced metric Kounterterm}
\end{equation}
 with normal vector:
 \begin{equation}
     n^{(3)}_z=-\frac{\ell_{eff}}{z}\sqrt{1+\frac{z^2}{4l^2}}
 \end{equation}
Now we compute the determinant of the metric of cod-3 and the extrinsic curvature needed to obtain the Kounterterm $B_2=tr(\kappa)$:
\begin{equation}
    \sqrt{\tilde{\sigma}}=\frac{\ell^2_{eff} l}{\delta^2}-\frac{\ell^2_{eff}}{4l}+O(\delta^2)  \:\:\:\:;\:\:\:\:Tr(\kappa)=\frac{2}{\ell_{eff}}+\frac{1}{4}\frac{\delta^2}{\ell_{eff}l^2}+O(\delta^4)\label{sqrt root and trace cylinder}\end{equation}
therefore:
\begin{equation}
 B_2=-2\int_{0}^{1}ds\int_{0}^{s}dtTr(\kappa)=-2\left(\frac{2}{\ell_{eff}}+\frac{1}{4}\frac{\delta^2}{\ell_{eff}l^2}+O(\delta^4)\right)
\end{equation}

Now we replace \eqref{sqrt root and trace cylinder} into the definition of what we obtained in Eq.\eqref{HEE in CCG Kounterterm}:
 \begin{align}
  S_{KT}&=\frac{c_4}{4G}\bigg\lfloor\frac{4+1}{2}\bigg\rfloor\int_{\partial\Sigma}d^2x\sqrt{\tilde{\sigma}}B_2\nonumber\\
  &=\frac{c_4}{2G}\int_0^Hdx_3\int_{0}^{2\pi}d\theta\left(\frac{\ell^2_{eff} l}{\delta^2}-\frac{\ell^2_{eff}}{4l}+O(\delta^2) \right)\nonumber\\&\times\left[-2\left(\frac{2}{\ell_{eff}}+\frac{1}{4}\frac{\delta^2}{\ell_{eff}l^2}+O(\delta^4)\right)\right]   \nonumber  \\
  &=-\frac{2c_4\pi H }{G}\frac{\ell_{eff}l}{\delta^{2}}+O(\delta),
 \end{align}
 and replacing $c_4=\frac{a_4\ell^2_{eff}}{8}$, we get
 \begin{equation}
  S_{KT}=  -\frac{a_4\pi H }{4G}\frac{\ell^3_{eff}l}{\delta^{2}}+O(\delta).
 \end{equation}
Finally, we can use this result to obtain the universal part of the EE in a cylindrical entangling region as:
 \begin{equation}
     S_{CCG}^{Univ}=S_{CCG}+S_{KT}=-b_4\frac{\pi H\ell^3_{eff}}{16l G}\ln\left(\frac{l}{\delta}\right).
 \end{equation}

\chapter{Explicit calculation: EE for the Deformed sphere} \label{Explicit calculation: EE for the Deformed sphere}

In this Appendix we will show the explicit calculation of the $C_T$ in a CCG theory by considering a deformed spherical entanglement region. We start using the embedding of Eq.\eqref{embedding deformed sphere} to compute the squared root of the determinant of the metric:
\begin{align}
   \sqrt{\sigma}&=\frac{\ell_{eff}^{2}}{z^{2}}\left[1+\sum_{l}\bigg(\frac{1}{(1-z^2)}\left(\frac{1-z}{z+1}\right)^{\frac{l}{2}} \left((l^{2}-1) z^{2}+l z+1\right) \left(a_l \cos\! \left(l \phi\right)+b_l \sin\! \left(l \phi\right)\right)  \epsilon\right.\nonumber\\ 
   &-\frac{1}{2 \left(1-z^2\right)^{2}} \left(\frac{1-z}{z+1}\right)^{l}\bigg(  \left(a_l^2-b_l^2\right)  \left(z^{2}+2lz+1\right) \left((1-2l^2)z^2+l^2\right) \cos\! \left(l \phi\right)^{2}\nonumber\\
   &+2 \left(z^2+2lz+1\right)  \left((1-2l^2)z^2+l^2\right) a_l\sin\! \left(l \phi\right) b_l \cos\! \left(l \phi\right)\\
   &+z^{2} \left(z^2-1\right) \left(a_l^{2}+b_l^{2}\right) l^{4}+2\left(\left( a_l^{2}- b_l^{2}\right) z^{3}-a_l^{2} z\right) l^{3}\nonumber\\&+\left(-2 b_l^{2} z^{4}+a_l^{2} z^{2}-a_l^{2}\right) l^{2}+2 b_l^{2} l \,z^{3}+b_l^{2} z^{2} \left(z^{2}+1\right)\bigg)\epsilon^2+ \mathrm{O}\! \left(\epsilon^{3}\right)\bigg)\bigg]\nonumber
\end{align}

We use the same embedding to compute the quantities of interest in the computation of the EE in CCG and the results are  the Eq. \eqref{SR2 deformed CT}-\eqref{SK4 deformed CT}. We will show them here for continuity: 
\begin{align}
    S_{R^2}&=-\frac{6 \left(2 \mu_\mathrm{1}+4 \mu_\mathrm{2}+6 \mu_\mathrm{3}+24 \mu_\mathrm{4}+9 \mu_\mathrm{5}+9 \mu_\mathrm{6}+36 \mu_\mathrm{7}+144 \mu_\mathrm{8}\right)}{\ell_{eff}^{4}}\\
    S_{K^2R}&= \frac{12z^4 (\mu_1 -4 \mu_2 -2 \mu_3 -8 \mu_4)}{\ell_{eff}^4(1-z^2)^2}\sum_{l}\left(\frac{1-z}{z +1}\right)^{l}(a_l^{2}+b_l^{2}) l^{2} (l^2 -1)^{2} \epsilon^{2}\\
    &+\mathrm{O}\! \left(\epsilon^{3}\right)\nonumber\\
    S_{K^4}^{min}&=\mathrm{O}\! \left(\epsilon^{4}\right)\\
     S_{K^4}^{non-min}&=\mathrm{O}\! \left(\epsilon^{4}\right)
\end{align}
We can see from here that at least at order $\epsilon^2$ the two splittings coincide, that is $S_{CCG}^{min}=S_{CCG}^{non-min}=S_{CCG}$. Replacing all this results in the computation for the EE:

\begin{align}
   &S_{CCG}=\frac{\mathcal{A}[\Sigma]}{4G}-\frac{1}{8G}\int_{\Sigma}dx^2\sqrt{\sigma}\left(S_{R2}+S_{K^2R}+S_{K^4}\right)\nonumber \\
   &=\frac{1}{4G}\int_{\Sigma}dx^2\sqrt{\sigma}\left(1-\frac{1}{2}\left(S_{R2}+S_{K^2R}+S_{K^4}\right)\right)\nonumber\\
   &=\frac{1}{4G}\int_{\Sigma}dx^2\left(\frac{\ell_{eff}^{2}}{z^{2}}\bigg[1+\sum_{l}\bigg(\frac{1}{(1-z^2)}\left(\frac{1-z}{z+1}\right)^{\frac{l}{2}} \left((l^{2}-1) z^{2}+l z+1\right) \left(a_l \cos\! \left(l \phi\right)\right.\right.\nonumber\\
   &\left.+b_l \sin\! \left(l \phi\right)\right)  \epsilon-\frac{1}{2 \left(1-z^2\right)^{2}} \left(\frac{1-z}{z+1}\right)^{l}\bigg(  \left(a_l^2-b_l^2\right)  \left(z^{2}+2lz+1\right) \left((1-2l^2)z^2\right.\nonumber\\
   &\left.+l^2\right) \cos\! \left(l \phi\right)^{2}+2 \left(z^2+2lz+1\right)  \left((1-2l^2)z^2+l^2\right) a_l\sin\! \left(l \phi\right) b_l \cos\! \left(l \phi\right)\nonumber\\
   &+z^{2} \left(z^2-1\right) \left(a_l^{2}+b_l^{2}\right) l^{4}+2\left(\left( a_l^{2}- b_l^{2}\right) z^{3}-a_l^{2} z\right) l^{3}+\left(-2 b_l^{2} z^{4}+a_l^{2} z^{2}-a_l^{2}\right) l^{2}+2 b_l^{2} l \,z^{3}\nonumber\\
   &+b_l^{2} z^{2} \left(z^{2}+1\right)\bigg)\epsilon^2\left.+ \mathrm{O}\! \left(\epsilon^{3}\right)\bigg)\bigg]\right)\bigg(a_3\nonumber\\
   &-\frac{6z^4(\mu_1 -4 \mu_2 -2 \mu_3 -8 \mu_4) }{\ell_{eff}^4(1-z^2)^2}\sum_{l}\left(\frac{1-z}{z +1}\right)^{l} (a_l^{2}+b_l^{2})l^{2} (l^2 -1)^{2} \epsilon^{2}+\mathrm{O}\! \left(\epsilon^{3}\right)\bigg)\nonumber\\
   &=\frac{\ell_{eff}^2}{4G}\int_0^{2\pi} d\phi\int_\delta^1dz\bigg[\frac{1}{z^2}\bigg(a_3+\frac{a_3}{(1-z^2)}\sum_{l}\left(\frac{1-z}{z+1}\right)^{\frac{l}{2}} \left((l^{2}-1) z^{2}+l z+1\right) \left(a_l \cos\! \left(l \phi\right)\right.\nonumber\\
   &+\left.b_l \sin\! \left(l \phi\right)\right)  \epsilon -\frac{1}{2 \left(1-z^2\right)^{2}} \sum_{l}\left(\frac{1-z}{z+1}\right)^{l}\bigg(a_3\bigg(  \left(a_l^2-b_l^2\right)  \left(z^{2}+2lz+1\right) \left((1-2l^2)z^2\right.\nonumber\\
   &\left.+l^2\right) \cos\! \left(l \phi\right)^{2}+2 \left(z^2+2lz+1\right)  \left((1-2l^2)z^2+l^2\right) a_l\sin\! \left(l \phi\right) b_l \cos\! \left(l \phi\right)\nonumber\\
   &+z^{2} \left(z^2-1\right) \left(a_l^{2}+b_l^{2}\right) l^{4} 
   +2\left(\left( a_l^{2}- b_l^{2}\right) z^{3}-a_l^{2} z\right) l^{3}+\left(-2 b_l^{2} z^{4}+a_l^{2} z^{2}-a_l^{2}\right) l^{2}+2 b_l^{2} l \,z^{3}\nonumber\\
   &+b_l^{2} z^{2} \left(z^{2}+1\right)\bigg)+\frac{12z^4(a_l^{2}+b_l^{2}) }{L^4}l^{2} (l^2 -1)^{2}(\mu_1 -4 \mu_2 -2 \mu_3 -8 \mu_4)\bigg)\epsilon^2\bigg)\bigg] \nonumber\\
   &=\frac{\pi \ell_{eff}^2}{2G}\int_\delta^1dz\bigg[\frac{1}{z^2}\bigg(a_3-\frac{1}{2 \left(1-z^2\right)^{2}} \sum_{l}\left(\frac{1-z}{z+1}\right)^{l}\bigg(a_3\bigg(  \frac{1}{2}\left(a_l^2-b_l^2\right)  \left(z^{2}+2lz+1\right)\nonumber\\
   &\times\left((1-2l^2)z^2+l^2\right)+z^{2} \left(z^2-1\right) \left(a_l^{2}+b_l^{2}\right) l^{4}+2\left(\left( a_l^{2}- b_l^{2}\right) z^{3}-a_l^{2} z\right) l^{3}\nonumber\\
   &+\left(-2 b_l^{2} z^{4}+a_l^{2} z^{2}-a_l^{2}\right) l^{2} +2 b_l^{2} l \,z^{3}+b_l^{2} z^{2} \left(z^{2}+1\right)\bigg)\nonumber\\
   &+\frac{12z^4(a_l^{2}+b_l^{2}) }{\ell_{eff}^4}l^{2} (l^2 -1)^{2}(\mu_1 -4 \mu_2 -2 \mu_3 -8 \mu_4)\bigg)\epsilon^2\bigg)
   +O(\epsilon^3)\bigg] \nonumber\\
   &=-\frac{a_3\pi \ell_{eff}^2}{2G}\left(1-\frac{1}{\delta}\right)-\frac{\pi \ell_{eff}^2}{8G}\sum_l\bigg[(a_l^2+b_l^2)l(l^2-1)\bigg(1+\frac{3}{\ell_{eff}^4}(4\mu_1-4\mu_2\nonumber\\
   &+2\mu_3+8\mu_4+9\mu_5+9\mu_6+36\mu_7+144\mu_8)\bigg)-\frac{a_3}{\delta}(a^2+b^2)l^2\bigg]\epsilon^2+O(\epsilon^3)
\end{align}
Now we are going to construct the Kouterterm. First we set $z=delta$ to obtain the metric for the entangling region in the boundary:
  \begin{align}
   ds_{\partial\Sigma}^2&=\frac{\ell_{eff}^2}{\delta^2}\bigg((1-\delta^2+O(\delta^3))+\sum_l2\left(a_l \cos\! \left(l \phi\right)+b_l \sin\! \left(l \phi\right)\right)\left(1-\frac{1}{2} l^{2} \delta^{2}+\mathrm{O}\! \left(\delta^{3}\right)\right) \epsilon\nonumber\\
  &+\sum_l\left(\left(a_l^{2}-b_l^{2}\right) \left(l^{2}-1\right) \cos\! \left(l \phi\right)^{2}+2 a_l b_l\left(l^2-1\right) \sin\! \left(l \phi\right)   \cos\! \left(l \phi\right)\right.\nonumber\\
  &\left.-a_l^{2} l^{2}-b_l^{2}\right)\left(-1+\left(l^{2}-1\right) \delta^{2}+\mathrm{O}\! \left(\delta^{3}\right)\right)\epsilon^2 +O(\epsilon^3)\bigg)d\phi^2  
\end{align}
We can use this result to obtain the squared root of the determinant of the codimension-3 metric (we need it to obtain the Kounterterm):
\begin{align}
\sqrt{\tilde{\sigma}}&=\frac{\ell_{eff}}{\delta}\bigg(1-\sum_l\left(a_l \cos\! \left(l \phi\right)+b_l \sin\! \left(l \phi\right)\right)\epsilon\nonumber\\
&-\sum_l\frac{l^2}{2} \left(2 \cos\! \left(l \phi\right) \sin\! \left(l \phi\right) a_l b_l+(a_l^{2}-b_l^{2}) \cos\! \left(l \phi\right)^{2}-a_l^{2}\right)\epsilon^2\bigg)+O(\epsilon^3,\delta)    \label{metric Kounterterm deformed CT}
    \end{align}
On the other hand for $D=4$ the $S_{KT}$ is defined in codimension-3 and therefore we will need to compute $B_1$:
\begin{equation}
 B_1=-2\int_{0}^{1}dsTr(\kappa)=-\frac{2}{\ell_{eff}}+O(\epsilon^3,\delta)  \label{B1 deformed CT}
\end{equation}
Now we can replace \eqref{metric Kounterterm deformed CT} and \eqref{B1 deformed CT} in Eq.\eqref{HEE in CCG Kounterterm}
 \begin{align}
  S_{KT}&=\frac{c_3}{4G}\bigg\lfloor\frac{3+1}{2}\bigg\rfloor\int_{\partial\Sigma}dx^2\sqrt{\tilde{\sigma}}B_1\\
  &=\frac{c_3}{2G}\int_{0}^{2\pi}d\phi\bigg[\frac{\ell_{eff}}{\delta}\bigg(1-\sum_l\left(a_l \cos\! \left(l \phi\right)+b_l \sin\! \left(l \phi\right)\right)\epsilon\\
  &-\sum_l\frac{l^2}{2} \left(2 \cos\! \left(l \phi\right) \sin\! \left(l \phi\right) a_l b_l+(a_l^{2}-b_l^{2}) \cos\! \left(l \phi\right)^{2}-a_l^{2}\right)\epsilon^2\bigg)\left(-\frac{2}{\ell_{eff}}+O(\delta^2)\right)\bigg]\\
  &=-\frac{2\pi c_3}{G}\frac{1}{\delta}\left(1+\sum_l\frac{l^2}{4}(a_l^2+b_l^2)\epsilon^2\right)+O(\epsilon^3,\delta)  
 \end{align}
 we can express $c_4=\frac{a_3\ell_{eff}^2}{4}$, therefore:
 \begin{equation}
  S_{KT}=  -\frac{\pi \ell_{eff}^2 a_3}{2G}\frac{1}{\delta}\left(1+\sum_l\frac{l^2}{4}(a_l^2+b_l^2)\epsilon^2\right)+O(\epsilon^3,\delta)  
 \end{equation}
Notice that this result is the same (with a negative sign) as the divergent part of the EE, therefore:
 \begin{align}
     S_{CCG}^{Univ}=S_{CCG}+S_{KT}&=-\frac{\pi \ell_{eff}^2}{2G}\bigg[a_3+\sum_l\frac{1}{4}(a_l^2+b_l^2)l^2(l^2-1)\bigg(1+\frac{3}{\ell_{eff}^2}(4\mu_1-4\mu_2+2\mu_3\nonumber\\&+8\mu_4+9\mu_5+9\mu_6+36\mu_7+144\mu_8)\bigg)\epsilon^2\bigg] +O(\epsilon^4)
 \end{align}
If we compare with Eq.\eqref{SEE eps} we can express the subleading term ($\epsilon^2$ term) as:
\begin{equation}
     S_{EE}^{Univ,CCG,(2)}=\frac{\pi^4 C_T^{CCG}}{24}\sum_{l}l(l^2-1)(a_l^2+b_l^2)
\end{equation}
where $C_T^{CCG}$ is given in  Eq.\eqref{Ct CCG}. Now, as an example, let us calculate the $C_T$ for Einstein cubic gravity. We set the couplings of the general CCG theory to be :
\begin{equation}
  \mu_1=-\frac{3}{2}\mu  \:\: ; \:\: \mu_2=-\frac{\mu}{8} \:\:  ;\:\:\mu_3= 0 \:\: ; \:\: \mu_4= 0 \:\: ; \:\:\mu_5= \frac{3}{2}\mu  \:\: ;\:\:  \mu_6= -\mu \:\: ;\:\:\mu_7= 0  \:\:;\:\:\mu_8=0
\end{equation}
replacing in  Eq.\eqref{Ct CCG} we obtain:
\begin{equation}
    C_T^{ECG}=\bigg(1-\frac{3}{\ell_{eff}^4}\mu\bigg)C_T\label{Ct ECG}
\end{equation}